\documentclass[a4paper,12pt]{article} 

\usepackage{jheppub}
\usepackage{hyperref}

\usepackage{bm, bbm, bbold}
\usepackage{latexsym}
\usepackage{dcolumn}
\usepackage[T1]{fontenc}
\usepackage{comment}
\usepackage{graphicx,epsfig}
\usepackage{environ}
\usepackage{mathtools}
\usepackage{braket}
\usepackage{graphicx,epstopdf}
\usepackage{tikz}
\usepackage{float}
\usepackage{framed}
\usepackage{soul}
\usetikzlibrary{positioning,decorations.markings, decorations.pathmorphing, calc}
\tikzset{snake it/.style={decorate, decoration=snake}}

\usepackage{stmaryrd}
\usepackage{slashed}
\usepackage{array,multirow}

\usepackage[framemethod=default]{mdframed}
\newmdenv[skipabove=7pt,
skipbelow=7pt,
rightline=false,
leftline=false,
topline=false,
bottomline=false,
backgroundcolor=gray!10,
linecolor=gray,
innerleftmargin=5pt,
innerrightmargin=5pt,
innertopmargin=5pt,
innerbottommargin=5pt,
leftmargin=0cm,
rightmargin=0cm,
linewidth=4pt]{eBox}

\def\a{\alpha}
\def\b{\beta}

\def\e{\epsilon}

\def\p{\partial}
\def\tt{\tilde{\tau}}

\newcommand{\re}{\mbox{Re}}

\newcommand{\be}{\begin{equation}}
\newcommand{\ee}{\end{equation}}
\newcommand{\bes}{\begin{equation*}}
\newcommand{\ees}{\end{equation*}}

\newcommand{\scrip}{\mathcal I^+}

\def\Mpl([#1],[#2]){\textrm{Li}_{#1}\left(#2\right)}

\NewEnviron{derivation}{
\begin{framed}
\begin{center}
{\bf-: Derivation :-}\\
\end{center}
  \BODY

\end{framed}
}

\NewEnviron{eqn}{
\begin{align}
\begin{split}
  \BODY
\end{split}
\end{align}
}

\NewEnviron{eqn*}{
\begin{align*}
\begin{split}
  \BODY
\end{split}
\end{align*}
}

\newcommand{\intinf}{\int_{-\infty}^{\infty}} 
\newcommand{\intsinf}{\int_{0}^{\infty}} 

\newcommand{\reference}[1]{{\fontfamily{lmtt}\selectfont #1}}

\usepackage{cancel}
\usepackage{tikz, pgf}
\usetikzlibrary{shapes.misc}
\usetikzlibrary{shapes}
\usetikzlibrary{fit}
\usetikzlibrary{arrows}

\usetikzlibrary{decorations.pathmorphing}	
\usetikzlibrary{decorations.markings}
\tikzset{
    sugra/.style={decorate, decoration={snake}, draw=black},
    scalarphi/.style={dashed,draw=black, postaction={decorate},
        },
    hwbou/.style={draw=blue, postaction={decorate}, ultra thick
        },
    vector/.style={draw=blue,decorate, decoration={snake}, draw},
	provector/.style={decorate, decoration={snake,amplitude=2.5pt}, draw},
	antivector/.style={decorate, decoration={snake,amplitude=-2.5pt}, draw},
   	 fermion/.style={draw=cyan, postaction={decorate},
        decoration={markings,mark=at position .55 with {\arrow[draw=black]{>}}}},
    fermionbar/.style={draw=cyan, postaction={decorate},
        decoration={markings,mark=at position .55 with {\arrow[draw=black]{<}}}},
    fermionnoarrow/.style={draw=black},
    gluon/.style={decorate, draw=red,
        decoration={coil,amplitude=4pt, segment length=5pt}},
    scalar/.style={dashed,draw=black, postaction={decorate},
        decoration={markings,mark=at position .55 with {\arrow[draw=black]{>}}}},
    scalarbar/.style={dashed,draw=black, postaction={decorate},
        decoration={markings,mark=at position .55 with {\arrow[draw=black]{<}}}},
    electron/.style={draw=black, postaction={decorate},
        decoration={markings,mark=at position .55 with {\arrow[draw=black]{>}}}},
    scalarnoarrow/.style={dashed, draw=black},
    electron/.style={draw=black, postaction={decorate},
        decoration={markings, mark=at position .55 with {\arrow[draw=black]{>}}}},
	bigvector/.style={decorate, decoration={snake, amplitude=4pt}, draw},
    photon/.style={draw=red, decorate, decoration={zigzag}, draw},
    higgs/.style={dashed, draw=black, postaction={decorate},
        },	
        goldstone/.style={draw=brown, postaction={decorate},
        },    
          ghost/.style={dashed, draw=magenta, postaction={decorate},
        decoration={markings, mark=at position .55 with {\arrow[draw=black]{>}}}
        },  
          antighost/.style={dashed, draw=magenta, postaction={decorate},
        decoration={markings, mark=at position .55 with {\arrow[draw=black]{<}}}
        },  
          mphoton/.style={decorate, decoration={snake}, draw=violet},
            realscalar/.style={draw=black}, 
           mgluon/.style={decorate, draw=blue,
        decoration={coil,amplitude=4pt, segment length=5pt}},
         weylfermion/.style={draw=orange, postaction={decorate},
        decoration={markings,mark=at position .55 with {\arrow[draw=black]{>}}}},
         weylfermionbar/.style={draw=orange, postaction={decorate},
        decoration={markings,mark=at position .55 with {\arrow[draw=black]{<}}}}, 
   	wboson/.style={draw=blue,decorate, decoration={snake,amplitude=4pt}, draw},  
    zboson/.style={draw=violet, decorate, decoration={snake}, draw},   
    lepton/.style={draw=black, postaction={decorate},
        decoration={markings,mark=at position .55 with {\arrow[draw=black]{>}}}},
    leptonbar/.style={draw=black, postaction={decorate},
        decoration={markings,mark=at position .55 with {\arrow[draw=black]{<}}}}, 
        graviton/.style={draw=blue,decorate, decoration={snake,amplitude=4pt}, draw},  
        gravitino/.style={draw=red, postaction={decorate},
        decoration={snake, markings, mark=at position .55 with {\arrow[draw=black]{>}}}},
    gravitinobar/.style={draw=red, postaction={decorate},
        decoration={snake, markings,mark=at position .55 with {\arrow[draw=black]{<}}} },    
}

\newcommand{\vertex}[1]{\begin{tikzpicture}[baseline]
\node at (0,0) {\textbullet};
\node at (0,-0.25) {$#1$};
\end{tikzpicture}}

\usepackage{tikz-feynman}

\tikzfeynmanset{/tikzfeynman/momentum/arrow shorten=0.25}

\hypersetup{pagebackref,urlcolor=brown,citecolor=blue,linkcolor=purple}
\usepackage{amsthm,amsmath,latexsym,amssymb,amsfonts,amscd,soul}
\usepackage{mathrsfs}
\usepackage{mathtools,commath}
\usepackage{soul}
\usepackage{comment}
\usepackage{subcaption}
\usepackage{array}
\definecolor{bubbles}{rgb}{0.91, 1.0, 1.0}
\definecolor{aquamarine}{rgb}{0.5, 1.0, 0.83}
\definecolor{bubblegum}{rgb}{0.99, 0.76, 0.8}
\definecolor{blackbell}{rgb}{0.64, 0.64, 0.82}
\definecolor{dollarbill}{rgb}{0.72, 0.93, 0.6}
\definecolor{CBPurple}{RGB}{204, 121, 167}
\definecolor{CBBlue}{RGB}{0, 114, 178}   
\definecolor{CBOrange}{RGB}{230, 159, 0}

\title{Cosmological Dressing Rules}

\author[a]{Chandramouli Chowdhury,}
\author[b]{Arthur Lipstein,} 
\author[b]{Joe Marshall,}
\author[c]{Jiajie Mei,}
\author[d]{and Ivo Sachs}

\affiliation[a]{Mathematical Sciences and STAG Research Centre, University of Southampton, Highfield, Southampton SO17 1BJ, United Kingdom}
\affiliation[b]{Department of Mathematical Sciences, Durham University, Stockton Road, DH1 3LE, Durham, United Kingdom}
\affiliation[c] {Institute of Physics, University of Amsterdam, Amsterdam, 1098 XH, The Netherlands}
\affiliation[d]{Arnold-Sommerfeld-Center for Theoretical Physics, Ludwig-Maximilians-Universit\"at M\"unchen, Theresienstr. 37, D-80333 Munich, Germany}

\abstract{
The basic observables in cosmology are known as in-in correlators. Recent calculations have revealed that in-in correlators in four dimensional de Sitter space exhibit hidden simplicity stemming from a close relation to scattering amplitudes in flat space. In this paper we explain how to make this property manifest by dressing flat space Feynman diagrams with certain auxiliary propagators. These dressing rules are derived for conformally coupled and massless scalar theories and we show that they reproduce the same infrared divergences predicted by the Schwinger-Keldysh formalism.

}

\begin{document}
\begin{flushright}
	LMU-ASC 03/25\\
\end{flushright}
\maketitle

\section{Introduction}
The calculation of correlation functions in curved space time is complicated by the absence of Poincare invariance on which many of the powerful tools of computing flat space amplitudes rely. On the other hand, one should be able to reach a similar level of sophistication for highly symmetric curved spacetimes such as de Sitter (dS) space since it has the same number of isometries as Minkowski space. This background is of particular interest in cosmology because it approximately describes the early Universe during inflation \cite{Guth:1980zm,Linde:1981mu,Albrecht:1982wi}. Yet the standard tools of momentum space still have to be modified due to the absence of time-translation invariance, implying, in particular, that $in$ and $out$ vacua are not identical. The canonical approach for in-in calculations with time-dependent backgrounds is the Schwinger-Keldysh formalism \cite{Keldysh:1964ud, Schwinger:1960qe, Maldacena:2002vr, Weinberg:2005vy}. Alternatively, one may start with the cosmological wavefunction \cite{Hartle:1983ai}, which can be perturbatively obtained from analytic continuation of Witten diagrams in Anti de Sitter (AdS) space \cite{Maldacena:2002vr,McFadden:2009fg,Maldacena:2011nz}, 
and then compute cosmological correlators as expectation values. In neither of the two approaches are tools of flat space Feynman integrals manifestly applicable. Nevertheless, there has been a great deal of work in recent years to bootstrap wavefunction coefficients using ideas inspired by the study of scattering amplitudes in flat space \cite{Raju:2012zr,Arkani-Hamed:2017fdk,Arkani-Hamed:2018kmz,Benincasa:2018ssx,Sleight:2019hfp,Baumann:2020dch,Goodhew:2020hob,Gomez:2021qfd,Jazayeri:2021fvk,Melville:2021lst,Arkani-Hamed:2015bza,Bonifacio:2022vwa,Arkani-Hamed:2023kig,Arkani-Hamed:2024jbp,Melville:2024ove,Baumann:2024ttn,Glew:2025otn, Cespedes:2025dnq, Goodhew:2024eup}.
In this paper we will argue that in-in correlators are actually simpler and more closely related to scattering amplitudes than wavefunction coefficients.

There is an equivalent way to calculate cosmological correlators in terms of an effective action in Euclidean AdS, at the cost of introducing a doubled set of extra ($shadow$) fields \cite{Sleight:2020obc, Sleight:2021plv,DiPietro:2021sjt}. 
For the conformally coupled $\phi^4$ theory, \cite{Chowdhury:2023arc} showed that individual loop-level diagrams in this formalism have the same complexity as a wavefunction coefficients, but after summing over all contributing diagrams, a flat space structure appears somewhat miraculously, allowing one to use standard methods for flat space integrals for calculation. In subsequent work \cite{Donath:2024utn}, it was shown that in-in correlators of certain IR-finite scalar theories can be mapped to in-out correlators. For other recent work relating cosmological correlators to scattering amplitudes see \cite{Bonifacio:2021azc,Benincasa:2024leu}.


One may then wonder if the close relation to scattering amplitudes is a generic feature of cosmological correlators. The aim of this paper is to argue that this is indeed the case and to propose a set of rules that allow one to describe correlators in a large class of scalar field theories by dressing the corresponding flat space diagrams with auxiliary propagators. The dressing rules are formulated at the integrand level and hold to all orders in perturbation theory. Quite remarkably, in all the cases where we are able to carry out the integrals, we find that the analytic structure of the final result can also be read off from the dressing rules in a way that we will make more precise later on.   
To illustrate the power of the dressing rules,
we evaluate the tree-level five-point correlator of conformally coupled $\phi^3$ theory which has not appeared in the literature before (moreover the methods of \cite{Donath:2024utn} do not apply to this theory even though it is IR finite). We find that its analytic structure is considerably simpler than that of the wavefunction \cite{Hillman:2019wgh} and is made manifest by the dressing rules. 

The dressing rules are also applicable to massless scalars in de Sitter, which is perhaps of more phenomenological interest. Such correlation functions are well known to feature infrared divergencies \cite{Birrell:1982ix, Maldacena:2002vr, Tsamis:1993ub,Beneke:2012kn} which need to be regularized. In contrast to a hard cut-off in conformal time often used in the literature\footnote{We should note that for phenomenological applications, such as a finite duration of the inflationary phase, a non-invariant cut-off may be physically motivated \cite{Senatore:2009cf}. It is also worth noting that there is a modification of the hard-cutoff, known as the {\it comoving cutoff}, which can be used to obtain dS invariant correlators \cite{Baumgart:2019clc}.} \cite{Anninos:2014lwa},  we will primarily use dimensional regularization. It turns out that a dimensional regularization is required to derive the dressing rules from the shadow formalism\footnote{We would like to the thank the authors of \cite{DiPietro:2021sjt} for suggesting to consider non-integer dimensions in the shadow formalism.} and we show that with this regularization the dressing rules apply and are indeed equivalent to Schwinger-Keldysh calculations. Note that a regulator is also required for UV divergences. In general, standard regulators like a cut-off or dimensional regularization will break the de Sitter symmetry but there is an alternative known as analytic regularization which preserves this property, although it arises from a non-local deformation of the action. Nevertheless, dressing rules can also be defined for this regulator, as explained in Appendix \ref{app:Green}. 

The ability to express in-in correlators in terms of dressed flat space Feynman diagrams allows one to make use of standard techniques for computing flat space Feynman integrals, as demonstrated in Appendix \ref{app:explicit}. While dressing rules can also be defined for wavefunction coefficients, they take a more complicated form, which reflects the underlying additional complexity of wavefunctions, as discussed in appendix \ref{app:WFdress}. 
 
The paper is organized as follows. In section \ref{sec:review} we review the standard formalisms for computing in-in correlators in dS. Next, in section \ref{invitation} we illustrate the main ideas of the paper through a few simple examples. This section also provides a detailed summary of the ``dressing rules'' developed in the rest of the paper. In section \ref{sec:Dressing} we derive the dressing rules from the shadow effective action for a variety of theories by considering numerous examples. In section \ref{sec:fivept} these rules are then applied to the  conformally coupled $\phi^3$ theory to obtain a compact expression for the tree level five-point correlator. In sections \ref{sec:SK} and \ref{sec:WF} we compare the dressing rules with Schwinger-Keldysh formalism and the wavefunction approach, respectively, and find complete agreement in particular for IR divergent examples. Finally, we present our conclusions in section \ref{sec:conclusions}. We also provide some Appendices with various technical details. In particular we discuss propagators in \ref{app:Green}, provide a general derivation for the dressing rules for power law potentials in \ref{app:MasslessAuxProp}, calculate various integrals in \ref{app:Integrals} and \ref{app:explicit}, and derive dressing rules for wavefunction coefficients in \ref{app:WFdress}.

\section{Overview}\label{sec:overview}

In this section, we will briefly review cosmological wavefunctions and in-in correlators, and how to compute the latter using the shadow formalism. We then provide an invitation to the dressing rules, illustrating how they arise in a few simple examples.

 
 \subsection{Review}\label{sec:review}

We will work in the Poincar\'e patch of dS$_4$ equipped with the metric 
\begin{equation}
{\rm d}s^{2}=\ell_{dS}^2{-{\rm d}\eta^{2}+{\rm d}\vec{x}^{2}\over \eta^{2}},
\label{metric}
\end{equation}
where $\ell_{dS}$ denotes the curvature radius (which is henceforth set to $1$), $-\infty<\eta<0$ is the conformal time, and $\vec{x}$ denotes the Euclidean boundary directions with individual components $x^{i}$, $i=1,2,3$. Cosmological (or in-in) correlators of a scalar field $\phi$ 
are given by
\begin{equation}\label{eq:corrfromWF}
\mathcal{A}_n=\frac{\int\mathcal{D}\phi \, \phi(\vec{k}_{1})\cdots\phi(\vec{k}_{n})\left|\Psi\left[\phi\right]\right|^{2}}{\int\mathcal{D}\phi\left|\Psi\left[\phi\right]\right|^{2}},
\end{equation}
where $\phi$ represents the value of a generic bulk field in the future boundary Fourier transformed to momentum space, $\vec{k}_a$ with $a\in\left\{ 1,...,n\right\}$ are boundary momenta, and $\Psi\left[\phi\right]$ is the cosmological wavefunction~\cite{Hartle:1983ai}, which is a functional of $\phi$. The wavefunction $\Psi\left[\phi\right]$ in turn, has the  perturbative expansion
\begin{equation}\label{eq:WFdefn0}
\ln\Psi\left[\phi\right]=-\sum_{n=2}^{\infty}\frac{1}{n!}\int\prod_{i=1}^{n}\frac{{\rm d}^{d}k_{i}}{(2\pi)^{d}}\delta^d(\sum_i^n \vec{k}_i)\Psi_{n}\left(\vec{k}_{1},\dots,\vec{k}_{n}\right)\phi(\vec{k}_{1})\cdots\phi(\vec{k}_{n})\,,
\end{equation}
where $d$ is the number of boundary dimensions. In practice, we set $d=3+2 \epsilon$ in dimensional regularization. The wavefunction coefficients $\Psi_n$ can be related to CFT correlators in the future boundary~\cite{Maldacena:2002vr,Maldacena:2011nz,McFadden:2009fg,Bzowski:2012ih,Bzowski:2013sza,Bzowski:2017poo,Bzowski:2018fql,Bzowski:2020kfw}. 
Note that in de Sitter momentum space, momentum is conserved along the boundary but the total energy, defined as 
\begin{equation}
E=\sum_{a=1}^n k_a\;;\quad k_a = |\vec{k}_a|, 
\end{equation}
is not conserved. In the following, we will not write out the delta functions which impose momentum conservation along the boundary.

Alternatively, { \textit{in-in}} correlators can also be computed using the Schwinger-Keldysh formalism~\cite{Keldysh:1964ud, Schwinger:1960qe}. 
We will spell out the detailed Feynman rules for Schwinger-Keldysh computations in section \ref{sec:SK}. It was shown that the Schwinger-Keldysh formalism can be mapped via analytic continuation to Witten diagrams in Euclidean anti-de Sitter space (EAdS)~\cite{Sleight:2019mgd, Sleight:2020obc, Sleight:2021plv}. For a single scalar field with polynomial interactions in dS, the resulting Feynman rules are conveniently encoded by a \textit{shadow} effective action ~\cite{DiPietro:2021sjt} containing two fields, $\phi_+$ and $\phi_-$, given by 
\begin{multline}
    \label{eq:EAdS_action_gen_pot}
	iS
=\int\limits_0^{\infty}\frac{{\rm d} z {\rm d}^dx}{z^{d+1}}\left[\sin\left(\pi(\frac{d}{2}-\Delta_+)\right)\left((\partial\phi_+)^2\!-\!m^2{\phi_+}^2\right)\right.\cr+\sin\left(\pi(\frac{d}{2}-\Delta_- )\right)\left((\partial\phi_-)^2\!-\!m^2{\phi_-}^2\right)\cr
	\left.+e^{i\pi\frac{d-1}{2}}V\left(e^{-i\frac{\pi}{2}\Delta_+}\phi_+ +e^{-i\frac{\pi}{2}\Delta^-}\phi_-\right)
	+e^{-i\pi\frac{d-1}{2}}V\left(e^{i\frac{\pi}{2}\Delta_+}\phi_+ +e^{i\frac{\pi}{2}\Delta_-}\phi_-\right)\right],
\end{multline}
where $0<z<\infty$ is the radial cooridinate of EAdS, $V$ is the scalar potential, and $\Delta_{\pm}$ are the scaling dimensions of the dual CFT operators, which are related to the mass of the scalar fields $\phi_{\pm}$ via 
\begin{align}
    \Delta_{\pm}=\frac{d}{2}\pm\sqrt{\left(\frac{d}{2}\right)^{2}-m^{2}}\,.
\label{scalingdim}
\end{align}
Note that the kinetic terms of $\phi_{\pm}$ have opposite sign. The propagators of ghost-like fields have an additional minus sign, as we will spell out below.

\begin{figure}
    \centering
    \begin{subfigure}{0.45\textwidth}
    \centering
        \begin{tikzpicture}[baseline=(b.base)]
            \begin{feynman}
            \vertex (a) ;
            \vertex [ right = 50pt of a       ] (i1) ;
            \vertex [ right = 50pt of i1      ] (i2);
            \vertex [right = 75pt of a      ] (b);
            \vertex [right = 75pt of b      ] (c);
            \vertex [above = 75pt of b](l1)
            ;
    
            \diagram* {
                (l1) -- [scalar](b),

            };
            \draw[thick] ($(a)!0.5!(c)$) circle [radius=75pt
            ];
            \end{feynman}
        \end{tikzpicture}
        \caption{$\phi^-$ bulk-boundary propagator.}
    \end{subfigure}
    \begin{subfigure}{0.45\textwidth}
    \centering
        \begin{tikzpicture}[baseline=(b.base)]
        \begin{feynman}
            \vertex (a) ;
            \vertex [ right = 50pt of a       ] (i1) ;
            \vertex [ right = 50pt of i1      ] (i2);
            \vertex [right = 75pt of a      ] (b);
            \vertex [right = 75pt of b      ] (c);
    
            \diagram* {
                (i1) --
                (i2)
            };
            \draw[thick] ($(a)!0.5!(c)$) circle [radius=75pt];
        \end{feynman}
    \end{tikzpicture}
    \caption{$\phi^+$ bulk-bulk propagator.}
    \end{subfigure}
    \begin{subfigure}{0.45\textwidth}
    \centering
        \begin{tikzpicture}[baseline=(b.base)]
        \begin{feynman}
            \vertex (a) ;
            \vertex [ right = 50pt of a       ] (i1) ;
            \vertex [ right = 50pt of i1      ] (i2);
            \vertex [right = 75pt of a      ] (b);
            \vertex [right = 75pt of b      ] (c);
    
            \diagram* {
                (i1) --[scalar, edge 
                ] (i2)
            };
            \draw[thick] ($(a)!0.5!(c)$) circle [radius=75pt];
        \end{feynman}
    \end{tikzpicture}
    \caption{$\phi^-$ bulk-bulk propagator.}
    \end{subfigure}
    \caption{Diagrammatic representation of propagators which feature in the shadow formalism, the formulae for which are given in \eqref{eq:EAdSProps}. $\phi_+$ fields have $\nu>0$ and $\phi_-$ fields have $\nu<0$. Only the latter appear as external legs. \label{fig:ShadowProps}} 
\end{figure}
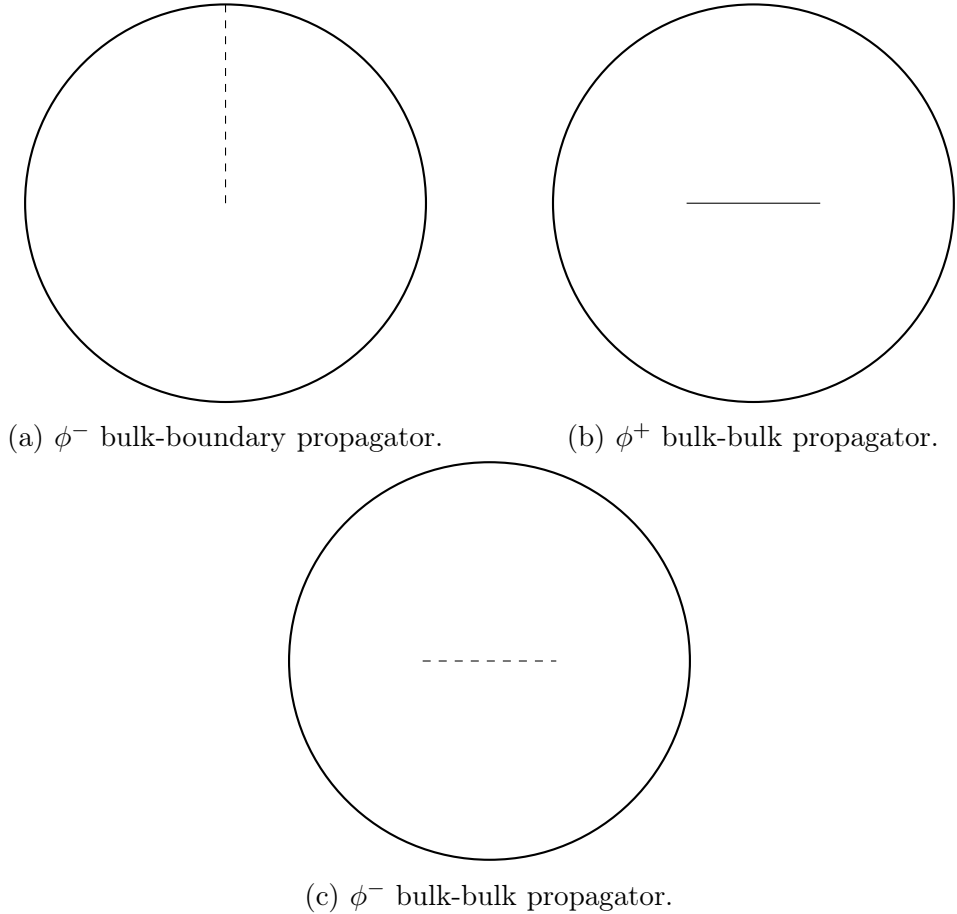

The effective action above describes how cosmological correlation functions in dS are expressed in terms of EAdS fields. The propagators arising from the EAdS action are given for conformally coupled and massless fields as \cite{Liu:1998ty,Raju:2011mp}
\begin{eqn}
\mathcal B_\nu(z, \vec k) &= -\sin\left(\pi \nu  \right)z^{d/2} \frac{2^{ -\nu} k^{\nu} }{\Gamma(1 + \nu)}  K_{\nu  }(k z)~,\\
G_\nu(z, z'; \vec k) &=-\sin\left(\pi \nu  \right)\int dp \, \frac{p}{p^2+\vec{k}^2}(z\, z')^{d/2}J_\nu (p z) J_\nu(p z') \\
&=-\sin(\pi \nu)(z z')^{d/2} \Bigg\{ \Theta(z - z') K_\nu(k z) I_\nu(k z') + \Theta(z' - z) I_\nu(k z) K_\nu(k z')  \Bigg\}~,
\label{eq:EAdSProps}
\end{eqn}
where $K_{\nu}$ and $J_{\nu}$ are Bessel functions. The overall sign implements whether the field is ghost-like or not. The propagators are depicted in Figure\ \ref{fig:ShadowProps}. In particular $\mathcal B_\nu(z, \vec k)$ denotes a bulk-boundary propagator and $ G_\nu(z, z'; \vec k)$ denotes the bulk-bulk propagator with $\nu = \Delta - \frac{d}{2}$. When performing dimensional regularization, we shift $d=3\rightarrow d=3+2 \epsilon$ and $\Delta \rightarrow \Delta+ \epsilon $ such that $\nu$ is preserved. For conformally coupled scalars $\nu=\pm 1/2$, and for massless scalars $\nu=\pm 3/2$. 
For $\nu=-3/2$, the integrand in \eqref{eq:EAdSProps} has a pole at $p=0$, but we define the contour of the $p$ integral to exclude this pole (see appendix \ref{app:Green} for more details).

In practice, we only consider diagrams with $\phi_-$ fields in the external legs, because these diagrams give the dominant contribution as the endpoints of the diagrams approach the future boundary, although subleading contributions are also of interest for studying the dS/CFT correspondence \cite{Heckelbacher:2022fbx}.

In the following, we will neglect a prefactor of $\eta_0^{\Delta_{-}} k_a^{2\Delta_--d}$ for each external leg (where $\eta_0$ denotes the end of inflation). It is understood that all of the formulae that follow should be dressed with this prefactor. The factor of $\eta_0^{\Delta_{-}}$ is not to be confused with a cutoff in the radial integrals which is typically introduced to regularize infrared divergences. Instead, as discuss below, we use dimensional regularisation to regularize such divergences.   
Throughout this paper we will use the following conventions:

\begin{align}
    k_i \quad \quad \quad &\text{Energy of $i$\textsuperscript{th} external leg.} \nonumber \\
    p_i \quad \quad \quad &\text{Energy of $i$\textsuperscript{th} internal non-auxiliary propagator.} \nonumber \\
    k_{ \rm ext} \quad \quad \quad  &\text{Sum of $k_i$'s at a vertex.} \nonumber \\
    p_{\rm tot} \quad \quad \quad &\text{Sum of $p_i$'s at a vertex.} \nonumber \\
    k_{ij\cdots} \quad \quad \quad &|\vec{k}_i|+|\vec{k}_j| + \cdots \nonumber \\ 
    \vec{y}_{ij\cdots} \quad \quad \quad &\vec{k}_i + \vec{k}_j +\cdots \nonumber\\
    \label{eq:Definitions}
\end{align}

\subsection{Invitation to dressing rules} \label{invitation}

In this paper, we will present a simple set of rules for obtaining in-in correlators of conformally coupled and massless scalars by dressing flat space Feynman diagrams with auxiliary propagators. We will then argue that this provides a systematic way to determine the analytic structure of these correlators. In this section we would like to first motivate the existence of such rules by looking at two simple examples.

It was shown in \cite{Chowdhury:2023arc} that the $s$-channel contribution to the 1-loop 4-point in-in correlator in conformally coupled $\phi^4$ theory is given by 
\begin{equation}
\mathcal{A}_4=\int\limits _{-\infty}^{\infty}dp\frac{k_{12}k_{34}}{(k_{12}^{2}+p^{2})(k_{34}^{2}+p^{2})}\int d^{4}L\frac{1}{L^{2}(L+K)^{2}}\,,
\label{1loopbubble}
\end{equation}
where $K^{\mu}=\left(p,\vec{y}_{12}\right)^\mu$, $\vec y_{12} = \vec k_1 + \vec k_2$, and $k_{ij}=k_i+k_j=|\vec{k}_i|+|\vec k_j|$. The loop integral needs to be regularized \cite{Chowdhury:2023arc}, but we keep this aside for illustration\footnote{We discuss the regularized expression in Appendices \ref{app:Green} and \ref{app:explicit}.}. Note that the integrand factorizes into a 4d Lorentz-invariant integrand dressed with an auxiliary integral with poles corresponding to external energies of the diagram. This simplicity arises from non-trivial cancellations after summing over two diagrams in the shadow formalism and does not occur for wavefunction coefficients (see appendix \ref{app:WFdress} for a discussion of wavefunction coefficients from this point of view). These simplifications can also be understood from non-trivial interference of the phases derived in \cite{Sleight:2020obc,Sleight:2021plv}. 

Let us now take a step back and think about how to directly arrive at \eqref{1loopbubble} without having to sum over different diagrams. 
We recall that energy is not conserved in de Sitter space since the metric has non-trivial time dependence. We may then naively guess that a physical observable can be obtained by taking a flat space Feynman diagram and dressing it with auxiliary propagators which encode the non-conservation of energy at each vertex:
 \begin{eqn}
 & \begin{tikzpicture}[baseline=(b.base)]
        \begin{feynman}
            \vertex (a);
            \vertex [right = 75pt of a       ] (b) ;
            \vertex [right = 75pt of b        ] (c) ;
            \vertex [above left = 26.39pt and 63.71pt of b](l1) {$\vec{k_1}$};
            \vertex [below left = 26.39pt and 63.71pt of b](l2) {$\vec{k_2}$};
            \vertex [above right = 26.39pt and 63.71pt of b](r1) {$\vec{k_3}$};
            \vertex [below right = 26.39pt and 63.71pt of b](r2) {$\vec{k_4}$};
            \vertex [right = 40pt of a       ] (i1);
            \vertex [left = 40pt of c       ](i2);
            \vertex [above = 50pt of b](e) ;
    
            \diagram* {
                (l1) -- (i1),
                (l2) -- (i1),
                (r1) -- (i2),
                (r2) -- (i2),
                (i1) --[ momentum' = $L$, half right,looseness=0.8  ] (i2),
                (i2) --[ momentum= $L+K$, half right, looseness=0.8  ] (i1),
                (e) --[scalar,momentum = $p$](i2),
                (i1) --[scalar, momentum = $p$](e),
            };
        \end{feynman}
    \end{tikzpicture}
\end{eqn}
where $p$ is a one-dimensional variable that encodes the energy of $K^\mu$ as defined below \eqref{1loopbubble}, and we impose energy conservation at the point where the two auxiliary propagators meet. At this point, it is not obvious that such a prescription will lead to anything physically sensible, but looking at \eqref{1loopbubble} we see that there is indeed a simple choice of auxiliary propagator which yields something observable, notably
\begin{equation}
\frac{k}{p^{2}+k^{2}},
\label{ccphi4prop}
\end{equation}
and the observable in question is precisely the in-in correlator! The main point of this paper is to show that this is not an accident of a particular diagram or the conformally coupled $\phi^4$ theory. Rather, we find that that in-in correlators for conformally coupled and massless scalar theories with more general polynomial interactions can be obtained by dressing flat space Feynman diagrams with simple auxiliary propagators, although the particular form of the propagators is theory-dependent. Note that the auxiliary propagators do not actually describe propagating degrees of freedom, but we nevertheless refer to them in this way because of their resemblance to one-dimensional Feynman propagators.

As another illustration of this point, let us next consider the tree-level 4-point in-in correlator in the conformally coupled $\phi^3$ theory. Using the Schwinger-Keldysh formalism, this was found to be a sum of two terms \cite{Arkani-Hamed:2015bza}: 
\begin{equation}
\mathcal{A}_4=2\mathcal{T}_{1}+\mathcal{T}_{2},
\end{equation}
where
\begin{align}
 \mathcal{T}_1 &= \frac{ 1}{2y_{12}} \Big[ \mbox{Li}_2 \left(  \frac{k_{34} - y_{12}}{k_{1234}} \right) + \mbox{Li}_2 \left(\frac{k_{12} - y_{12}}{k_{1234}} \right) + \log \frac{k_{12} + y_{12}}{k_{1234}}\log \frac{k_{34} + y_{12}}{k_{1234}} - \frac{\pi^2}{6} \Big],\\
\mathcal{T}_2 &=\frac{\pi^2}{2y_{12}},
\label{4ptphi3tree}
\end{align}
where $y_{12}=\left|\vec{y}_{12}\right|$  For illustration let us focus on $\mathcal{T}_1$ and show that it can be obtained by dressing a flat space Feynman diagram with certain auxiliary propagators. We will see in the next section that $\mathcal{T}_2$ arises by dressing of a flat space diagram as well. In particular, $\mathcal{T}_1$ is given by the following dressed flat space Feynman diagram:
\begin{eqn}
  \mathcal{T}_{1} &{}
       =\begin{tikzpicture}[baseline=(b.base)]
        \begin{feynman}
            \vertex (a) ;
            \vertex [ right = 50pt of a       ] (i1) ;
            \vertex [ right = 50pt of i1      ] (i2);
            \vertex [right = 75pt of a      ] (b);
            \vertex [right = 75pt of b      ] (c);
            \vertex [above left = 26.39pt and 63.71pt of b](l1) {\(\vec{k_1}\)};
            \vertex [below left = 26.39pt and 63.71pt of b](l2) {\(\vec{k_2}\)};
            \vertex [below right = 26.39pt and 63.71pt of b](r1) {\(\vec{k_3}\)};
            \vertex [above right = 26.39pt and 63.71pt of b](r2) {\(\vec{k_4}\)};
            \vertex [above = 50pt of b](e);
    
            \diagram* {
                (l1) --(i1),
                (l2) -- (i1),
                (r1)  --(i2),
                (r2)  --(i2),
                (i1) --[edge label=$p {,}\;\vec{y}_{12}$] (i2),
                (i1) --[scalar, edge label=$p$] (e),
                (i2) --[scalar, edge label'=$p$] (e),
            };
        \end{feynman}
    \end{tikzpicture}
\label{dressed4ptphi3tree}\\
&=\int_{-\infty}^{\infty}dp\int_0^{\infty}ds_{1}\int_0^{\infty}ds_{2}\frac{p}{p^{2}+(s_1+k_{12})^{2}}\frac{p}{p^{2}+(s_2+k_{34})^{2}}\frac{1}{p^{2}+y_{12}^{2}}\,.
\end{eqn}
From this we read off the the following form of the auxiliary propagators required for the conformally coupled $\phi^3$ theory:
\begin{equation}
\int_0^{\infty} ds \,\frac{p}{p^{2}+(s+k_{\rm ext})^{2}}\,.
\end{equation}
The integral over $s$ leads to a log and since the diagram in \eqref{dressed4ptphi3tree} contains two such auxiliary propagators this evaluates to a dilog. This suggests that the transcendentality of the in-in correlator can be read off from the number of log-type auxiliary propagators. More precisely, it is given by the transcendentality of the flat space Feynman diagram plus the number of log-type auxiliary propagators. 
In the conformally coupled $\phi^4$ theory, the auxiliary propagators do not have an $s$ integral so one may expect that the in-in correlators have the same transcendentality as flat space Feynman diagrams and this is indeed the case for known examples \cite{Chowdhury:2023arc}. Hence, the dressing rules not only provide the relation of in-in correlators to scattering amplitudes, but make their analytic structure manifest. 

\section{Dressing rules \label{sec:Dressing}}

In this section, we will spell out the dressing rules more systematically and illustrate how they are derived from the shadow formalism for a number of simple examples. A general derivation of the dressing rules can be found in appendix \ref{app:MasslessAuxProp}. 

We claim that in-in correlators can be obtained by summing over flat space Feynman diagrams dressed with certain one-dimensional auxiliary propagators attached to the interactions vertices. As we will see shortly, a dressed flat space diagram with $L$ loops arises from summing over $2^L$ diagrams in EAdS using the shadow formalism. Hence, at tree-level there are an equal number of dressed diagrams and shadow diagrams but at loop-level the dressed diagrams arise from very non-trivial cancellations. The precise form of the auxiliary propagators will be theory-dependent, but the following algorithm for computing in-in correlators from dressed flat space diagrams can be used in general:
\begin{enumerate}
    \item Write down a flat space Feynman diagram and relax energy conservation, i.e. $k_{\rm ext}+p_{\rm tot}=0$ at the vertices, where $k_{\rm ext}$ and $p_{\rm tot}$ are defined in \eqref{eq:Definitions} \footnote{We will use the convention where each flat space propagator comes with a factor $\frac{1}{\pi}$}.
    \item Attach an auxiliary propagator to each vertex and set the energy of each auxiliary propagator to be $p_{\rm tot}$ of the vertex it is attached to. 
    \item Attach the other end of all the auxiliary propagators to a single point, at which energy is conserved. 
    \item In general there will be two types of auxiliary propagators (except in the conformally coupled $\phi^4$ theory, where there is only one type). We will denote the two types using dashed and dotted lines, respectively. 
    Sum over all ways of attaching the two types of auxiliary propagators subject to simple constraints that we specify in Table \ref{tab:AuxPropSummary}. 
    \item Integrate over all the unfixed energy variables.
\end{enumerate}
In the final step, we can treat the integrals over the unfixed energy variables as contour integrals, summing over residues of the poles in the upper half of the complex plane. The remaining integrals will be over the boundary components of loop momenta. 
In what follows we will illustrate these rules for a variety of scalar field theories. We provide a summary of the dressing rules in Table \ref{tab:AuxPropSummary} for the benefit of the reader.

\subsection{Conformally coupled $\phi^4$}
For $V(\phi) = \frac{\lambda}{4!}\phi^4$, the shadow action given in \eqref{eq:EAdS_action_gen_pot} becomes 
\begin{eqn}
    \label{eq:CCphi4Action}
	L=-\left((\partial\phi_+)^2-m^2{\phi_+}^2\right)+\left((\partial\phi_-)^2-m^2{\phi_-}^2\right)-\frac{\lambda}{12} (\phi_-^4+\phi_+^4)+\frac{\lambda}{2} \phi_+^2\phi_-^2.
\end{eqn}
From this we can identify the Feynman rules for this theory. We find it convenient to absorb the symmetry factors due to the $m!$ ways of contracting identical fields at a $\phi^m$ vertex, leading to the Feynman vertex rules shown in Table \ref{tab:vertex-factorsCC4}.
\begin{table}[h]
    \centering
    \begin{tabular}{c|ccc}
        Vertex & $\phi_-^4$ & $\phi_+^2 \phi_-^2$ & $\phi_+^4$ \\
        \hline
        Factor  & $-2 \lambda$ & $2 \lambda$ & $-2 \lambda$       
    \end{tabular}
    \caption{The Feynman vertex rules for conformally coupled $\phi^4$. \label{tab:vertex-factorsCC4}}
\end{table}

For this particular theory, the auxiliary propagator takes the following simple form
    \begin{equation}
        \begin{tikzpicture}[baseline=($(b)!0.25!(w)$.base)]
            \begin{feynman}
                \vertex (a);
                \vertex [ right = 50pt of a , dot     ] (b) {};
                \vertex [ right = 40pt  of b](w);
        
                \diagram* {
                    (b)  -- [momentum= $p_{\rm tot}$, scalar] (w),
                };
            \end{feynman}
        \end{tikzpicture} =-\frac{2k_{\rm ext}}{p_{\rm tot}^2+k_{\rm ext}^2}\,,
        \label{eq:CC4AuxProp}
    \end{equation}
where the dot refers to an interaction vertex to which four non-auxiliary propagators are attached, $p_{tot}$ is the sum of the internal energies flowing into the vertex, and $k_{ext}$ is the sum of external energies flowing into the vertex. See \eqref{eq:Definitions} for more details.

We have already seen how this rule emerges at 1-loop in section \ref{invitation}, so let us now apply it to a more complicated example, notably the 2-loop necklace depicted below. In this case there are three vertices so we must attach three auxiliary propagators. Since the middle vertex is not attached to any external propagators, we have $k_{\rm ext}=0$ so the auxiliary propagator attached to it reduces to $\pi \delta(p_{\rm tot})$, i.e. energy is conserved at this vertex just like in flat space. We subsequently obtain (modulo regularization)

\begin{tikzpicture}[baseline=(b.base)]
        \begin{feynman}
            \vertex (a) ;
            \vertex [right = 75pt of a       ] (b);
            \vertex [right = 75pt of b        ] (c);
            \vertex [above left = 26.39pt and 63.71pt of b](l1) {$\vec{k_1}$};
            \vertex [below left = 26.39pt and 63.71pt of b](l2) {$\vec{k_2}$};
            \vertex [above right = 26.39pt and 63.71pt of b](r1) {$\vec{k_3}$};
            \vertex [below right = 26.39pt and 63.71pt of b](r2) {$\vec{k_4}$};
            \vertex [right = 30pt of a       ] (i1);
            \vertex [left = 30pt of c       ](i2);
            \vertex [above = 50pt of b](e);
    
            \diagram* {
                (l1) -- (i1),
                (l2) -- (i1),
                (r1) -- (i2),
                (r2) -- (i2),
                (i1) --[half right, looseness=0.8, momentum' = $\omega_1 +p$  ] (b),
                (b)  --[ half right, looseness=0.8, momentum = $\omega_1$   ] (i1),
                (b)  --[ half right, looseness=0.8, momentum' = $\omega_2 +p$  ] (i2),
                (i2) --[ half right , looseness=0.8, momentum = $\omega_2$  ] (b),
                (e) --[scalar, momentum' = $p$](i1),
                (i2) --[scalar, momentum' = $p$](e),
                (b) -- [scalar, momentum' = $0$](e),

            };
        \end{feynman}
    \end{tikzpicture}
\be 
\begin{split}
=&-\frac{ 8}{\pi^3 }\int_{-\infty}^\infty dp \frac{k_{12}k_{34}}{(p^2+k_{12}^2)(p^2+k_{34}^2)}\int d^4 L_1 \,d^4 L_2 \frac{1}{L_1^2(L_1+P)^2L_2^2(L_2+P)^2}\,,
\end{split}
\label{eq:CC4Necklace}
\ee
 where $P^\mu=(p,\vec{y}_{12})^\mu$  and $L_i^\mu=(\omega_i, \vec{q}_i)^\mu$. The dashed propagator attaching to the middle vertex has zero energy since energy-momentum is conserved at that vertex. The modification of \eqref{eq:CC4Necklace} for a de Sitter-invariant UV-regularization can be found in equation \eqref{bubbleanareg}.

This result was derived using the shadow formalism in \cite{Chowdhury:2023arc}, where the simple flat space integrand was obtained from nontrivial cancellations after summing over four different diagrams. This illustrates the power of the dressing rules.

\subsection{Conformally coupled $\phi^3$}

In this case, there are two types of auxiliary propagators:
\begin{itemize}
    \item dashed auxiliary propagator
        \begin{equation}
        \begin{aligned}
        \begin{tikzpicture}[baseline=(b.base)]
            \begin{feynman}
                \vertex (a);
                \vertex [ right = 50pt of a, dot      ] (b) {};
                \vertex [ right = 40pt  of b](w);
        
                \diagram* {
                    (b)  -- [momentum= $p_{\rm tot}$, scalar] (w),
                };
            \end{feynman}
        \end{tikzpicture} 
        \label{eq:CC3UnsuppAuxProp}
    \end{aligned}=-2i \int_0^\infty ds\, \frac{p_{\rm tot}}{p_{\rm tot}^2+(s+k_{\rm ext})^2}
    \end{equation}
of which there must be an even number.
    
    \item dotted auxiliary propagator
    \begin{equation}
        \begin{tikzpicture}[baseline=($(b)!0.25!(w)$.base)]
            \begin{feynman}
                \vertex (a);
                \vertex [ right = 50pt of a , dot      ] (b){};
                \vertex [right = 50pt of b        ] (c);
                \vertex [above left = 35.36pt and 35.36pt of b](oi);
                \vertex [below left = 35.36pt and 35.36pt of b](oj) ;
                \vertex [below = 7pt of b](z) ;
                \vertex [ right = 40pt of b](w);
        
                \diagram* {
                    (b)  -- [momentum= $p_{\rm tot}$, ghost] (w),
                };
            \end{feynman}
        \end{tikzpicture} = -\pi 
        \label{eq:cc3SuppAuxProp}
    \end{equation}
of which there can be any number. 
\end{itemize}
Even though the propagator in \eqref{eq:cc3SuppAuxProp} has no energy dependence, we must still associate an energy to it that will enter after imposing energy conservation at each vertex following the algorithm in the beginning of this section. Note that the propagator in \eqref{eq:CC3UnsuppAuxProp} gives rise to a logarithm after performing the $s$-integral. We may therefore expect that attaching such a propagator to a flat space Feynman diagram increases the transcendentality by one. We will see below and in later sections that this is indeed the case. For example, adding two such auxiliary propagators to a tree-level exchange diagram in flat space gives a dilogarithm. The auxiliary propagator in \eqref{eq:cc3SuppAuxProp} also increases the transcendentality by one, but in a more trivial sense. When we discuss the transcendentality of correlators in what follows, we refer to their functional form rather than overall factors of $\pi$.

Let us illustrate how these dressing rules arise for a few simple examples starting from the shadow effective action.

\subsubsection*{Tree-level 3-point} 

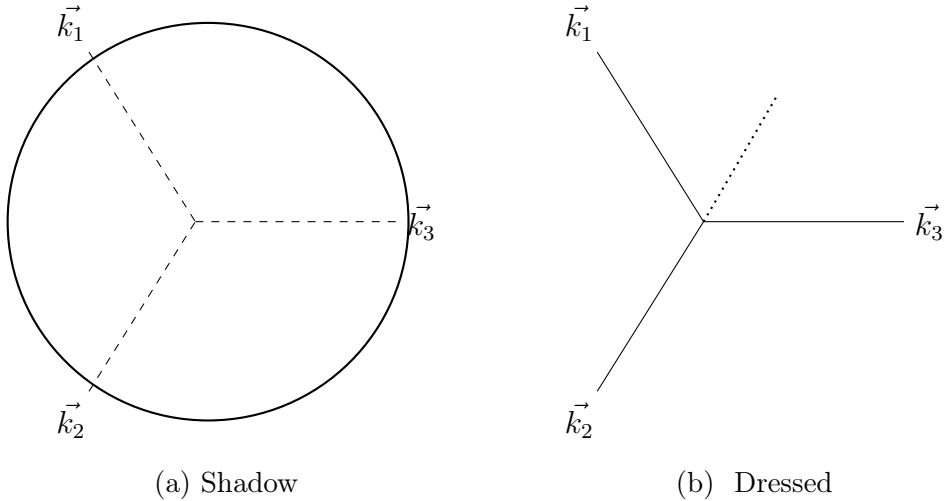
\begin{figure}[H]
    \centering    
    \begin{subfigure}{0.45\textwidth}
    \centering
    \begin{tikzpicture}[baseline=(b.base)]
        \begin{feynman}
            \vertex (a) ;
            \vertex [ right = 75pt of a       ] (b);
            \vertex [right = 75pt of b        ] (c) {\(\vec{k_3}\)};
            \vertex [above left = 64pt and 37pt of b](oi) {\(\vec{k_1}\)};
            \vertex [below left = 64pt and 37pt of b](oj) {\(\vec{k_2}\)};
    
            \diagram* {
                (oi) --[scalar] (b),
                (oj) --[scalar] (b),
                (c)  --[scalar] (b),
            };
            \draw[thick] ($(a)!0.5!(c)$) circle [
            radius=75pt
            ];
        \end{feynman}
    \end{tikzpicture}
    \caption{\label{sfig:CC3ContactShadow}Shadow}
    \end{subfigure}
    \begin{subfigure}{0.45\textwidth}
    \centering
    \begin{tikzpicture}[baseline=(b.base)]
        \begin{feynman}
            \vertex (a) ;
            \vertex [ right = 75pt of a       ] (b);
            \vertex [right = 75pt of b        ] (c) {\(\vec{k_3}\)};
            \vertex [above left = 64pt and 37pt of b](oi) {\(\vec{k_1}\)};
            \vertex [below left = 64pt and 37pt of b](oj) {\(\vec{k_2}\)};
            \vertex [above = 7pt of b](x);
            \vertex [above right  = 48pt and 27.75pt of b](e);
    
            \diagram* {
                (oi) -- (b),
                (oj) -- (b),
                (c)  -- (b),
                (b)  --[ghost] (e),
            };
        \end{feynman}
    \end{tikzpicture}
    \caption{\label{sfig:CC3ContactDressed} Dressed }
    \end{subfigure}
    \caption{Three-point contact diagram in conformally coupled $\phi^3$ from the shadow formalism and the dressing rules. \label{fig:CC3ContactDressed}}
\end{figure}

For $V(\phi)=\frac{\lambda}{3!}\phi^3$, the shadow action \eqref{eq:EAdS_action_gen_pot} features no $\phi_-^3$ vertex in $d=3$ dimensions; therefore, this diagram does not naively exist. However, if one perturbs the boundary dimension to $d\to d+2\epsilon$ and the scaling dimension $\Delta_\pm\to \Delta_\pm + \e$\footnote{\label{fn:dimShift}We perturb the scaling dimension to ensure that $\Delta_{\pm} - \frac{d}{2}$ is unchanged. This combination appears as the order of the Bessel functions in the propagators (see appendix \ref{app:Green}). For convenience, in a $\phi^v$ potential we shift to $d=3+\frac{2}{v-2}\epsilon$ such that the overall power of $z$ has a coefficient $+1$ on $\epsilon$. We henceforth leave this implicit.} then the Lagrangian in \eqref{eq:EAdS_action_gen_pot} becomes
\begin{eqn}
    \label{eq:EAdS_action_gen_pot_epsilon}
	L=-\left((\partial\phi_+)^2-m^2{\phi_+}^2\right)&+\left((\partial\phi_-)^2-m^2{\phi_-}^2\right)\\
    &-\frac{\lambda \pi \e}{2} \big(\frac{1}{3}\phi_-^3-\phi_- \phi_+^2\big)+\lambda \big(\frac{1}{3}\phi_+^3-\phi_-^2 \phi_+\big).
\end{eqn}
As in the $\phi^4$ case, we obtain the Feynman rules after accounting for symmetry factors at the vertex, which  are given in Table \ref{tab:vertex-factorsCC3}.
The contact Witten diagram is displayed in Fig.\ \ref{sfig:CC3ContactShadow} and, using the propagators given in \eqref{eq:EAdSProps}, evaluates to

\be \label{eq:3pc}
\mathcal{A}_{3}= -\pi \epsilon \lambda \int_0^\infty dz \, \frac{(z^{1+\epsilon})^3}{z^{4+2\epsilon}}  e^{-(k_1+k_2+k_3)z}= -\pi \lambda k^{-\epsilon}   \epsilon \Gamma(\epsilon) \xrightarrow[]{\epsilon \rightarrow 0} - \pi\lambda\,,
\ee 
which agrees with the result obtained in \cite{Arkani-Hamed:2015bza}. Note that despite the vertex being a suppressed one, we obtain a non-zero result because the $\epsilon$ from the interaction vertex cancels the $\epsilon^{-1}$ from the $z$-integral.

\begin{table}
    \centering
    \begin{tabular}{c|cccc}
        Vertex & $(\phi^+)^3$ & $(\phi^+)^2 \phi^-$ & $\phi^+ (\phi^-)^2$ & $(\phi^-)^3$ \\
        \hline
        Factor  & $2 \lambda$ & $\pi \epsilon \lambda$ & $-2 \lambda$ & $-\pi \epsilon \lambda$        
    \end{tabular}
    \caption{The Feynman vertex rules of conformally coupled $\phi^3$ in $d=3+2\epsilon$ with $\Delta_+=2+\e$ and $\Delta_-=1+\e$. \label{tab:vertex-factorsCC3}}
\end{table}

On the other hand, we may dress the flat space vertex applying (\ref{eq:cc3SuppAuxProp}) as in Figure \ref{sfig:CC3ContactDressed} and can apply the dressing rules. We can only use the auxiliary propagator \eqref{eq:cc3SuppAuxProp} since the one in \eqref{eq:CC3UnsuppAuxProp} must occur an even number of times.
Hence the dressed three-point diagram is simply   
\be 
\mathcal{A}_3=- \pi \lambda\,,
\ee 
in agreement with \eqref{eq:3pc}.

\subsubsection*{Tree-level 4-point}\label{sec:dress-conf-4pt-ex}

\begin{figure}[b]
    \centering
    \begin{subfigure}{0.4\textwidth}
    \centering
    \begin{tikzpicture}[baseline=(b.base)]
        \begin{feynman}
            \vertex (a) ;
            \vertex [ right = 50pt of a       ] (i1) ;
            \vertex [ right = 50pt of i1      ] (i2);
            \vertex [right = 75pt of a      ] (b);
            \vertex [right = 75pt of b      ] (c);
            \vertex [above left = 28pt and 63pt of b](l1) {\(\vec{k_1}\)};
            \vertex [below left = 28pt and 63pt of b](l2) {\(\vec{k_2}\)};
            \vertex [below right = 28pt and 63pt of b](r1) {\(\vec{k_3}\)};
            \vertex [above right = 28pt and 63pt of b](r2) {\(\vec{k_4}\)};
            \vertex [above = 50pt of b](e);
    
            \diagram* {
                (l1) -- [scalar](i1),
                (l2) -- [scalar](i1),
                (r1) -- [scalar](i2),
                (r2) -- [scalar](i2),
                (i1) --[edge label=$p {,}\;\vec{y}_{12}$] (i2)
            };
            \draw[thick] ($(a)!0.5!(c)$) circle [
            radius=75pt
            ];
        \end{feynman}
    \end{tikzpicture}
    \caption{Unsuppressed\label{fig:CC3exchangeShadowUnsuppressed}}
    \end{subfigure}
    \hspace{1cm}
    \begin{subfigure}{0.4\textwidth}
        \centering
        \begin{tikzpicture}[baseline=(b.base)]
        \begin{feynman}
            \vertex (a) ;
            \vertex [ right = 50pt of a       ] (i1) ;
            \vertex [ right = 50pt of i1      ] (i2);
            \vertex [right = 75pt of a      ] (b);
            \vertex [right = 75pt of b      ] (c);
            \vertex [above left = 28pt and 63pt of b](l1) {\(\vec{k_1}\)};
            \vertex [below left = 28pt and 63pt of b](l2) {\(\vec{k_2}\)};
            \vertex [below right = 28pt and 63pt of b](r1) {\(\vec{k_3}\)};
            \vertex [above right = 28pt and 63pt of b](r2) {\(\vec{k_4}\)};
            \vertex [above = 50pt of b](e);
    
            \diagram* {
                (l1) -- [scalar](i1),
                (l2) -- [scalar](i1),
                (r1) -- [scalar](i2),
                (r2) -- [scalar](i2),
                (i1) -- [scalar, edge label=$p {,}\;\vec{y}_{12}$] (i2),
            };
            \draw[thick] ($(a)!0.5!(c)$) circle [radius=75pt];
        \end{feynman}
    \end{tikzpicture}
    \caption{Suppressed \label{fig:CC3exchangeShadowSuppressed}}
    \end{subfigure}
    \caption{Shadow diagrams contributing to the four-point correlator at tree level in conformally coupled $\phi^3$. Here $\vec{y}_{12}=\vec{k}_1+\vec{k}_2.$}
\end{figure}
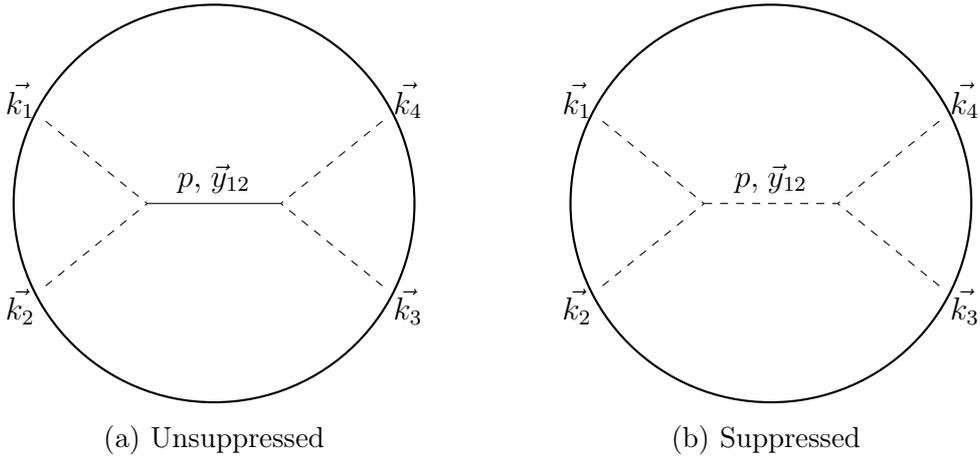

Now let us consider the tree-level 4-point correlator. Focusing on the s-channel, there are two diagrams to consider in the shadow formalism, which are depicted in Figure 3. 
In Figure \ref{fig:CC3exchangeShadowUnsuppressed}, the internal propagator is $\phi^+$, and therefore we have two vertices of the type $(\phi^-)^2 \phi^+$. Table \ref{tab:vertex-factorsCC3} tells us that the vertex factors are unsuppressed for $\epsilon\to  0$ and are simply $-2 \lambda$. Using the propagators given in \eqref{eq:EAdSProps} then gives 
\be 
    \mathcal{A}^{(1)}_{4} = -\frac{(-2\lambda)^2 }{\pi }\int_0^\infty dz_1 \, dz_2 \int_{-\infty}^\infty dp \frac{1}{p^2+\vec{y}_{12}^2} \frac{e^{-k_{12} z_1}e^{-k_{34} z_2}}{z_1 z_2} \sin(p z_1) \sin(p z_2)\,,
    \label{eq:CC3ExchangeShadow}
\ee 

where $\vec{y}_{ij}=\vec{k}_i+\vec{k}_j$ and $k_{ij}=|\vec{k}_i|+|\vec{k}_j|$.

We shall now show that this expression arises from dressing a flat space Feynman diagram 
by transforming the $z$-integral into an $s$-integral via the identities
\be 
z^{-n} e^{-k z} = \frac{1}{\Gamma(n)} \int_0^\infty ds \, s^{n-1} e^{-(s+k) z}
\label{eq:sIntIdentity}
\ee 
and 
\be 
\int_0^\infty dz \, e^{-(s+k) z} \sin(p z) = \frac{p}{p^2 +(s+k)^2}.
\ee 
The identity (\ref{eq:sIntIdentity}) will be used repeatedly below. Applying each of these identities in turn (with $n=1$) gives
\be 
\int_0^\infty dz \, \frac{e^{- k z}}{z} \sin(p z) = \intsinf ds \intsinf dz e^{- (k + s) z} \sin(p z) =  \int_0^\infty ds \, \frac{p}{(s+k)^2+p^2}~,
\ee 
and therefore (\ref{eq:CC3ExchangeShadow}) becomes
\be\label{eq:conformalphi3exshadow}
\mathcal{A}^{(1)}_{4} = -\frac{4 \lambda^2 }{\pi }\int_{-\infty}^\infty dp \int_0^\infty ds_1 \int_0^\infty ds_2  \frac{p}{p^2+(s_1+k_{12})^2}\frac{p}{p^2+(s_2+k_{34})^2} \frac{1}{p^2+y_{12}^2}.
\ee 
As shown in appendix \ref{app:Integrals}, this representation allows us to express the integrals in a form which can be recognized as Hypergeometric functions. 

\begin{figure}
    \centering
    \begin{subfigure}{0.45\textwidth}
    \centering
    \begin{tikzpicture}[baseline=(b.base)]
        \begin{feynman}
            \vertex (a) ;
            \vertex [ right = 50pt of a       ] (i1) ;
            \vertex [ right = 50pt of i1      ] (i2);
            \vertex [right = 75pt of a      ] (b);
            \vertex [right = 75pt of b      ] (c);
            \vertex [above left = 26.39pt and 63.71pt of b](l1) {\(\vec{k_1}\)};
            \vertex [below left = 26.39pt and 63.71pt of b](l2) {\(\vec{k_2}\)};
            \vertex [below right = 26.39pt and 63.71pt of b](r1) {\(\vec{k_3}\)};
            \vertex [above right = 26.39pt and 63.71pt of b](r2) {\(\vec{k_4}\)};
            \vertex [above = 50pt of b](e);
    
            \diagram* {
                (l1) --(i1),
                (l2) -- (i1),
                (r1)  --(i2),
                (r2)  --(i2),
                (i1) --[edge label=$p {,}\;\vec{y}_{12}$] (i2),
                (i1) --[scalar, edge label=$p$] (e),
                (i2) --[scalar, edge label'=$p$] (e),
            };
        \end{feynman}
    \end{tikzpicture}
    \caption{Dashed dressing, corresponding to  shadow diagram with unsuppressed vertices. \label{fig:CC3exchangeDressedUnsuppressed}}
    \end{subfigure}
    \hspace{1cm}
    \begin{subfigure}{0.45\textwidth}
        \centering
        \begin{tikzpicture}[baseline=(b.base)]
        \begin{feynman}
            \vertex (a) ;
            \vertex [ right = 50pt of a       ] (i1) ;
            \vertex [ right = 50pt of i1      ] (i2);
            \vertex [right = 75pt of a      ] (b);
            \vertex [right = 75pt of b      ] (c);
            \vertex [above left = 26.39pt and 63.71pt of b](l1) {\(\vec{k_1}\)};
            \vertex [below left = 26.39pt and 63.71pt of b](l2) {\(\vec{k_2}\)};
            \vertex [below right = 26.39pt and 63.71pt of b](r1) {\(\vec{k_3}\)};
            \vertex [above right = 26.39pt and 63.71pt of b](r2) {\(\vec{k_4}\)};
            \vertex [above = 50pt of b](e);
    
            \diagram* {
                (l1) -- (i1),
                (l2) -- (i1),
                (r1)  --(i2),
                (r2)  --(i2),
                (i1) --[edge label=$p {,}\;\vec{y}_{12}$] (i2),
                (i1) --[ghost, edge label= $p$] (e),
                (i2) --[ghost, edge label' =$p$] (e),
            };
        \end{feynman}
    \end{tikzpicture}
    \caption{Dotted dressing, corresponding to  shadow diagram with suppressed vertices. \label{fig:CC3exchangeDressedSuppressed}}
    \end{subfigure}
    \caption{The two dressed diagrams which contribute to the four-point conformally coupled correlator at tree level. Recall that dotted auxiliary propagators only occur in pairs, so there are no diagrams with both types.}
\end{figure}
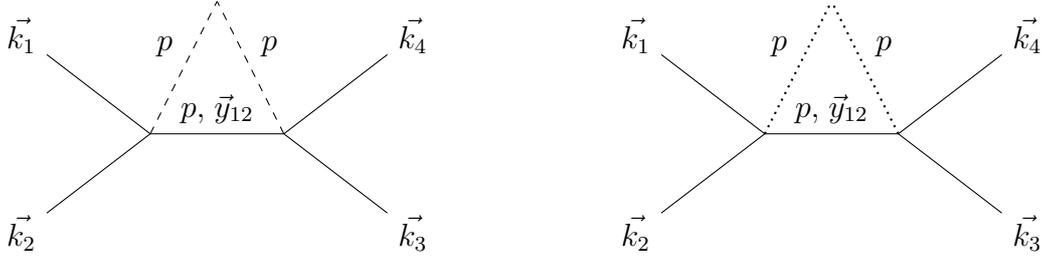

We can now see that this is equivalent to the dressed diagram shown in Fig.\ \ref{fig:CC3exchangeDressedUnsuppressed} with the choice of auxiliary propagator put forward in (\ref{eq:CC3UnsuppAuxProp}). Indeed, applying the dressing rules yields 
\be 
\begin{split}
\mathcal{A}^{(1)}_{4} =& \lambda^2 \int_{-\infty}^\infty dp\bigg(-2i\int_0^\infty ds_1 \,  \frac{p}{p^2+(s_1+k_{12})^2}\bigg)\bigg(-2i\int_0^\infty ds_2 \frac{p}{p^2 + (s_2+k_{34})^2}\bigg) \\
&\times \Big(\frac{1}{\pi}\frac{1}{p^2+y_{12}^2}\Big),
\end{split}
\ee 
which is clearly the same as \eqref{eq:conformalphi3exshadow}. In this instance, one may integrate to obtain 
\be 
\mathcal{A}^{(1)}_{4} = \frac{4 \lambda^2 }{ y_{12}}\Bigg[\text{Li}_2\bigg(\frac{k_{12}-y_{12}}{k_{12}+k_{34}}\bigg)+\text{Li}_2\bigg(\frac{k_{34}-y_{12}}{k_{12}+k_{34}}\bigg)+\ln\bigg|\frac{k_{12}+y_{12}}{k_{12}+k_{34}}\bigg|\ln \bigg|\frac{k_{34}+y_{12}}{k_{12}+k_{34}}\bigg|-\frac{\pi^2}{6}\Bigg].
\ee 

We now turn to the second exchange diagram contributing in the shadow formalism given in Fig.\ \ref{fig:CC3exchangeShadowSuppressed}. This diagram requires careful handling of $\epsilon$ factors. On the one hand, each vertex factor is linear in $\epsilon$. 
On the other, each $z$ integral gives a $1/\epsilon$ which exactly cancels the factors of $\epsilon$ in the vertices. 
Applying the Feynman rules to the diagram in Figure \ref{fig:CC3exchangeShadowSuppressed} yields 
\be 
\mathcal{A}^{(2)}_{4} = \pi \epsilon^2 \lambda^2  \int_{-\infty}^\infty dp \int_0^\infty dz_1 \, dz_2 \frac{e^{-k_{12} z_1}e^{-k_{34} z_2}}{(z_1 z_2)^{1-\epsilon}}\cos(p z_1)\cos(p z_2)\frac{1}{p^2+y_{12}^2}\,.
\ee 
By performing the $z$-integrals we see that the leading contribution is just $\frac{1}{\epsilon^2}$. This all we require due to the $\epsilon^2$ prefactor.   
The resulting integral in the $\epsilon \rightarrow 0$ limit is therefore simply
\be \label{eq:A4tree2shadow}
\mathcal{A}^{(2)}_{4} = \lambda^2 \pi  \int_{-\infty}^\infty dp \, \frac{1}{p^2+y_{12}^2}.
\ee 
We recognise this to be a tree-level Feynman diagram integrated over the energy flowing through the propagator.
    

On the other hand, we have the dressed diagram with dotted auxiliary propagators, shown in Fig.\ \ref{fig:CC3exchangeDressedSuppressed}. Applying the dressing rules here gives
\be 
\mathcal{A}^{(2)}_{4} = \lambda^2 \intinf dp \,(-\pi)(-\pi)\Big(\frac{1}{\pi}\frac{1}{p^2 + y_{12}^2}\Big)
\ee 
which matches the result from the shadow computation, \eqref{eq:A4tree2shadow}. This integral is simple and yields 
\be 
\mathcal{A}^{(2)}_{4} = \lambda^2  \frac{\pi^2}{ y_{12}}.
\ee 
The sum $\mathcal{A}_4=\mathcal{A}^{ (1)}_{4}+\mathcal{A}^{ (2)}_{4}$ indeed matches the result found in \cite{Arkani-Hamed:2015bza}.


\subsubsection*{3-point 1-loop }

\begin{figure}[H]
    \centering
    
    \begin{tikzpicture}[baseline=(b.base)]
        \begin{feynman}
            \vertex (a) ;
            \vertex [right = 37.5pt of a       ] (b);
            \vertex [right = 37.5pt of b        ] (c);
            \vertex [below = 75pt of b          ] (ok) {\(\vec{k}_2\)};
            \vertex [ below = 25pt of b          ] (ik) ;
            \vertex [ above right = 37.5pt and 65pt of b](oj) {\(\vec{k}_1\)};
            \vertex [ above right = 17.34pt and 27pt of b](ij);
            \vertex [above left = 37.5pt and 65pt of b](oi) {\(\vec{k}_3\)};
            \vertex [ above left = 17.34pt and 27pt of b](ii) ;
            \vertex [below right = 0pt and 0pt of ii] (li) {\(z_3\)};
            \vertex [below left = 0pt and 0pt of ij] (lj) {\(z_1\)};
            \vertex [above = 0pt of ik] (lk) {\(z_2\)};
    
            \diagram* {
                (oi) --[scalar] (ii),
                (oj) --[scalar] (ij),
                (ok) --[scalar] (ik),
                (ii) --[ quarter left, looseness=1.3, edge label = $p_3 {,}\;\vec{q}$] (ij),
                (ij) --[scalar, quarter left, looseness=1.3, edge label = \footnotesize{$p_1 {,}\;\vec{k}_{1}+\vec{q}$}, inner sep=1pt] (ik),
                (ik) --[scalar, quarter left, looseness=1.3, edge label = \footnotesize{$p_2 {,}\;\vec{y}_{12}+\vec{q}$}, inner sep=1pt] (ii),
            };
            \draw[thick] ($(a)!0.5!(c)$) circle [
            radius=75pt
            ];
        \end{feynman}
    \end{tikzpicture}
    \hspace{.5cm}
    +
    \hspace{.5cm}
    \begin{tikzpicture}[baseline=(b.base)]
        \begin{feynman}
            \vertex (a) ;
            \vertex [right = 37.5pt of a       ] (b);
            \vertex [right = 37.5pt of b        ] (c);
            \vertex [below = 75pt of b          ] (ok) {\(\vec{k}_2\)};
            \vertex [ below = 25pt of b          ] (ik) ;
            \vertex [ above right = 37.5pt and 65pt of b](oj) {\(\vec{k}_1\)};
            \vertex [ above right = 17.34pt and 27pt of b](ij);
            \vertex [above left = 37.5pt and 65pt of b](oi) {\(\vec{k}_3\)};
            \vertex [ above left = 17.34pt and 27pt of b](ii) ;
            \vertex [below right = 0pt and 0pt of ii] (li) {\(z_3\)};
            \vertex [below left = 0pt and 0pt of ij] (lj) {\(z_1\)};
            \vertex [above = 0pt of ik] (lk) {\(z_2\)};
    
            \diagram* {
                (oi) --[scalar] (ii),
                (oj) --[scalar] (ij),
                (ok) --[scalar] (ik),
                (ii) --[scalar, quarter left, looseness=1.3, edge label = $p_3 {,}\;\vec{q}$] (ij),
                (ij) --[ quarter left, looseness=1.3, edge label = \footnotesize{$p_1 {,}\;\vec{k}_{1}+\vec{q}$}, inner sep=1pt] (ik),
                (ik) --[ quarter left, looseness=1.3, edge label = \footnotesize{$p_2 {,}\;\vec{y}_{12}+\vec{q}$}, inner sep=1pt] (ii),
            };
            \draw[thick] ($(a)!0.5!(c)$) circle [
            radius=75pt
            ];
        \end{feynman}
    \end{tikzpicture}
    \caption{Shadow diagrams with a single suppressed vertex in conformally coupled $\phi^3$. \label{fig:CC3ThreePointRingShadow1}}
\end{figure}
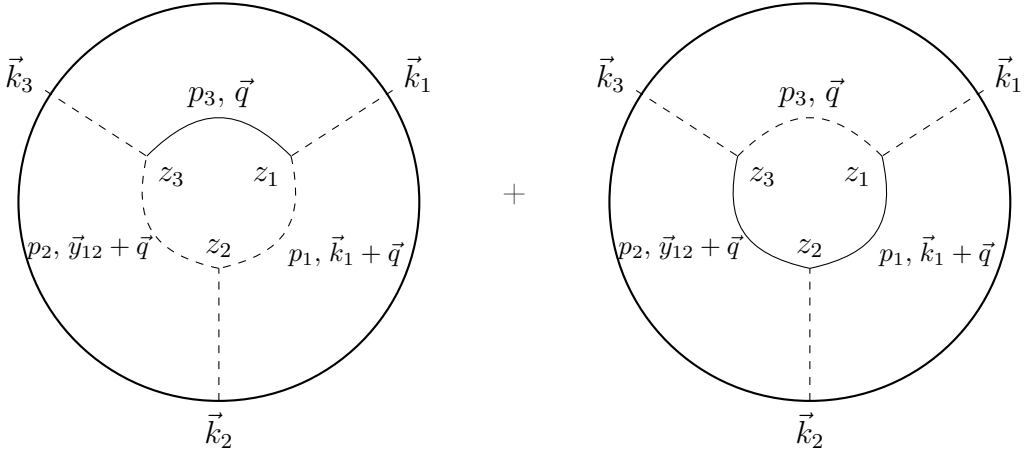

Finally let us consider the 1-loop 3-point correlator.
We have four shadow diagrams to consider.
The first two are shown in Fig.\ \ref{fig:CC3ThreePointRingShadow1}, with permutations of the external legs giving other channels.

Using the propagators in \eqref{eq:EAdSProps}
the sum of these diagrams is\footnote{To avoid excessive notation, and since we shall not use them together, $\mathcal{A}_3$ is used to refer both to the one-loop diagram here and contact diagram above. It shall be clear from the context which diagram we mean.} 
\be 
\begin{split}
\mathcal{A}^{(1)}_{3} &= \frac{4 \epsilon \lambda^3 }{\pi^2} \int d^3 q \, \int_{-\infty}^\infty dp_1 \, dp_2 \, dp_3 \int_0^\infty dz_1 \, dz_2 \, dz_3\, (z_1 z_2 z_3)^{\epsilon-1}e^{-k_1 z_1}e^{-k_2 z_2}e^{-k_3 z_3}\\
&\times\Bigg(\sin(p_3 z_1)\sin(p_3 z_3)\cos(p_1 z_1)\cos(p_1 z_2)\cos(p_2 z_2)\cos(p_2 z_3)+(-1)^2\big(\sin \leftrightarrow \cos \big)\Bigg)\\
&\times\frac{1}{p_3^2+\vec{q}^2}\frac{1}{p_1^2+(\vec{q}+\vec{k_1})^2}\frac{1}{p_2^2+(\vec{q}+\vec{y}_{12})^2},
\end{split}\,,
\label{eq:CC3ThreePointRingStart}
\ee 
where we have made explicit that the sign difference due to ghost-like propagators was required to cancel the sign difference due to different vertex factors. We can simplify the expression in the brackets by noticing the integrand is invariant under sign change of $p_i$ and using standard trigonometric identities to obtain
\be 
\Bigg( \cdots\Bigg)= \sin\big((p_3-p_2)z_3\big)\sin\big((p_3+p_1)z_1\big)\cos\big((p_1+p_2)z_2\big).
\ee 
From this form it is easy to see that only the $z_2$ integral in \eqref{eq:CC3ThreePointRingStart} is divergent. Evaluating this we have 
\be 
\begin{split}
\epsilon \int_0^\infty dz_2 \, z_2^{-1+\epsilon} e^{-k z} \cos((p z_2) = \frac12 \epsilon \Gamma(\epsilon) \big((k-i p)^\epsilon + (k+i p)^\epsilon \big)\xrightarrow[]{\epsilon \rightarrow 0} 1.
\end{split}
\label{eq:CCSuppressedIntegral}
\ee 
Then, using (\ref{eq:sIntIdentity}) and
\be 
\int_0^\infty dz \, e^{-s z} \sin(p z) = \frac{p}{p^2+s^2}\,,
\ee 
the remaining integrals over $z_1$ and $z_3$ give
\be 
\begin{split}
&\mathcal{A}^{(1)}_{3}\\
&=\frac{4\lambda^3}{\pi^2 }\int d^3q \int_{-\infty}^\infty dp_1 \, dp_2 \, dp_3 \int_0^\infty ds_1 \int_0^\infty ds_2 \frac{p_3+p_1}{(p_3+p_1)^2+(s_1+k_1)^2}\frac{p_3-p_2}{(p_3-p_2)^2+(s_2+k_3)^2}\\
&\times\frac{1}{p_3^2+\vec{q}^2}\frac{1}{p_1^2+(\vec{q}+\vec{k_1})^2}\frac{1}{p_2^2+(\vec{q}+\vec{y}_{12})^2}.
\end{split}
\label{eq:CC3ThreePointRingFromShadow1}
\ee

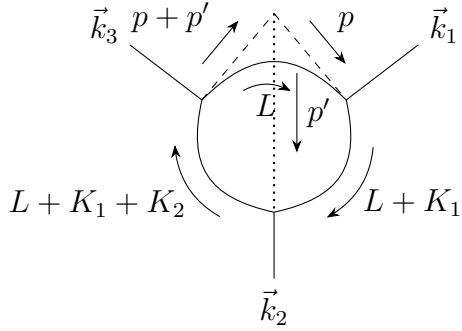
\begin{figure}[H]
    \centering
    \begin{tikzpicture}[baseline=(b.base)]
        \begin{feynman}
            \vertex (a) ;
            \vertex [right = 37.5pt of a       ] (b);
            \vertex [right = 37.5pt of b        ] (c);
            \vertex [below = 50pt of b          ] (ok) {\(\vec{k}_2\)};
            \vertex [ below = 25pt of b          ] (ik) ;
            \vertex [ above right = 34.67pt and 54pt of b](oj) {\(\vec{k}_1\)};
            \vertex [ above right = 17.34pt and 27pt of b](ij);
            \vertex [above left = 34.67pt and 54pt of b](oi) {\(\vec{k}_3\)} ;
            \vertex [ above left = 17.34pt and 27pt of b](ii) ;
            \vertex [above = 50pt of b](e);
    
            \diagram* {
                (oi) -- (ii),
                (oj) -- (ij),
                (ok) -- (ik),
                (ii) --[ quarter left, looseness=1.3, momentum' ={[xshift=-.1cm] \(L\)}] (ij),
                (ij) --[ quarter left, looseness=1.3, momentum = \(L+K_1\), inner sep=1pt] (ik),
                (ik) --[ quarter left, looseness=1.3, momentum = \(L+K_1+K_2\), inner sep=1pt] (ii),
                (ii) --[scalar, near end, momentum=\(p+p'\)](e),
                (e)--[scalar, momentum=\(p\)](ij),
                (e)--[ghost, momentum={[arrow shorten=0.3]\(p'\)},near end](ik),
            };
        \end{feynman}
    \end{tikzpicture}
    \caption{Dressed diagram with two dashed propagators which corresponds to the sum in Fig.\ \ref{fig:CC3ThreePointRingShadow1}. \label{fig:CC3ThreePointRingDressed1}}
\end{figure}
We now consider the dressed diagram in Fig.\ \ref{fig:CC3ThreePointRingDressed1} which we claim is equivalent to the sum of shadow diagrams which gave \eqref{eq:CC3ThreePointRingFromShadow1}. Applying the dressing rules we get
\be 
\begin{split}
    &\mathcal{A}^{(1)}_{3}\\
    &= \lambda^3 \int dp \, dp' \int d^3 q\, d\omega \;(-\pi)\big(-2i\;\int_0^\infty ds_1 \, \frac{p}{p^2+(s_1+k_1)^2}\big)\;\big(-2i\int_0^\infty ds_2 \, \frac{p+p'}{(p+p')^2+(s_2+k_3)^2}\big)\\
    & \times \Big(\frac{1}{\pi} \frac{1}{\omega^2+\vec{q}^2}\Big)\Big(\frac{1}{\pi}\frac{1}{(\omega+p)^2+(\vec{q}+\vec{k}_1)^2}\Big)\Big(\frac{1}{\pi}\frac{1}{(\omega+p+p')^2+(\vec{q}+\vec{y}_{12})^2}\Big)\\
    &=\lambda^3\int dp \, dp' (-\pi)\big(-2i\;\int_0^\infty ds_1 \, \frac{p}{p^2+(s_1+k_1)^2}\big)\;\big(-2i\int_0^\infty ds_2 \, \frac{p+p'}{(p+p')^2+(s_2+k_3)^2}\big)\\
    &\times \frac{1}{\pi^3}\int d^4 L \frac{1}{L^2(L+K_1)^2(L+K_1+K_2)^2}
\end{split}
\ee 
where $L^\mu=(\omega, \vec{q})^\mu$, $K_1^\mu=(p,\vec{k}_1)^\mu$, and $K_2^\mu=(p', \vec{k}_2)^\mu$. This matches \eqref{eq:CC3ThreePointRingFromShadow1} after making the following replacements in that equation:
\begin{eqn}
   p_1&\rightarrow -p-\omega, \\
   p_2&\rightarrow p+p'+\omega,\\
   p_3&\rightarrow\omega,
   \label{eq:3ptVarShift}
\end{eqn}
such that $p_3 +p_1 \rightarrow -p$ and $p_3-p_2 \rightarrow -p-p'$. 

\begin{figure}[H]
    \centering
\begin{tikzpicture}[baseline=(b.base)]
        \begin{feynman}
            \vertex (a) ;
            \vertex [right = 37.5pt of a       ] (b);
            \vertex [right = 37.5pt of b        ] (c);
            \vertex [below = 75pt of b          ] (ok) {\(\vec{k}_2\)};
            \vertex [ below = 25pt of b          ] (ik) ;
            \vertex [ above right = 37.5pt and 65pt of b](oj) {\(\vec{k}_1\)};
            \vertex [ above right = 17.34pt and 27pt of b](ij);
            \vertex [above left = 37.5pt and 65pt of b](oi) {\(\vec{k}_3\)};
            \vertex [ above left = 17.34pt and 27pt of b](ii) ;
            \vertex [below right = 0pt and 0pt of ii] (li) {\(z_3\)};
            \vertex [below left = 0pt and 0pt of ij] (lj) {\(z_1\)};
            \vertex [above = 0pt of ik] (lk) {\(z_2\)};
    
            \diagram* {
                (oi) --[scalar] (ii),
                (oj) --[scalar] (ij),
                (ok) --[scalar] (ik),
                (ii) --[ scalar, quarter left, looseness=1.3, edge label = $p_3 {,}\;\vec{q}$] (ij),
                (ij) --[scalar, quarter left, looseness=1.3, edge label = \footnotesize{$p_1 {,}\;\vec{k}_{1}+\vec{q}$}, inner sep=1pt] (ik),
                (ik) --[scalar, quarter left, looseness=1.3, edge label = \footnotesize{$p_2 {,}\;\vec{y}_{12}+\vec{q}$}, inner sep=1pt] (ii),
            };
            \draw[thick] ($(a)!0.5!(c)$) circle [
            radius=75pt
            ];
        \end{feynman}
    \end{tikzpicture}
\hspace{.5cm}
+
\hspace{.5cm}
\begin{tikzpicture}[baseline=(b.base)]
        \begin{feynman}
            \vertex (a) ;
            \vertex [right = 37.5pt of a       ] (b);
            \vertex [right = 37.5pt of b        ] (c);
            \vertex [below = 75pt of b          ] (ok) {\(\vec{k}_2\)};
            \vertex [ below = 25pt of b          ] (ik) ;
            \vertex [ above right = 37.5pt and 65pt of b](oj) {\(\vec{k}_1\)};
            \vertex [ above right = 17.34pt and 27pt of b](ij);
            \vertex [above left = 37.5pt and 65pt of b](oi) {\(\vec{k}_3\)};
            \vertex [ above left = 17.34pt and 27pt of b](ii) ;
            \vertex [below right = 0pt and 0pt of ii] (li) {\(z_3\)};
            \vertex [below left = 0pt and 0pt of ij] (lj) {\(z_1\)};
            \vertex [above = 0pt of ik] (lk) {\(z_2\)};
    
            \diagram* {
                (oi) --[scalar] (ii),
                (oj) --[scalar] (ij),
                (ok) --[scalar] (ik),
                (ii) --[ quarter left, looseness=1.3, edge label = $p_3 {,}\;\vec{q}$] (ij),
                (ij) --[ quarter left, looseness=1.3, edge label = \footnotesize{$p_1 {,}\;\vec{k}_{1}+\vec{q}$}, inner sep=1pt] (ik),
                (ik) --[ quarter left, looseness=1.3, edge label = \footnotesize{$p_2 {,}\;\vec{y}_{12}+\vec{q}$}, inner sep=1pt] (ii),
            };
            \draw[thick] ($(a)!0.5!(c)$) circle [
            radius=75pt
            ];
        \end{feynman}
    \end{tikzpicture}

\caption{The two shadow diagrams in which every vertex is suppressed for the three-point ring in conformally coupled $\phi^3$. \label{fig:CC3ThreePointRingShadow2}}
\end{figure}
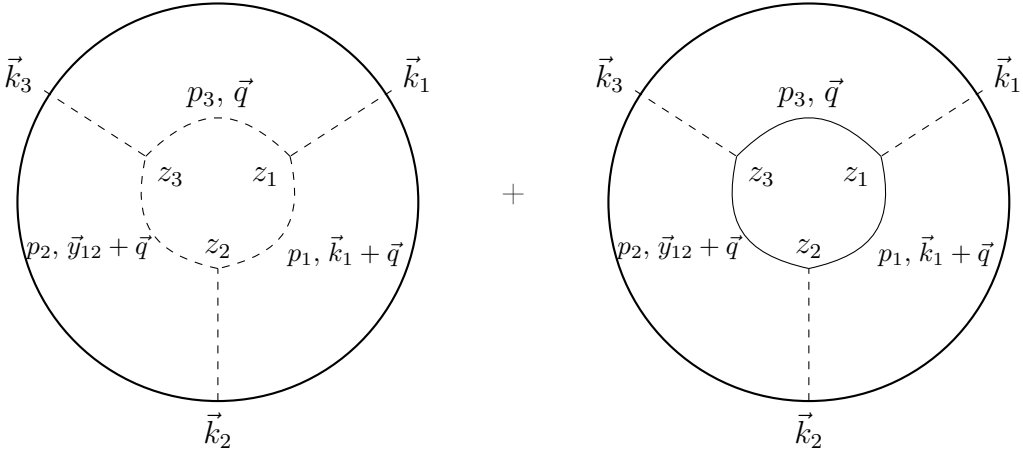

Finally, we consider the pair of shadow diagrams shown in Fig.\  \ref{fig:CC3ThreePointRingShadow2}, in which all vertices are suppressed. From the shadow action \eqref{eq:EAdS_action_gen_pot_epsilon} we get their sum as,
\be \label{eq:A3c1loop2}
\begin{split}
\mathcal{A}^{(2)}_{3} &= -\epsilon^3\lambda^3\int d^3q \int_{-\infty}^\infty dp_1 \, dp_2 \, dp_3 \int_0^\infty dz_1 \, dz_2 \, dz_3 \, (z_1 z_2 z_3)^{\epsilon-1} e^{-k_1 z_1}e^{-k_2 z_2}e^{-k_3 z_3}\\
&\times \Bigg(\cos\big((p_3-p_2)z_3\big)\cos\big((p_3+p_1)z_1\big)\cos\big((p_1+p_2)z_2\big)\Bigg)\\
&\times\;\frac{1}{p_3^2+\vec{q}^2}\frac{1}{p_1^2+(\vec{q}+\vec{k_1})^2}\frac{1}{p_2^2+(\vec{q}+\vec{y}_{12})^2}.
\end{split}
\ee 
This time all three $z$ integrals are divergent. Using (\ref{eq:CCSuppressedIntegral}), we obtain
\be 
\begin{split}
\mathcal{A}^{(2)}_{3} = -\lambda^3\int d^3q \int_{-\infty}^\infty dp_1 \, dp_2\, dp_3\,  \frac{1}{p_3^2+\vec{q}^2}\frac{1}{p_1^2+(\vec{q}+\vec{k_1})^2}
\frac{1}{p_2^2+(\vec{q}+\vec{y}_{12})^2}.
\label{eq:CC3RingSuppFin}
\end{split}
\ee

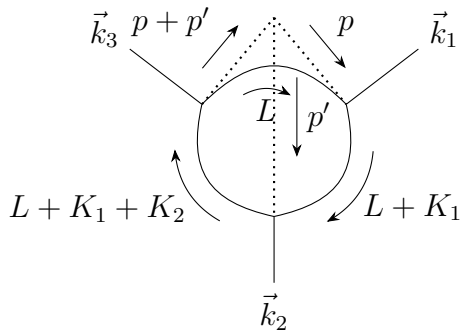
\begin{figure}[H]
    \centering
        \begin{tikzpicture}[baseline=(b.base)]
        \begin{feynman}
            \vertex (a) ;
            \vertex [right = 37.5pt of a       ] (b);
            \vertex [right = 37.5pt of b        ] (c);
            \vertex [below = 50pt of b          ] (ok) {\(\vec{k}_2\)};
            \vertex [ below = 25pt of b          ] (ik) ;
            \vertex [ above right = 34.67pt and 54pt of b](oj) {\(\vec{k}_1\)};
            \vertex [ above right = 17.34pt and 27pt of b](ij);
            \vertex [above left = 34.67pt and 54pt of b](oi) {\(\vec{k}_3\)} ;
            \vertex [ above left = 17.34pt and 27pt of b](ii) ;
            \vertex [above = 50pt of b](e);
    
            \diagram* {
                (oi) -- (ii),
                (oj) -- (ij),
                (ok) -- (ik),
                (ii) --[ quarter left, looseness=1.3, momentum' ={[xshift=-.1cm] \(L\)}] (ij),
                (ij) --[ quarter left, looseness=1.3, momentum = \(L+K_1\), inner sep=1pt] (ik),
                (ik) --[ quarter left, looseness=1.3, momentum = \(L+K_1+K_2\), inner sep=1pt] (ii),
                (ii) --[ghost,  momentum=\(p+p'\),near end](e),
                (e)--[ghost, momentum=\(p\)](ij),
                (e)--[ghost, momentum={[arrow shorten=0.3]\(p'\)},near end](ik),
            };
        \end{feynman}
    \end{tikzpicture}
    \caption{Dressed diagram with three dotted auxiliary propagators, which corresponds to the sum in Figure \ref{fig:CC3ThreePointRingShadow2} \label{fig:CC3ThreePointRingDressed2}.
    }
\end{figure}
The equivalent dressed diagram is shown in Fig.\ \ref{fig:CC3ThreePointRingDressed2} and the dressing rule gives the following expression,
\be 
\begin{split}
\mathcal{A}^{(2)}_{3}=& \lambda^3 \int dp \, dp' \int d^3q \, d\omega (-\pi)(-\pi)(-\pi)\\
&\times\Bigg(\frac{1}{\pi}\frac{1}{\omega^2+\vec{q}^2}\Bigg)\Bigg(\frac{1}{\pi}\frac{1}{(\omega+p)^2+(\vec{q}+\vec{k}_1^2)}\Bigg)\Bigg(\frac{1}{\pi}\frac{1}{(\omega+p+p')^2+(\vec{q}+\vec{y}_{12})^2}\Bigg)\\
&=-\lambda^3 \int dp \, dp' \int d^4L\frac{1}{L^2(L+K_1)^2(L+K_1+K_2)^2}
\end{split}
\ee 
which agrees with \label{eq:CC3RingSuppFin} after applying the variable shifts given in \eqref{eq:3ptVarShift} to the former.

\subsection{Massless theories} \label{sec:dressmassless}
Here we focus on correlators of massless fields with $\phi^3$ and $\phi^4$ interactions. These correlators exhibit IR divergences, which we will regulate using 
dimensional regularization, setting the boundary dimension to be $d=3+ 2\epsilon$. For a simplified example of applying IR-divergent dressing rules and performing all integrals, the interested reader should see the supplementary Mathematica file available in \cite{github}. For a derivation of the auxiliary propagators presented here, see appendix \ref{app:MasslessAuxProp}.

Apart from the $\epsilon$-dependence, the auxiliary propagators for these theories take a similar form to those of conformally coupled $\phi^3$ except that they have nontrivial numerators. We therefore define the following numerators: 
\be 
\mathcal{N}_v(s,\{k\},\{p\},\epsilon)=  \frac{s^{-\epsilon}}{a_M}\sum_{i,j=0}^{i+j=v} \frac{a_j b_i s^{3-i-j}}{\Gamma(4-i-j-\epsilon)}\big((s+k_{\rm ext}) \cos(\frac{\pi}{2} j)+ p_{\rm tot} \sin(\frac{\pi}{2} j)\big)
\label{eq:t0Integrand}
\ee 
and 
\be 
\widetilde{\mathcal{N}}_v(s,\{k\},\{p\},\epsilon) =  \frac{ s^{-\epsilon}}{a_M}\sum_{i,j=0}^{i+j=v} \frac{a_j b_i s^{3-i-j}}{\Gamma(4-i-j-\epsilon)}\big( p_{\rm tot} \cos(\frac{\pi}{2} j)-(s+k_{\rm ext}) \sin(\frac{\pi}{2} j)\big),
\label{eq:t1Integrand}
\ee 
where $v$ is the valency of the interaction vertex, 
\be 
a_j= \sum_{l_1<\ldots <l_j} p_{l_1}\ldots p_{l_j},\,\,\,\,
b_j= \sum_{l_1<\ldots <l_j} k_{l_1}\ldots k_{l_j}\,,
\label{eq:aParametersDef}
\ee 
and the $p$ and $k$ variables are energies of internal and external propagators attached to the vertex, respectively. If a vertex has $i$ bulk-boundary propagators and $j$ bulk-bulk propagators attached, $b_{i+m}=a_{j+m}=0$ for $m>0$.
Moreover $M$ is the number of internal propagators at the vertex and 
\be 
a_M = \prod_{j=1}^M p_j ; \qquad a_j =0 \, \text{ if } \, j>M.
\ee 

Using the above definitions, the auxiliary propagators for a massless $\phi^v$ theory are
\begin{itemize}
    \item dashed auxiliary propagator
        \begin{equation}
        \begin{aligned}
        \begin{tikzpicture}[baseline=($(b)!0.25!(w)$.base)]
            \begin{feynman}
                \vertex (a);
                \vertex [ right = 50pt of a, dot    ] (b) {};
                \vertex [ right = 40pt  of b](w);
        
                \diagram* {
                    (b)  -- [momentum= $p_{\rm tot}$, scalar] (w),
                };
            \end{feynman}
        \end{tikzpicture} =-2 \cos\left(\frac{\pi}{2} \epsilon \right) \int_0^\infty ds\, \frac{\mathcal{N}_v}{p_{\rm tot}^2+(s+k_{\rm ext})^2},
        \label{eq:M3SuppAuxProp}
    \end{aligned}
    \end{equation}
    of which there can be any number.

    \item dotted auxiliary propagator\footnote{As in the conformally coupled case, one of the auxiliary propagators has a factor of $i$ which may seem unusual. The auxiliary propagators do not have a lagrangian description, hence the normalisation has no physical significance. Rather, it is used to ensure the correct relative factors between contributions to the same correlator, without having to introduce additional confusing rules.}
    \begin{equation}
        \begin{aligned}
        \begin{tikzpicture}[baseline=($(b)!0.25!(w)$.base)]
            \begin{feynman}
                \vertex (a);
                \vertex [ right = 50pt of a, dot      ] (b){};
                \vertex [ right = 40pt  of b](w);
        
                \diagram* {
                    (b)  -- [momentum= $p_{\rm tot}$, ghost] (w),
                };
            \end{feynman}
        \end{tikzpicture} =+2i  \sin\left(\frac{\pi}{2} \epsilon\right) \int_0^\infty ds\, \frac{\widetilde{\mathcal{N}}_v}{p_{\rm tot}^2+(s+k_{\rm ext})^2}\,,
        \label{eq:M3UnsuppAuxProp}
    \end{aligned}
    \end{equation}
of which there must be an even number.
\end{itemize}
Note that the coefficient of the auxiliary propagator in \eqref{eq:M3UnsuppAuxProp} vanishes as $\epsilon \rightarrow 0$, but this will be balanced by IR divergences coming from the $s$ integral, giving a non-zero finite contribution. On the other hand, the auxiliary propagator in \eqref{eq:M3SuppAuxProp} will give rise to $1/\epsilon$ terms, and therefore encodes IR divergences arising from the expansion of spacetime.

The dressing rules for all the theories we consider are summarised in table \ref{tab:AuxPropSummary}. Analogous dressing rules for the wavefunction are discussed in Appendix \ref{app:WFdress}. As shown in the appendix, the dressing rules for the wave function contain denominators that explicitly violate Lorentz invariance and hence are not as simple as the dressing rules for in-in correlators\footnote{See \cite{Lee:2023kno} for a similar observation which pointed out that the poles of the in-in correlator overlap with the poles of the S-matrix.}. We will conclude this section by illustrating the massless dressing rules for a simple example and matching it with the shadow formalism. 
\renewcommand{\arraystretch}{1.5}

\begin{table}[h]
\centering
\begin{tabular}{| >{\centering\arraybackslash}m{2cm} | >{\centering\arraybackslash}m{2cm} | >{\centering\arraybackslash}m{2cm} | >{\centering\arraybackslash}m{7cm} |}
        \hline
        Theory & Propagator Type & Requires an Even Number? & Schematic Propagator Form  \\
        \hline
        conformally coupled $\phi^4$  & Dashed & No & $
        \frac{-2k_{\rm ext}}{p_{\rm tot}^2+k_{\rm ext}^2}$ \\
        \hline
        conformally coupled $\phi^3$ & Dashed & Yes & $
        2i\int_0^\infty ds \, \frac{p_{\rm tot}}{p_{\rm tot}^2+(s+k_{\rm ext})^2}$ \\
        &Dotted & No & $
        -\pi $
        \\
        \hline 
        massless $\phi^4$ & Dashed & No & $-2 \cos(\frac{\pi}{2}\epsilon)\int_0^\infty ds \, \frac{\mathcal{N}_4}{p_{\rm tot}^2+(s+k_{\rm ext})^2}$ \\ 
         & Dotted & Yes & $ 2i\sin(\frac{\pi}{2} \epsilon)\int_0^\infty ds \, \frac{\widetilde{\mathcal{N}}_4}{p_{\rm tot}^2+(s+k_{\rm ext})^2}$ \\ \hline 
         massless $\phi^3$ & Dashed & No & $-2 \cos(\frac{\pi}{2}\epsilon)\int_0^\infty ds \, \frac{\mathcal{N}_3}{p_{\rm tot}^2+(s+k_{\rm ext})^2}$  \\ 
          & Dotted & Yes & $ 2i\sin(\frac{\pi}{2} \epsilon)\int_0^\infty ds \, \frac{\widetilde{\mathcal{N}}_3}{p_{\rm tot}^2+(s+k_{\rm ext})^2}$  \\ \hline
    \end{tabular}
    \caption{Schematic dressing rules for conformally coupled and massless  $\phi^3$ and $\phi^4$. In the massless theories, the numerators are polynomials in $s$, demonstrating IR divergences. These are defined in (\ref{eq:t0Integrand}) and (\ref{eq:t1Integrand}). }
    \label{tab:AuxPropSummary}
\end{table}
For completeness, let us state the shadow action and subsequent Feynman rules for massless $\phi^3$ and $\phi^4$ theory. Similarly to \eqref{eq:EAdS_action_gen_pot_epsilon}, we now set $d=3+2 \epsilon$ which gives $\Delta_- = \epsilon$, and $\Delta_+=3+\e$. Previously there were IR divergences only at intermediate steps but the final result was finite. Here we shall see that the IR divergences persist and are regulated by $\e$ in the final result. Using these values in \eqref{eq:EAdS_action_gen_pot} with $V(\phi)=\frac{\lambda}{v!} \phi^v$ gives the
$\phi^3$ action,
\begin{eqn}
    \label{eq:Mphi3Action}
	L=(\partial\phi_+)^2-(\partial\phi_-)^2&+\lambda\cos\left(\frac{\pi \e}{2}\right)\big(\phi_- \phi_+^2 - \frac{1}{3}\phi_-^3 \big)\\
    &+\lambda \sin\left(\frac{ \pi \e}{2} \right)\big(\frac{1}{3}\phi_+^3 -\phi_- \phi_+^2 \big).
\end{eqn}
with Feynman rules shown in Table \ref{tab:vertex-factorsM3}. Similarly, the action for the massless $\phi^4$ theory  takes a similar form,
\begin{eqn}
    \label{eq:Mphi4Action}
	L=(\partial\phi_+)^2-(\partial\phi_-)^2&-\frac{\lambda\cos\left(\frac{\pi \e}{2}\right)}{2}\big(\frac{1}{6} \phi_-^4 -\phi_-^2 \phi_+^2 + \frac{1}{6} \phi_+^4 \big)\\
    &+\frac{\lambda \sin\left(\frac{ \pi \e}{2} \right)}{3}\big(\phi_-^3 \phi_+ - \phi_- \phi_+^3 \big).
\end{eqn}
for which the Feynman rules are shown in Table \ref{tab:vertex-factorsM4}. Note that we do not take $\e \to 0$ at this step. Due to the IR divergences in this theory, subleading terms in $\cos(\frac{\pi \e}{2})$ and $\sin(\frac{\pi \e}{2})$ can lead to finite contributions when one eventually expands in small $\e$. 
\begin{table}[H]
    \centering
    \begin{tabular}{c|cccc}
        Vertex & $(\phi^+)^3$ & $(\phi^+)^2 \phi^-$ & $\phi^+ (\phi^-)^2$ & $(\phi^-)^3$ \\
        \hline
        Factor  & $-2 \lambda\sin\big(\frac{\pi \epsilon}{2}\big)$ & $2\lambda \cos\big(\frac{\pi \epsilon}{2}\big)$ & $2 \lambda \sin\big(\frac{\pi \epsilon}{2}\big)$ & $-2 \lambda \cos\big(\frac{\pi \epsilon}{2}\big)$        
    \end{tabular}
    \caption{The Feynman vertex rules of massless $\phi^3$ in dimensional regularisation. \label{tab:vertex-factorsM3}}
\end{table}
\begin{table}[H]
    \centering
    \begin{tabular}{c|ccccc}
        Vertex & $\phi_-^4$ & $\phi_-^3 \phi_+$ & $\phi_-^2 \phi_+^2$ & $\phi_- \phi_+^3$ & $\phi_+^4$\\
        \hline
        Factor  & $-2 \lambda\cos\big(\frac{\pi \epsilon}{2}\big)$ & $2 \lambda\sin\big(\frac{\pi \epsilon}{2}\big)$ & $2 \lambda\cos\big(\frac{\pi \epsilon}{2}\big)$ & $-2 \lambda\sin\big(\frac{\pi \epsilon}{2}\big)$   &$-2 \lambda\cos\big(\frac{\pi \epsilon}{2}\big)$     
    \end{tabular}
    \caption{The Feynman vertex rules of massless $\phi^4$ in dimensional regularisation. \label{tab:vertex-factorsM4}}
\end{table}

Comparing the conformally coupled actions in \eqref{eq:CCphi4Action} and \eqref{eq:EAdS_action_gen_pot_epsilon} to the massless actions in \eqref{eq:Mphi3Action} and \eqref{eq:Mphi4Action}, we see that the $\phi_-$ and $\phi_+$ fields have swapped roles. For example, the $\phi_-^3$ vertex is suppressed in the conformally coupled theory but unsuppressed in the massless theory. Furthermore, the signs of the kinetic terms have swapped between the two cases. This is why the rules governing the number of dashed versus dotted auxiliary propagators given in Table \ref{tab:AuxPropSummary} are opposite for massless and conformally coupled theories.

\subsection*{Tree-level 3-point}

\begin{figure}
    \centering    
    \begin{subfigure}{0.45\textwidth}
    \centering
    \begin{tikzpicture}[baseline=(b.base)]
        \begin{feynman}
            \vertex (a) ;
            \vertex [ right = 75pt of a       ] (b);
            \vertex [right = 75pt of b        ] (c) {\(\vec{k_3}\)};
            \vertex [above left = 64pt and 37pt of b](oi) {\(\vec{k_1}\)};
            \vertex [below left = 64pt and 37pt of b](oj) {\(\vec{k_2}\)};
    
            \diagram* {
                (oi) --[scalar] (b),
                (oj) --[scalar] (b),
                (c)  --[scalar] (b),
            };
            \draw[thick] ($(a)!0.5!(c)$) circle [
            radius=75pt
            ];
        \end{feynman}
    \end{tikzpicture}
    \caption{\label{sfig:M3ContactShadow}Shadow}
    \end{subfigure}
    \begin{subfigure}{0.45\textwidth}
    \centering
    \begin{tikzpicture}[baseline=(b.base)]
        \begin{feynman}
            \vertex (a) ;
            \vertex [ right = 75pt of a       ] (b);
            \vertex [right = 75pt of b        ] (c) {\(\vec{k_3}\)};
            \vertex [above left = 64pt and 37pt of b](oi) {\(\vec{k_1}\)};
            \vertex [below left = 64pt and 37pt of b](oj) {\(\vec{k_2}\)};
            \vertex [above = 7pt of b](x);
            \vertex [above right  = 41.6pt and 41.6pt of b](e);
    
            \diagram* {
                (oi) -- (b),
                (oj) -- (b),
                (c)  -- (b),
                (b)  --[scalar, momentum = $p{=}0$] (e),
            };
        \end{feynman}
    \end{tikzpicture}
    \caption{\label{sfig:M3ContactDressed} Dressed }
    \end{subfigure}
    \caption{The three-point contact diagram in massless $\phi^3$ from the shadow formalism and the dressing rule.
    \label{fig:M3ContactDressed}}
\end{figure}
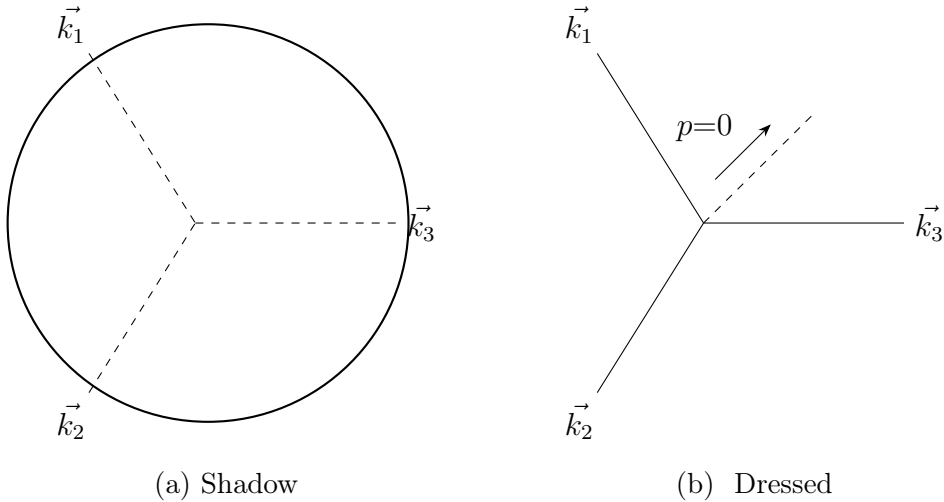

Let us consider the tree-level 3-point correlator in massless $\phi^3$ theory. From the Feynman rules in Table \ref{tab:vertex-factorsM3} we see that there is a single shadow diagram, which is shown in Figure \ref{fig:M3ContactDressed}.


Using \eqref{eq:EAdSProps} and the Feynman rule for the $\left(\phi^{-}\right){}^{3}$ vertex in Table \ref{tab:vertex-factorsM3}, we find
\be 
\begin{split}
\mathcal{A}_{3}&=-2 \lambda \cos\left(\frac{\pi\epsilon}{2} \right) \int dz \, z^{-4+\epsilon} (1+k_1 z)(1+k_2 z)(1+k_3 z)e^{-k_t z}\\
&=-2 \lambda \cos\left(\frac{\pi\epsilon}{2} \right) \int dz \, \sum_{i=0}^3 b_i z^{-4+i+\epsilon} e^{-k_{123} z}.
\end{split}
\label{eq:M3ContactShadow}
\ee 
Using \eqref{eq:sIntIdentity} then gives  
\be \label{eq:massless3ptcontactshadow}
\mathcal{A}_{3} = -2 \lambda \cos\left(\frac{\pi\epsilon}{2} \right) \int_{0}^\infty ds\,\sum_{i=0}^3 \frac{1}{s+k_{123}}\frac{b_i s^{3-i-\epsilon}}{\Gamma(4-i-\epsilon)}.
\ee 
On the other hand, if we consider the dressed diagram in Figure \ref{fig:M3ContactDressed} with the auxiliary propagator given by (\ref{eq:t0Integrand}) and \eqref{eq:M3SuppAuxProp} with $p_j=0$ we obtain
\be 
\mathcal{A}_{3} = \lambda \Big(-2\cos\left(\frac{\pi\epsilon}{2} \right)\int_0^\infty ds \, \sum_{i=0}^3 \frac{1}{(s+k_{123})^2} \frac{b_i s^{3-i-\epsilon}}{\Gamma(4-i-\epsilon)}(s+k_{123})\Big)~,
\ee 
which agrees with the result from shadow \eqref{eq:massless3ptcontactshadow}. We will explain how to evaluate this integral and integrals for other massless tree-level diagrams in section \ref{sec:SK}. A general derivation of the dressing rules for massless correlators is given in Appendix \ref{app:MasslessAuxProp}.

\section{Conformally coupled 5-point} \label{sec:fivept}
To illustrate the power of the dressing rules, in this section we will compute for the first time the tree-level 5-point in-in correlator in conformally coupled $\phi^3$ theory. The Feynman diagrams contributing to the corresponding 5-point amplitude have three interactions vertices, and therefore according to the dressing rules we expect to have four dressings for each flat space Feynman diagram. In particular, since dashed auxiliary propagators must appear in pairs, we can dress a flat space Feynman diagram with three dotted auxiliary propagators, or we can dress it in three different ways with two dashed auxiliary propagators and one dotted auxiliary propagator.

The first dressed diagram we consider is
\begin{eqn}
&\mathcal{A}_{5}^{(1)} \equiv
\begin{tikzpicture}[baseline=(b.base)]
        \begin{feynman}
            \vertex (a) ;
            \vertex [ right = 50pt of a       ] (i1) ;
            \vertex [ right = 50pt of i1      ] (i2);
            \vertex [right = 75pt of a      ] (b);
            \vertex [right = 75pt of b      ] (c);
            \vertex [above left = 26.39pt and 63.71pt of b](l1) {\(\vec{k_2}\)};
            \vertex [below left = 26.39pt and 63.71pt of b](l2) {\(\vec{k_1}\)};
            \vertex [below right = 26.39pt and 63.71pt of b](r1) {\(\vec{k_4}\)};
            \vertex [above right = 26.39pt and 63.71pt of b](r2) {\(\vec{k_3}\)};
            \vertex [below = 30pt of b](m) {\(\vec{k_5}\)};
            \vertex[above = 50pt of b](e);
            \diagram* {
                (l1) --(i1),
                (l2) -- (i1),
                (r1)  --(i2),
                (r2)  --(i2),
                (b) --(m),
                (i1) --[momentum'=$p_1$] (b),
                (b) --[momentum'=$p_2$](i2),
                (e) --[scalar, out=220,in=90,  momentum' = $p_1$] (i1),
                (i2) --[scalar, out=90,in=320,momentum' = $p_2$] (e),
                (b) --[ghost,momentum = {[label distance=-1cm]$\substack{p_1\,\\-p_2}$}] (e),
            };
        \end{feynman}
    \end{tikzpicture}\\
    &= 
\frac{\lambda^3}{\pi}\int_0^\infty ds_1 \,ds_3\int_{-\infty}^\infty dp_1 \, dp_2  \frac{p_1}{p_1^2+(s_1+k_{12})^2}\frac{p_2}{p_2^2+(s_3+k_{34})^2}\frac{1}{p_1^2+y_{12}^2}\frac{1}{p_2^2+y_{34}^2}=0
\end{eqn}
where $s_1$ is the variable associated to the left auxiliary propagator, and $\vec{y}_{ij}=\vec{k}_i +\vec{k}_j$. This integral is zero because each $p$ integrand is an odd function. The second dressed diagram we consider is
\be
\begin{split}
    & \mathcal{A}_{5}^{ (2)}\equiv 
    \begin{tikzpicture}[baseline=(b.base)]
        \begin{feynman}
            \vertex (a) ;
            \vertex [ right = 50pt of a       ] (i1) ;
            \vertex [ right = 50pt of i1      ] (i2);
            \vertex [right = 75pt of a      ] (b);
            \vertex [right = 75pt of b      ] (c);
            \vertex [above left = 26.39pt and 63.71pt of b](l1) {\(\vec{k_2}\)};
            \vertex [below left = 26.39pt and 63.71pt of b](l2) {\(\vec{k_1}\)};
            \vertex [below right = 26.39pt and 63.71pt of b](r1) {\(\vec{k_4}\)};
            \vertex [above right = 26.39pt and 63.71pt of b](r2) {\(\vec{k_3}\)};
            \vertex [below = 30pt of b](m) {\(\vec{k_5}\)};
            \vertex[above = 50pt of b](e);
            \diagram* {
                (l1) --(i1),
                (l2) -- (i1),
                (r1)  --(i2),
                (r2)  --(i2),
                (b) --(m),
                (i1) --[momentum'=$p_1$] (b),
                (b) --[momentum'=$p_2$](i2),
                (e) --[ghost, out=220,in=90,  momentum' = $p_1$] (i1),
                (i2) --[scalar, out=90,in=320,momentum' = $p_2$] (e),
                (b) --[scalar, momentum = {[label distance=-1cm]$\substack{p_1\,\\-p_2}$}] (e),
            };
        \end{feynman}
    \end{tikzpicture} \\
    &=
    \frac{\lambda^3}{\pi}\int_0^\infty ds_2 \,ds_3\int_{-\infty}^\infty dp_1 \, dp_2 \frac{p_1-p_2}{(s_2+k_5)^2+(p_1-p_2)^2}  \frac{p_2}{(s_3+k_{34})^2+p_2^2}\frac{1}{p_1^2+y_{12}^2}\frac{1}{p_2^2+y_{34}^2}
\end{split}
\ee
which can be straightforwardly integrated as there are no IR divergences and the form of the integral is similar to 4-pt function (see \eqref{eq:conformalphi3exshadow}).

The third diagram is related to the previous one by relabelling:
\begin{eqn}
\mathcal{A}_{5}^{(3)}\equiv 
\begin{tikzpicture}[baseline=(b.base)]
        \begin{feynman}
            \vertex (a) ;
            \vertex [ right = 50pt of a       ] (i1) ;
            \vertex [ right = 50pt of i1      ] (i2);
            \vertex [right = 75pt of a      ] (b);
            \vertex [right = 75pt of b      ] (c);
            \vertex [above left = 26.39pt and 63.71pt of b](l1) {\(\vec{k_2}\)};
            \vertex [below left = 26.39pt and 63.71pt of b](l2) {\(\vec{k_1}\)};
            \vertex [below right = 26.39pt and 63.71pt of b](r1) {\(\vec{k_4}\)};
            \vertex [above right = 26.39pt and 63.71pt of b](r2) {\(\vec{k_3}\)};
            \vertex [below = 30pt of b](m) {\(\vec{k_5}\)};
            \vertex[above = 50pt of b](e);
            \diagram* {
                (l1) --(i1),
                (l2) -- (i1),
                (r1)  --(i2),
                (r2)  --(i2),
                (b) --(m),
                (i1) --[momentum'=$p_1$] (b),
                (b) --[momentum'=$p_2$](i2),
                (e) --[scalar, out=220,in=90,  momentum' = $p_1$] (i1),
                (i2) --[ghost,  out=90,in=320,momentum' = $p_2$] (e),
                (b) --[scalar, momentum = {[label distance=-1cm]$\substack{p_1\,\\-p_2}$}] (e),
            };
        \end{feynman}
    \end{tikzpicture} =\mathcal{A}_5^{(2)}\Bigg|_{(12)\leftrightarrow (34)}.
\end{eqn}
Hence we do not need to evaluate another integral. The sum of these two diagrams is
\begin{eqn}
&\mathcal{A}_{5}^{(2)}+\mathcal{A}_{5}^{(3)} =\frac{\lambda^3\pi}{2 y_{12} y_{34}}\Big(\log \frac{1-v_-}{u_+} \log(1-v_--u_+)+\log\frac{1-u_-}{v_+}\log(1-u_--v_+)\\
    &\qquad\qquad+\log(1-u_--v_-)\log\frac{u_+ v_+}{(1-u_-)(1-v_-)} \\
    &\qquad \qquad+\mbox{Li}_2 \frac{u_-}{1-v_-}+\mbox{Li}_2\frac{v_-}{1-u_-}-\mbox{Li}_2\frac{v_+}{1-u_-}-\mbox{Li}_2\frac{u_+}{1-v_-}\Big)
\end{eqn}
where $u_\pm$ and $v_\pm$ are analogues of the cross ratios at five-point, defined by
\begin{eqn}\label{5ptcross}
2 k_{12} &= k_{12345}(u_+ + u_-), \qquad 2k_{34} =k_{12345}(v_+ + v_-),  \\
  2y_{12} &= k_{12345}(u_+ - u_-), \qquad 2 y_{34} = k_{12345}(v_+ - v_-)~.
\end{eqn}
The final diagram we consider has only trivial auxiliary propagators and is easily evaluated:

\be 
\begin{split}
&\mathcal{A}_{5}^{(4)}\equiv
\begin{tikzpicture}[baseline=(b.base)]
        \begin{feynman}
            \vertex (a) ;
            \vertex [ right = 50pt of a       ] (i1) ;
            \vertex [ right = 50pt of i1      ] (i2);
            \vertex [right = 75pt of a      ] (b);
            \vertex [right = 75pt of b      ] (c);
            \vertex [above left = 26.39pt and 63.71pt of b](l1) {\(\vec{k_2}\)};
            \vertex [below left = 26.39pt and 63.71pt of b](l2) {\(\vec{k_1}\)};
            \vertex [below right = 26.39pt and 63.71pt of b](r1) {\(\vec{k_4}\)};
            \vertex [above right = 26.39pt and 63.71pt of b](r2) {\(\vec{k_3}\)};
            \vertex [below = 30pt of b](m) {\(\vec{k_5}\)};
            \vertex[above = 50pt of b](e);
            \diagram* {
                (l1) --(i1),
                (l2) -- (i1),
                (r1)  --(i2),
                (r2)  --(i2),
                (b) --(m),
                (i1) --[momentum'=$p_1$] (b),
                (b) --[momentum'=$p_2$](i2),
                (e) --[ghost, out=220,in=90,  momentum' = $p_1$] (i1),
                (i2) --[ghost, out=90,in=320,momentum' = $p_2$] (e),
                (b) --[ghost, momentum = {[label distance=-1cm]$\substack{p_1\,\\-p_2}$}] (e),
            };
        \end{feynman}
    \end{tikzpicture} 
    =- \lambda^3\pi \int_{-\infty}^\infty dp_1 \, dp_2 \frac{1}{(p_1^2+y_{12}^2)(p_2^2+y_{34}^2)}\\
&=-\frac{\lambda^3 \pi^3}{y_{12}y_{34}}.
\end{split}
\ee 

We see that although there were four diagrams, we only actually had to evaluate two integrals, one of which was trivial. The total correlator is 
\begin{eqn}
&\mathcal{A}_{5}=\sum_i \mathcal{A}_{5}^{(i)}+\text{perms} \\
    =&\frac{\pi}{2 y_{12} y_{34}}\Big(\log \frac{1-v_-}{u_+} \log(1-v_--u_+)+\log\frac{1-u_-}{v_+}\log(1-u_--v_+)\\
    &\qquad\qquad+\log(1-u_--v_-)\log\frac{u_+ v_+}{(1-u_-)(1-v_-)} \\
    &\qquad \qquad+\mbox{Li}_2 \frac{u_-}{1-v_-}+\mbox{Li}_2\frac{v_-}{1-u_-}-\mbox{Li}_2\frac{v_+}{1-u_-}-\mbox{Li}_2\frac{u_+}{1-v_-}-2 \pi^2\Big)+\text{perms} 
\end{eqn}
where $+$ perms means permutations over the external legs. We shall see section \ref{sec:WF} that obtaining this result via the  wavefunction is highly non-trivial, with large and complicated expressions vanishing in the analytic continuation. In particular, we shall find that the in-in correlator has lower transcendentality than the wavefunction, even at tree-level, and can be computed in a much simpler way by the dressing rules. Note that one could have guessed the appearance of the dilogarithm before evaluating the integral by the fact at most two non-trivial auxiliary propagators appear in any diagram.

\section{Comparison to Schwinger-Keldysh formalism} \label{sec:SK}
In this section we will compare the correlators obtained via dressing rules to the correlators computed using Schwinger-Keldysh formalism, finding complete agreement, irrespective of the presence of IR-divergences.

\subsection{Propagators}
The two kinds of fields in Schwinger-Keldysh will be referred to as the time-ordered (T) and anti-time-ordered (A) fields.  There are four kinds of bulk-bulk propagators which are often referred to as the Wightman functions. These are represented as $\Lambda_{TT}^{(\Delta)}(\eta, \eta'; \vec k)$, $\Lambda^{(\Delta)}_{AA}(\eta, \eta'; \vec k)$, $\Lambda^{(\Delta)}_{AT}(\eta, \eta'; \vec k)$ and $\Lambda_{TA}^{(\Delta)}(\eta, \eta'; \vec k)$ with $\Delta$ denoting the conformal dimension of the exchanged fields. These arise via the Wick contractions between the time-ordered and anti-time-ordered fields. Their explicit expressions are given as 
\begin{eqn}
\Lambda_{TT}^{(\Delta)}(\eta, \eta'; \vec k) &=  (\eta \eta')^{d/2} \Big[  \Theta(\eta - \eta')  H_{\b}^{(2)}(k \eta) H_{\b}^{(1)}(k \eta') +  \Theta(\eta' - \eta)  H_{\b}^{(2)}(k \eta') H_{\b}^{(1)}(k \eta)\Big], \\
\Lambda_{AA}^{(\Delta)}(\eta, \eta'; \vec k) &=  (\eta \eta')^{d/2} \Big[  \Theta(\eta - \eta')  H_{\b}^{(2)}(k \eta') H_{\b}^{(1)}(k \eta) +  \Theta(\eta' - \eta)  H_{\b}^{(2)}(k \eta) H_{\b}^{(1)}(k \eta')  \Big], \\
\Lambda_{TA}^{(\Delta)}(\eta, \eta'; \vec k) &=   (\eta \eta')^{d/2} \Big[  H_{\b}^{(2)}(k \eta') H_{\b}^{(1)}(k \eta)\Big], \\
\Lambda_{AT}^{(\Delta)}(\eta, \eta'; \vec k) &=  (\eta \eta')^{d/2} \Big[  H_{\b}^{(2)}(k \eta) H_{\b}^{(1)}(k \eta') \Big]~,
\end{eqn}
where $H$ is a Hankel function, $\vec{k}$ is the boundary momentum flowing through the propagators, and 
\begin{eqn}
\beta^2 = \frac{d^2}{4} + \Delta(\Delta - d)~.
\end{eqn}
These will be denoted via the following diagrams 
\begin{eqn*}
\Lambda_{TT}^{(\Delta)}(\eta, \eta'; \vec k) &= \begin{tikzpicture}[baseline]
\draw[fermion, very thick, black] (-1, 0) -- (1, 0);
\node at (0, -0.4) {$\Delta$};
\node at (0, 0.4) {$\vec k$};
\node at (-1, -0.25) {$\eta$};
\node at (1, -0.25) {$\eta'$};
\end{tikzpicture}, \qquad 
\Lambda_{AA}^{(\Delta)} (\eta, \eta'; \vec k) = \begin{tikzpicture}[baseline]
\draw[fermion, very thick, dotted,  black] (-1, 0) -- (1, 0);
\node at (0, -0.4) {$\Delta$};
\node at (0, 0.4) {$\vec k$};
\node at (-1, -0.25) {$\eta$};
\node at (1, -0.25) {$\eta'$};
\end{tikzpicture},\\
\Lambda_{TA}^{(\Delta)}(\eta, \eta'; \vec k) &= \begin{tikzpicture}[baseline]
\draw[very thick, black] (-1, 0) -- (0, 0);
\draw[fermion, very thick, dotted, black] (-1, 0) -- (1, 0);
\node at (0, -0.4) {$\Delta$};
\node at (0, 0.4) {$\vec k$};
\node at (-1, -0.25) {$\eta$};
\node at (1, -0.25) {$\eta'$};
\end{tikzpicture}, \qquad 
\Lambda_{AT}^{(\Delta)} (\eta, \eta'; \vec k) = \begin{tikzpicture}[baseline]
\draw[fermion, very thick, dotted, black] (-1, 0) -- (1, 0);
\draw[very thick, black] (0, 0) -- (1, 0);

\node at (0, -0.4) {$\Delta$};
\node at (0, 0.4) {$\vec k$};
\node at (-1, -0.25) {$\eta$};
\node at (1, -0.25) {$\eta'$};
\end{tikzpicture}~. 
\end{eqn*}
The bulk-boundary propagators are obtained by taking a limit of the bulk-bulk propagators. For example, 
\begin{eqn}\label{bulk-bndy-SK}
\Lambda^{(\Delta)}_T(\eta; \vec k) &= \lim_{\eta' \to 0} \eta'^{\beta-d/2} \Lambda^{(\b)}_{TT}(\eta, \eta'; \vec k) \propto \eta^{d/2} H_{\b}^{(1)}(k \eta), \\
\Lambda^{(\Delta)}_A(\eta; \vec k) &= \lim_{\eta' \to 0} \eta'^{\beta-d/2} \Lambda^{(\b)}_{AA}(\eta, \eta'; \vec k) \propto  \eta^{d/2} H_{\b}^{(2)}(k \eta).
\end{eqn}
These are diagrammatically represented by
\begin{eqn*}
\Lambda^{(\Delta)}_T(\eta, \vec k) = 
\begin{tikzpicture}[baseline]
\draw (-2, 1) -- (2, 1);
\draw[very thick] (-1, 1) -- (0, 0);
\node at (0, -0.25) {$\Delta$};
\end{tikzpicture}, 
\qquad 
\Lambda^{(\Delta)}_A(\eta, \vec k) = 
\begin{tikzpicture}[baseline]
\draw (-2, 1) -- (2, 1);
\draw[very thick, dotted] (-1, 1) -- (0, 0);
\node at (0, -0.25) {$\Delta$};
\end{tikzpicture}~.
\end{eqn*}

We consider the three- and four-point functions for theories consisting of massless fields. For these fields with $d=3$ and $\beta=\frac{3}{2}$, and the bulk-boundary propagators take the following form:
\begin{eqn}
\Lambda_T^{(2)}(\eta; \vec k) &= \eta e^{i k \eta} , \qquad 
\Lambda_A^{(2)}(\eta; \vec k) = \eta e^{-i k \eta} , \\
\Lambda_T^{(3)}(\eta; \vec k) &=  (1-i \eta  k) e^{i \eta  k}, \qquad 
\Lambda_A^{(3)}(\eta; \vec k) =   (1+i \eta  k) e^{- i \eta  k}~.
\end{eqn}
Note that we are dropping an overall normalisation $k^{d-2\Delta_-}$ in order to match the convention used in the previous section. The bulk-bulk propagators are 
\begin{eqn}\label{SKbulkbulk}
\Lambda_{TT}^{(3)}(\eta, \eta'; \vec k) &=  \Big[ \Theta(\eta - \eta') (1 + i k \eta) (1 - i k \eta')e^{-i k (\eta - \eta')} + \Theta(\eta' - \eta) (1 - i k \eta) (1 + i k \eta') e^{i k (\eta - \eta')} \Big] ~,\\
\Lambda_{AA}^{(3)}(\eta, \eta'; \vec k) &=  \Big[ \Theta(\eta - \eta') (1 + i k \eta') (1 - i k \eta) e^{i k (\eta - \eta')} + \Theta(\eta' - \eta) (1 + i k \eta) (1 - i k \eta') e^{- i k (\eta - \eta')} \Big] ~,\\
\Lambda_{TA}^{(3)}(\eta, \eta'; \vec k) &=  (1+ i k \eta') (1- i k \eta)e^{i k (\eta - \eta')}  ~,\\
\Lambda_{AT}^{(3)}(\eta, \eta'; \vec k) &=  (1- i k \eta') (1+ i k \eta)e^{-i k (\eta - \eta')}  ~.
\end{eqn}

\subsection{Massless $\phi^3$ 3-point}

Using the Schwinger-Keldysh propagators and vertex rules, we obtain the following
\begin{eqn}\label{eq:corr333}
&\mathcal{A}_{3}
= \begin{tikzpicture}[baseline]
\draw (-2, 1) -- (2, 1);
\draw[very thick] (-1, 1) -- (0, -0);
\draw[very thick] (0, 1) -- (0, -0);
\draw[very thick] (1, 1) -- (0, -0);
\node at (-1, 1.25) {$3$};
\node at (0, 1.25) {$3$};
\node at (1, 1.25) {$3$};
\end{tikzpicture}
+ \begin{tikzpicture}[baseline]
\draw (-2, 1) -- (2, 1);
\draw[very thick, dotted] (-1, 1) -- (0, -0);
\draw[very thick, dotted] (0, 1) -- (0, -0);
\draw[very thick, dotted] (1, 1) -- (0, -0);
\node at (-1, 1.25) {$3$};
\node at (0, 1.25) {$3$};
\node at (1, 1.25) {$3$};
\end{tikzpicture} \\
&=  -i \lambda \int_{-\infty}^0 \frac{d\eta}{\eta^4} \eta^\e \Bigg[\Lambda^{(3)}_{TT}(\eta; \vec k_1) \Lambda^{(3)}_{TT}(\eta; \vec k_2) \Lambda^{(3)}_{TT}(\eta; \vec k_3) -  \Lambda^{(3)}_{AA}(\eta; \vec k_1) \Lambda^{(3)}_{AA}(\eta; \vec k_2) \Lambda^{(3)}_{AA}(\eta; \vec k_3) \Bigg]   \\
&= -\frac{\lambda}{3\e} k_{123c} + \frac{\lambda}{3 } \Big[ k_{123c} \left(\gamma_E + \log k_{123}-1\right)- \frac13 k_{123}^3+ 3k_1 k_2 k_3 \Big],
\end{eqn}
where we introduce the notation
\begin{eqn}
k_{12\cdots n c} = k_{1}^3 + k_{2}^3 + \cdots + k_n^3~.
\end{eqn}
We use the same dimensional regularization\footnote{The deformation of $\Delta_\pm$ and $d$ enters the propagators only via powers of $\eta$, in analogy with \eqref{eq:EAdSProps}. Furthermore, the arguments of the logarithms can be made dimensionless by accounting for the scale introduced by the regulator, which we suppress. } scheme as previous sections and, as will be shown below, the answer matches with the result obtained via dressing rules. 

Using the dressing rules given in section \ref{sec:Dressing}, there is only one diagram to consider,

\be 
\begin{tikzpicture}[baseline=(b.base)]
        \begin{feynman}
            \vertex (a) ;
            \vertex [ right = 75pt of a       ] (b);
            \vertex [below right = 10pt and 70pt of b        ] (c) {\(\vec{k_3}\)};
            \vertex [above left = 26.39pt and 63.71pt of b](oi) {\(\vec{k_1}\)};
            \vertex [below left = 26.39pt and 63.71pt of b](oj) {\(\vec{k_2}\)};
            \vertex [above = 7pt of b](x);
            \vertex [above right  = 41.6pt and 41.6pt of b](e);
    
            \diagram* {
                (oi) -- (b),
                (oj) -- (b),
                (c)  -- (b),
                (b)  --[scalar] (e),
            };
        \end{feynman}
    \end{tikzpicture}= -2 \lambda \int_0^\infty ds \, \sum_{i=0}^3 \frac{1}{s+k_{123}} \frac{b_i s^{3-i-\epsilon}}{\Gamma(4-i-\epsilon)}
\label{eq:3ptContactIRStart}
\ee 
where the $b_i$'s are defined in (\ref{eq:aParametersDef}). For this case we get 
\begin{eqn}
    b_0=1, \quad b_1 = k_{123}, \quad 
    b_2 =k_1 k_2 + k_1 k_3 + k_2 k_3, \quad   b_3=k_1 k_2 k_3~. 
\end{eqn}
We may evaluate the $s$-integral directly 
to obtain 
\be 
\mbox{\eqref{eq:3ptContactIRStart}}
=2\lambda \sum_{i=0}^3 \frac{b_i k_{123}^{3-i-\epsilon}\pi \; \text{cosec}(\pi(i+\epsilon)) }{\Gamma(4-i-\epsilon)}.
\label{eq:3ptContactIRInt}
\ee 
One may then easily expand in small $\epsilon$ and perform the sum to find that it agrees with \eqref{eq:corr333} up to an unimportant normalisation factor. 
The divergent term in \eqref{eq:3ptContactIRInt} is 
\begin{eqn}
    \sum_{i=0}^3 (-1)^i\frac{b_i b_1^{3-i}}{ (3-i)!\e } = -\frac{b_1^3-3b_1 b_2 +3 b_3}{3 \epsilon}.
\end{eqn}
Substituting in the values of $b_i$ in terms of external energies reduces the numerator to $k_{123c}$.
We note that the divergent piece indeed agrees with that obtained using cutoff regularisation, for example using eqs.\ (3.8) and (2.22)
of \cite{cespedes2024ir}, if one associates $\frac{1}{\epsilon}$ with a logarithm of the cutoff.

\subsection{Massless $\phi^4$ 4-point}
The computation of the four point function 
proceeds in a similar way.
\begin{eqn}
&\mathcal{A}_{4}
= \begin{tikzpicture}[baseline]
\draw (-2, 1) -- (2, 1);
\draw[very thick] (-1.25, 1) -- (0, -0);
\draw[very thick] (-0.35, 1) -- (0, -0);
\draw[very thick] (0.35, 1) -- (0, -0);
\draw[very thick] (1.25, 1) -- (0, -0);
\node at (-1.25, 1.25) {$3$};
\node at (-0.35, 1.25) {$3$};
\node at (0.35, 1.25) {$3$};
\node at (1.25, 1.25) {$3$};
\end{tikzpicture}
+ \begin{tikzpicture}[baseline]
\draw (-2, 1) -- (2, 1);
\draw[very thick, dotted] (-1.25, 1) -- (0, -0);
\draw[very thick, dotted] (-0.35, 1) -- (0, -0);
\draw[very thick, dotted] (0.35, 1) -- (0, -0);
\draw[very thick, dotted] (1.25, 1) -- (0, -0);
\node at (-1.25, 1.25) {$3$};
\node at (-0.35, 1.25) {$3$};
\node at (0.35, 1.25) {$3$};
\node at (1.25, 1.25) {$3$};
\end{tikzpicture} \\
&=  -i \lambda \int_{-\infty}^0 \frac{d\eta}{\eta^4} \eta^\e \Bigg[\Lambda^{(3)}_{TT}(\eta; \vec k_1)\cdots \Lambda^{(3)}_{TT}(\eta; \vec k_4) -  \Lambda^{(3)}_{AA}(\eta; \vec k_1) \cdots  \Lambda^{(3)}_{AA}(\eta; \vec k_4) \Bigg]   \\
&= -\frac{ \lambda}{3}\Big(\frac{k_{1234c}}{\epsilon}+\big(1-\gamma_E-\log(k_{1234})\big)k_{1234c}+\frac{k_{1234}^3}{3}-3 \sum_{i<j<l}k_i k_j k_l+\frac{3k_1 k_2 k_3 k_4}{k_{1234}}\Big).
\label{eq:corr3333}
\end{eqn}
From the dressing rule perspective the four-point contact calculation is very similar to the three-point.
The dressed diagram is

\begin{eqn}
\begin{tikzpicture}[baseline=(b.base)]
            \begin{feynman}
                \vertex (a);
                \vertex [ right = 50pt of a       ] (b);
                \vertex [right = 50pt of b        ] (c);
                \vertex [above left = 35.36pt and 35.36pt of b](l1);
                \vertex [below left = 35.36pt and 35.36pt of b](l2) ;
                \vertex [above right = 35.36pt and 35.36pt of b](r1);
                \vertex [below right = 35.36pt and 35.36pt of b](r2);
                \vertex [below = 7pt of b](z);
                \vertex [above = 35pt of b](w);
        
                \diagram* {
                    (l1) -- (b),
                    (l2) -- (b),
                    (r1)  -- (b),
                    (r2)  -- (b),
                    (b)  -- [ scalar] (w),
                };
            \end{feynman}
        \end{tikzpicture} =
  -2\lambda \int_0^\infty ds \sum_{i=0}^4\frac{b_i s^{3-i-\epsilon}}{\Gamma(4-i-\epsilon)(s+k_{1234})}.
\end{eqn}
By performing the integral manner similar  to \eqref{eq:3ptContactIRStart} we find that this agrees with \eqref{eq:corr3333}. Furthermore, the leading divergence again agrees with the cutoff result; see eqs.\ (3.7) and (2.22) of \cite{cespedes2024ir}.

\subsection{Massless $\phi^3$ 4-point} \label{sec:masslessexchSK}
We now evaluate tree-level exchange graphs using Schwinger-Keldysh and compare them with the expressions using the dressing rules. This also requires us to use explicit expressions for the bulk-bulk propagators given in \eqref{SKbulkbulk}. The four diagrams contributing to the correlator are
\begin{eqn}\label{SKmasslessexch}
\mathcal{A}_4
&= \begin{tikzpicture}[baseline]
\draw (-2, 1) -- (2, 1);
\draw[very thick] (-1.75, 1) -- (-1.5, -0);
\draw[very thick] (-1.25, 1) -- (-1.5, -0);

\draw[very thick] (1.75, 1) -- (1.5, -0);
\draw[very thick] (1.25, 1) -- (1.5, -0);

\draw[very thick] (-1.5, 0) -- (1.5, 0);

\node at (0, -0.25) {$3$};

\node at (-1.75, 1.25) {$3$};
\node at (-1.25, 1.25) {$3$};
\node at (1.25, 1.25) {$3$};
\node at (1.75, 1.25) {$3$};
\end{tikzpicture} 
+ 
\begin{tikzpicture}[baseline]
\draw (-2, 1) -- (2, 1);
\draw[very thick, dotted] (-1.75, 1) -- (-1.5, -0);
\draw[very thick, dotted] (-1.25, 1) -- (-1.5, -0);

\draw[very thick, dotted] (1.75, 1) -- (1.5, -0);
\draw[very thick, dotted] (1.25, 1) -- (1.5, -0);

\draw[very thick, dotted] (-1.5, 0) -- (1.5, 0);

\node at (0, -0.25) {$3$};

\node at (-1.75, 1.25) {$3$};
\node at (-1.25, 1.25) {$3$};
\node at (1.25, 1.25) {$3$};
\node at (1.75, 1.25) {$3$};
\end{tikzpicture}  \\
&\quad+ \begin{tikzpicture}[baseline]
\draw (-2, 1) -- (2, 1);
\draw[very thick] (-1.75, 1) -- (-1.5, -0);
\draw[very thick] (-1.25, 1) -- (-1.5, -0);

\draw[very thick, dotted] (1.75, 1) -- (1.5, -0);
\draw[very thick, dotted] (1.25, 1) -- (1.5, -0);

\draw[very thick, dotted] (1.5, 0) -- (0, 0);
\draw[very thick] (-1.5, 0) -- (0, 0);

\node at (0, -0.25) {$3$};

\node at (-1.75, 1.25) {$3$};
\node at (-1.25, 1.25) {$3$};
\node at (1.25, 1.25) {$3$};
\node at (1.75, 1.25) {$3$};
\end{tikzpicture} 
+ 
\begin{tikzpicture}[baseline]
\draw (-2, 1) -- (2, 1);
\draw[very thick, dotted] (-1.75, 1) -- (-1.5, -0);
\draw[very thick, dotted] (-1.25, 1) -- (-1.5, -0);

\draw[very thick] (1.75, 1) -- (1.5, -0);
\draw[very thick] (1.25, 1) -- (1.5, -0);

\draw[very thick, dotted] (-1.5, 0) -- (0, 0);
\draw[very thick] (1.5, 0) -- (0, 0);

\node at (0, -0.25) {$3$};

\node at (-1.75, 1.25) {$3$};
\node at (-1.25, 1.25) {$3$};
\node at (1.25, 1.25) {$3$};
\node at (1.75, 1.25) {$3$};
\end{tikzpicture}
\end{eqn}
Using the techniques described in Appendix \ref{app:Integrals}, we can evaluate the integrals appearing in these diagrams to obtain
 \begin{eqn}
 &\mathcal{A}_{4}
 = - \frac{1}{\e^2} \frac{k_{1234c}y_{12}^3 +  2 k_{12c} k_{34c}}{ y_{12}^3 } \\
 &\quad + \frac{1}{\e y_{12}^3 } \Bigg[
-\left(y_{12}^2+k_3 k_4\right) k_{34} k_{12 c} -\left(y_{12}^2+k_1 k_2\right) k_{12} k_{34 c}-\frac{8}{3} k_{34 c} k_{12 c}
-\frac{y_{12} E k_{12 c} k_{34 c}}{k_{12} k_{34}}\\
&\qquad - \frac{7}{6}  y_{12}^3 \left(k_{12 c}+k_{34 c}\right)  + \left(k_{12 c}+ y_{12}^3\right) k_{34 c} \log (k_{12} + y_{12}) + \left(k_{34 c}+ y_{12}^3\right)k_{12 c}  \log (k_{34} + y_{12}) \\
&\qquad +\gamma_E  \big\{ y_{12}^3 \left(k_{12 c}+k_{34 c}\right)+2 k_{12 c} k_{34 c}\big\}  \Bigg] + O(\e^0),
 \end{eqn}
where we have only quoted the divergent parts. The finite part can be found in Appendix 
\ref{app:shadowmassless}.

 Next we consider how one would obtain this result from the dressing rules. As in the conformally coupled case, there are two possible dressed diagrams. The first is 

 \begin{eqn}
     &\begin{tikzpicture}[baseline=(b.base)]
        \begin{feynman}
            \vertex (a) ;
            \vertex [ right = 50pt of a       ] (i1) ;
            \vertex [ right = 50pt of i1      ] (i2);
            \vertex [right = 75pt of a      ] (b);
            \vertex [right = 75pt of b      ] (c);
            \vertex [above left = 26.39pt and 63.71pt of b](l1) {\(\vec{k_1}\)};
            \vertex [below left = 26.39pt and 63.71pt of b](l2) {\(\vec{k_2}\)};
            \vertex [below right = 26.39pt and 63.71pt of b](r1) {\(\vec{k_3}\)};
            \vertex [above right = 26.39pt and 63.71pt of b](r2) {\(\vec{k_4}\)};
            \vertex [above = 50pt of b](e);
    
            \diagram* {
                (l1) -- (i1),
                (l2) -- (i1),
                (r1)  --(i2),
                (r2)  --(i2),
                (i1) --[momentum'=$p {,}\;\vec{y}_{12}$] (i2),
                (e) --[scalar, momentum'= $p$] (i1),
                (i2) --[scalar, momentum' =$p$] (e),
            };
        \end{feynman}
    \end{tikzpicture} \\
    &=\frac{4 \lambda^2 \cos^2(\frac{\pi}{2}\epsilon)}{\pi}\int_0^\infty ds_1 \,ds_2\int_{-\infty}^\infty dp 
    \frac{\mathcal{N}_3(s_1;k_1,k_2;p)}{p^2+(s_1+k_{12})^2} \frac{\mathcal{N}_3(s_2;k_3,k_4;p)}{p^2+(s_2+k_{34})^2}\frac{1}{p^2+\vec{y}_{12}^2},
    \label{eq:M4Dressed}
 \end{eqn}
which will give the bulk of the structure, including all of the divergent pieces, logarithms, and dilogarithms in the finite piece. $\mathcal{N}_3$ in the above is defined in \eqref{eq:t0Integrand}. The $p$ integral can be evaluated by summing over the residues of poles in the upper half plane (note that the pole at $p=0$ is not included in the contour. See Appendix \ref{app:Green} for more details). We are then left with integrals over $s$ parameters. See appendix \ref{app:Integrals} and the Mathematica code in \cite{github} for a discussion of how to evaluate integrals of this form. Note that the finite piece has a transcendentality two, as expected since the dressed diagram has two dashed auxiliary propagators.

As in the conformally coupled case, the diagram with dotted propagators provides a finite contribution to the correlator. The diagram and its integral expression are:

\begin{eqn}\label{eq:masslessdottedexchange}
     &\begin{tikzpicture}[baseline=(b.base)]
        \begin{feynman}
            \vertex (a) ;
            \vertex [ right = 50pt of a       ] (i1) ;
            \vertex [ right = 50pt of i1      ] (i2);
            \vertex [right = 75pt of a      ] (b);
            \vertex [right = 75pt of b      ] (c);
            \vertex [above left = 26.39pt and 63.71pt of b](l1) {\(\vec{k_1}\)};
            \vertex [below left = 26.39pt and 63.71pt of b](l2) {\(\vec{k_2}\)};
            \vertex [below right = 26.39pt and 63.71pt of b](r1) {\(\vec{k_3}\)};
            \vertex [above right = 26.39pt and 63.71pt of b](r2) {\(\vec{k_4}\)};
            \vertex [above = 50pt of b](e);
    
            \diagram* {
                (l1) -- (i1),
                (l2) -- (i1),
                (r1)  --(i2),
                (r2)  --(i2),
                (i1) --[momentum'=$p {,}\;\vec{y}_{12}$] (i2),
                (e) --[ghost, momentum'= $p$] (i1),
                (i2) --[ghost, momentum' =$p$] (e),
            };
        \end{feynman}
    \end{tikzpicture} \\
    &=\frac{\lambda^2 \sin^2(\frac{\pi}{2}\epsilon)}{\pi}\int_0^\infty ds_1 \,ds_2\int_{-\infty}^\infty dp 
    \frac{\widetilde{\mathcal{N}}_3(s_1;k_1,k_2;p)}{p^2+(s_1+k_{12})^2} \frac{\widetilde{\mathcal{N}}_3(s_2;k_3,k_4;p)}{p^2+(s_2+k_{34})^2}\frac{1}{p^2+y_{12}^2}.
\end{eqn}
where $\widetilde{\mathcal{N}}_3$ is defined in \eqref{eq:t1Integrand}. 
Using the method of Appendix \ref{app:Integrals}, we find that this diagram is given by
\be 
\mbox{\eqref{eq:masslessdottedexchange}} = 2\pi^2(2y_{12}^3+k_{1234c}) + O(\e),
\ee 
which is finite. In contrast to the finite part of \eqref{eq:M4Dressed}, this diagram does not contribute any terms of transcendental weight two. This further reinforces the observation that one can read off the transcendentality of tree-level correlators by counting the number of dashed auxiliary propagators, even in the massless case.

The sum of these diagrams matches the result obtained from the Schwinger-Keldysh formalism in \eqref{SKmasslessexch}. 
Hence, there is complete agreement between regularized correlators evaluated via dressing rules and the Schwinger-Keldysh formalism. 

\section{Comparison to wavefunction formalism\label{sec:WF}}
We review the computation of cosmological correlators from cosmological wavefunctions \cite{Hartle:1983ai}. As reviewed in \eqref{eq:corrfromWF}, this is defined by the following path integral:
\begin{eqn}\label{cosmocorr}
\mathcal A_n = \frac{\int D\phi  |\Psi[\phi]|^2  \phi(\vec k_1) \cdots \phi(\vec k_n)}{\int D\phi  |\Psi[\phi]|^2 }, 
\end{eqn}
where $\phi(\vec k)$ denotes field insertions at $\scrip$ in dS. The coefficients of the wavefunction $\Psi[\phi]$ in \eqref{eq:WFdefn0} can be computed 
by analytically continuing Witten diagrams in AdS \cite{Maldacena:2002vr,Maldacena:2011nz,McFadden:2009fg}. 
Plugging  \eqref{eq:WFdefn0} into  \eqref{cosmocorr}, the correlator is given as
\begin{eqn}
\mathcal A_2  &= \frac{1}{\re \Psi_2(\vec k)}, \\
\mathcal A_3  &= \frac{\re \Psi_3(\vec k_1, \vec k_2, \vec k_3)}{\re \Psi_2(\vec k_1)  \re \Psi_2(\vec k_2) \re \Psi_2(\vec k_3)}, \\
\mathcal A_4  &= \frac{1}{\re \Psi_2(\vec k_1) \cdots \re \Psi_2(\vec k_4)} \Bigg\{ \re \Psi_4(\vec k_1, \cdots, \vec k_4) \\
&\quad+ \frac{\re \Psi_3(\vec k_1, \vec k_2, \vec k_1 + \vec k_2) \re \Psi_3(\vec k_3, \vec k_4, -\vec k_1 - \vec k_2)}{\re \Psi_2(\vec k_1 + \vec k_2)} + (1 \leftrightarrow 4) + (1 \leftrightarrow 3) \Bigg\},
\label{ininfromwafunction}
\end{eqn}
with similar expressions at higher points. In this section we match various correlators obtained via the wavefunction formalism and the dressing rules. We also demonstrate the simplicity of the cosmological correlator compared to the wavefunction coefficient at five points in the conformally coupled $\phi^3$ theory and explain how this simplicity extends to any odd point function at tree level (see section \ref{5ptcc}). 

\subsection{Conformally coupled $\phi^3$ 4-point}\label{phi34pt}

First we review the 4-point in-in correlator in conformally coupled $\phi^3$ theory obtained from the wavefunction \cite{Arkani-Hamed:2015bza} and show how it relates to the result obtained from dressing rules.
The wavefunction can be computed from the following propagators in AdS:
\begin{eqn}
\mathcal B_{\frac12}(z, k) &= z^{3/2} K_{1/2}(k z) , \\
G_{\frac12}(z_1, z_2, k) &= \frac{(z_1 z_2)^{3/2}}{2k}  \Big[ \Theta(z_1 - z_2 )K_{\frac12}(k z_1) I_{\frac12}(k z_2) + \Theta(z_2 - z_1) I_{\frac12}(k z_2) K_{\frac12}(k z_1)  \Big]~.
\end{eqn}
We can then obtain the wavefunction in dS 
by performing the following analytical continuation:
\begin{eqn}\label{ancont}
l_{AdS} \to i l_{dS}, \qquad k \to -i k~.
\end{eqn}
Since we have set $l_{AdS} = 1$, the first analytical continuation can be absorbed by simply analytically continuing the coupling constant, $\lambda \to i \lambda$. Having performed the integrals and the analytical continuation we obtain the following expressions for $\Psi_3(\vec k_1, \vec k_2, \vec k_3)$ and the $s$-channel contribution to $\Psi_4(\vec k_1, \cdots, \vec k_4)$: 
\begin{eqn}\label{wf-cc}
\Psi_4^{(s)}&= \frac{1}{2y_{12}} \Big[ \mbox{Li}_2 \left(  \frac{k_{34} - y_{12}}{k_{1234}} \right) + \mbox{Li}_2 \left(  \frac{k_{12} - y_{12}}{k_{1234}} \right) + \log \frac{k_{12} + y_{12}}{k_{1234}}\log \frac{k_{34} + y_{12}}{k_{1234}} - \frac{\pi^2}{6} \Big], \\
\Psi_3 &= -i   \Big[ \frac1\e + \log(i k_{123}) \Big],
\end{eqn}
where  $\e$ is the regularization parameter. The expressions for other channels can be obtained by exchanging the momenta. To compute the cosmological correlator we require the real part of the wavefunction coefficients
\begin{eqn}\label{wf-cc2}
\re \Psi_4^{(s)} &=  \frac{ 1}{2y_{12}} \Big[ \mbox{Li}_2 \left(  \frac{k_{34} - y_{12}}{k_{1234}} \right) + \mbox{Li}_2 \left(\frac{k_{12} - y_{12}}{k_{1234}} \right) + \log \frac{k_{12} + y_{12}}{k_{1234}}\log \frac{k_{34} + y_{12}}{k_{1234}} - \frac{\pi^2}{6} \Big], \\
\re \Psi_3 &= \frac{\pi}{2}~.
\end{eqn}
Note that $\Psi_3$ is divergent but upon computing the real part, the divergence cancels out and the only contribution we receive is from the discontinuity in the logarithm, as given in \eqref{wf-cc2}. We will see a higher-point analogue of this mechanism in section \ref{5ptcc}.  

The real part of the 2-point wavefunction coefficient 
is given by
\begin{eqn}
\re\Psi_2(\vec k) = k.
\end{eqn}
Using the above formulae and \eqref{ininfromwafunction} we obtain the familiar expression for the 4-point cosmological correlator 
\begin{eqn}
\mathcal A_4
&= \frac{ 1}{2y_{12}} \Big[ \mbox{Li}_2 \left( \frac{k_{34} - y_{12}}{k_{1234}} \right) + \mbox{Li}_2 \left( \frac{k_{12} - y_{12}}{k_{1234}} \right) + \log \frac{k_{12} + y_{12}}{k_{1234}}\log \frac{k_{34} + y_{12}}{k_{1234}} + \frac{\pi^2}{3} \Big].
\end{eqn}
By comparing with the computation via dressing rules in section \ref{sec:dress-conf-4pt-ex} we see that $\re\Psi_4^{(s)}$ matches the diagram with two dashed auxiliary propagators whereas $\re\Psi_3^{(s)} \re\Psi_3^{(s)}$ matches the diagram with two dotted auxiliary propagators.


\subsection{Conformally coupled $\phi^3$ 5-point}\label{5ptcc}
The five point correlator in $\phi^3$ theory can be expressed in terms of the wavefunction coefficients by using the path integral representation given in \eqref{cosmocorr},
\begin{eqn}
&\mathcal A_5 = \re\Psi_5(\vec k_1, \cdots, \vec k_5)\\
&+ \sum_{\{\pi\}} \frac{\re \Psi_4(\vec k_{\pi_1}, \vec k_{\pi_2}, \vec k_{\pi_3}, \vec y_{\pi_1 \pi_2}) \re \Psi_3(-\vec y_{\pi_1 \pi_2}, \vec k_{\pi_4}, \vec k_{\pi_5})}{\re \Psi_2(\vec y_{\pi_1 \pi_2})} \\
&+ \sum_{\{\pi\}}   \frac{\re\Psi_3(\vec k_{\pi_1}, \vec k_{\pi_2}, \vec y_{\pi_1 \pi_2}) \re \Psi_3(-\vec y_{\pi_1 \pi_2}, \vec y_{\pi_4 \pi_5}, \vec k_{\pi_3}) \re \Psi_3(-\vec y_{\pi_4, \pi_5},  \vec k_{\pi_4}, \vec k_{\pi_5}) }{\re \Psi_2(\vec y_{\pi_1 \pi_2} ) \re \Psi_2(\vec y_{\pi_4 \pi_5})}~,
\end{eqn}
where $\vec y_{\pi_1\pi_2} = \vec k_{\pi_1} + \vec k_{\pi_2}$ and the sum is over various permutation of momenta denoting different channels.

The wavefunction coefficients appearing in the final answer have been previously computed \cite{Arkani-Hamed:2015bza, Hillman:2021bnk} and can be expressed in terms of polylogs with maximal weight 3. In particular, $\Psi_3 $ and $\Psi_4$ are computed in section \ref{phi34pt} and $\Psi_5$ is given in equation 5.8 of \cite{Hillman:2021bnk} and is computed using symbols \cite{Goncharov:2010jf}. This corresponds to the following Witten diagram 
\begin{eqn}
\Psi_5 &= 
\begin{tikzpicture}[baseline]
\draw (0, 0) circle (2.25);
\draw(-2, 1) -- (-1, 0);
\draw(-2, -1) -- (-1, 0);
\draw(2, 1) -- (1, 0);
\draw(2, -1) -- (1, 0);
\draw(-1, 0) --(1,0);
\draw (0, 0) -- (0,2.25);

\node at (-2.25, 1.25) {$\vec k_1$};
\node at (-2.25, -1.25) {$\vec k_2$};
\node at (2.25, -1.25) {$\vec k_3$};
\node at (2.25, 1.25) {$\vec k_4$};
\node at (0, 2.5) {$\vec k_5$};

\end{tikzpicture} + 
\mbox{ perms}
\end{eqn}
with perms denoting the sum over other channels obtained by exchanging the momenta. The final result of this diagram involves polylogs of weight 3, namely, Li$_3$,  Li$_2$ Li$_1$ and Li$_1^3$ along with polylogs of lower weight. In particular, it can be checked that the part of the answer that contains Li$_3$ simplifies to 
 \begin{eqn}\label{psi5Li3}
\Psi_5 =  -i \lambda^3 \sum_{a, b = \pm}(-1)^{ab} \Big( {\rm Li}_3(1 - \frac{1 - v_{a}}{u_{b}}) +  {\rm Li}_3(1 - \frac{1 - u_{b}}{v_{a}}) + {\rm Li}_3(1 -\frac{(1 - v_{a})(1 - u_{b})}{u_{b}v_{a}} )\Big) + \cdots~,
 \end{eqn}
where the factor of $i$ appears via the analytical continuation described in \eqref{ancont},  $v_{\pm}, u_{\pm}$ are defined in \eqref{5ptcross}, and $\cdots$ denote the other terms in the answer, which contain polylogs of lower weight (see equation 5.8 of \cite{Hillman:2019wgh} for the complete expression). The coefficients of the other weight 3 polylogs (i.e., Li$_2$ Li$_1$ and Li$_1^3$) also contain an overall factor of $i \lambda^3$.

Since the wavefunction contains functions with transcendentality weight three, one would naively expect the cosmological correlator to also have such functions. However, the correlator obtained in section \ref{sec:fivept} was found to contain functions of maximum weight two. This can be understood as follows.
Note that for physical momenta we have $\frac{1 - v_{a}}{u_b}, \frac{1 - u_{b}}{v_a}  > 0$, which ensures that the arguments of Li$_3$ in \eqref{psi5Li3} are $< 1$ and hence evaluate to real values. A similar statement is also true for the other polylogs of weight 3, i.e., ${\rm Li}_2 {\rm Li}_1$ and ${\rm Li}_1^3$. Since each of these polylogs have an imaginary coefficient, $i \lambda^3$, upon computing 
$\re\Psi_5$, polylogs of weight 3 drop out! A similar simplification was also found for the 3-point function in equation \eqref{wf-cc2} where the contribution to the in-in correlator appeared from the discontinuity of $\log$ in equation \eqref{wf-cc}. 
The relative simplicity of the three and five-point tree-level in-in correlators compared to wavefunction coefficients can be easily understood from the dressing rules given in Section \ref{sec:fivept}. Similar simplicity extends to all odd-point functions at tree-level.

\subsection{Massless 2-point \label{ssec:massless2pt}}

Next we consider massless two-point functions and use the dressing rules to reproduce previous results obtained using more conventional methods. In doing so, we will see how free correlators arise from the dressing rules. We will also consider a 1-loop example, and show that the dressing rules reproduce all of the expected IR divergences.

\subsubsection*{Free 2-point function}

The free 2-point wavefunction for massless scalar fields was computed in \cite{Maldacena:2002vr} by inserting two bulk-to-boundary propagators into the action and putting a cut-off in conformal time to regulate an IR divergence. While the divergence can be removed by adding a local counterterm in the CFT, it is imaginary and cancels out when computing the 2-point in-in correlator, which is obtained by taking the real part of the wavefunction and inverting it. The resulting expression is proportional to $k^{-3}$, where $k$ is the energy of the external legs.

This result can be trivially obtained from the dressing rules in the following way. We first start with a trivial Feynman diagram in flat space which has no interaction vertices and a single external leg, which gives a factor of $1$.  
The lack of vertices means there are no auxiliary propagators to attach. 
We then recall from section \ref{sec:review} that we must restore a factor of $\eta_0^{\Delta_-} k^{2 \Delta_- -d}$ for each external leg in order to obtain the in-in correlator. In the present case, $\Delta_-=0$ which yields $\mathcal{A}_2^{\rm free} =k^{-3}$, as expected. Using similar reasoning, we find that the free 2-point correlator in the conformally coupled case is proportional to $1/k$, in agreement with standard results.

    
To understand why we consider a Feynman diagram with only a single leg, note that we can compute the free 2-point correlator by starting with a single bulk-boundary propagator and dragging the other end to the boundary, as depicted below. 
    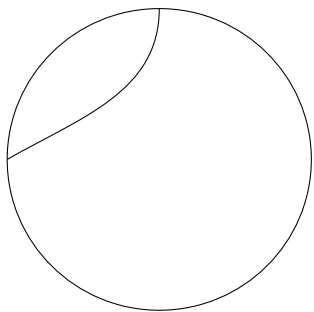
\begin{figure}[H]
    \centering
\begin{tikzpicture}
\draw (0, 0) circle (2);
\draw (-2, 0) to[out = 30, in= -90] (0,2);
\end{tikzpicture}
\caption{Free 2-point function}
\end{figure}

\subsubsection*{Massless $\phi^4$ tadpole}

\begin{figure}
    \centering
    \begin{tikzpicture}[baseline=(b.base)]
        \begin{feynman}
            \vertex (a) {$\vec{k}$} ;
            \vertex [right = 75pt of a       ] (b);
            \vertex [ above = 37.5pt of b       ] (i);
            \vertex [right = 75pt of b        ] (c) {$\vec{k}$};
            \vertex [above right = 26.39pt and 63.71pt of b](e) {\(p_{tot}=0\)};
    
            \diagram* {
                (a) -- (b),
                (b) -- (c),
                (b)  --[ min distance = 0.5cm, out=135, in=180](i),
                (i)  --[min distance = 0.5cm, out=0, in=45](b),
                (b) --[scalar] (e),
            };
        \end{feynman}
    \end{tikzpicture}
    \caption{Dressed tadpole diagram. \label{fig:M4TadpoleDressed}}
\end{figure}
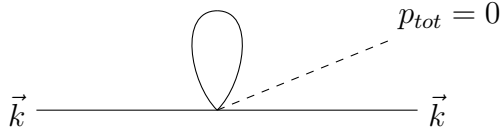

Let us now consider the 1-loop 2-point correlator in massless $\phi^4$ theory, which can be derived from the dressed flat space diagram 
in Fig.\ \ref{fig:M4TadpoleDressed}. In equation (3.15) of \cite{cespedes2024ir}, the leading IR divergence of this diagram using a cutoff was found to be
\begin{eqn}
    \mathcal{A}_2 \sim \frac{\lambda}{k^3} \log(k L_{IR}) \log(-2 k \eta_0)+\cdots\,,
    \label{eq:2ptMasslessTadpole}
\end{eqn}
where $L_{IR}$ is the loop momentum cutoff and $\eta_0$ encodes the end of inflation. The dressing rules explicitly use dimensional regularisation, so to compare to this result we make the association $\log(\# \eta_0) \sim \frac{1}{\e}$. 

The integral to be performed from the dressing rules in section \ref{sec:dressmassless} is 
\begin{eqn}
    &\lambda\int\!\! dp \int\limits_{L_{IR}}\!\! \frac{d^3 y}{\pi p^2 (p^2+\vec{y}^2)} \int\limits_0^\infty \frac{ds s^{-\e}}{s + 2k } \\
&\quad\times\left(\frac{k^2 p^2}{s \Gamma (-\epsilon )}+\frac{k^2 s}{\Gamma (2-\epsilon )}+\frac{2 k p^2}{\Gamma (1-\epsilon )}+\frac{2 k s^2}{\Gamma (3-\epsilon )}+\frac{p^2 s}{\Gamma (2-\epsilon )}+\frac{s^3}{\Gamma (4-\epsilon )}\right).
\end{eqn}
Note that this is both UV and IR divergent. The UV divergence can be regulated using a UV cutoff or analytic regularisation (see Appendix \ref{app:analyticReg} for more details on the latter), but we will neglect this divergence for simplicity. Performing the $p$ integral via residues and regulating the $\vec{y}$ integral with a cutoff then gives the leading divergence
\begin{eqn}
 \sim   \lambda \frac{k^3 \log(k L_{IR})}{\e}\,,
    \label{eq:2ptTadpoleLeading}
\end{eqn}
which matches \eqref{eq:2ptMasslessTadpole} after restoring a factor of $k^{2\Delta_--d}$ for each external leg.

\subsection{Massless $\phi^3$ 4-point}

Let us now compute the 4-point tree-level correlator in massless $\phi^3$ theory from the wavefunction and compare it to the dressing rules. For this purpose, we need to compute 3-point contact diagrams and 4-point exchange diagrams using the following propagators:
\begin{eqn}
\mathcal B_{\frac32}(z, k) &= z^{3/2} K_{3/2}(k z) , \\
G_{\frac32}(z_1, z_2, k) &= \frac{1}{2k} (z_1 z_2)^{3/2} \Big[ \Theta(z_1 - z_2 )K_{\frac32}(k z_1) I_{\frac32}(k z_2) + \Theta(z_2 - z_1) I_{\frac32}(k z_2) K_{\frac32}(k z_1)  \Big]~.
\end{eqn}
The integrals that arise from these diagrams are discussed in Appendix \ref{app:Integrals}. Since the expressions for the final results are long, we have presented them in Appendix \ref{app:WFmassless} and only quote the leading divergent part here:
\begin{eqn}\label{WFmassless-ex-div1}
\Psi_4^{(s)}&= \scalebox{0.75}{\begin{tikzpicture}[baseline]
\draw (0, 0) circle (2.25);
\draw(-2, 1) -- (-1, 0);
\draw(-2, -1) -- (-1, 0);
\draw(2, 1) -- (1, 0);
\draw(2, -1) -- (1, 0);
\draw(-1, 0) --(1,0);
\node at (-2.25, -1.25) {$\vec k_1$};
\node at (-2.25, 1.25) {$\vec k_2$};
\node at (2.25, -1.25) {$\vec k_4$};
\node at (2.25, 1.25) {$\vec k_3$};

\end{tikzpicture}} = -  \frac{1}{\e^2} (2 y_{12}^3+k_{1234c}) + O\left(\frac1\e\right),
\end{eqn}
\begin{eqn}\label{WFmassless-ex-div2}
\re \Psi_3^{(s)} \re \Psi_3^{(s)} &= - \frac{1}{\e^2} \left( y_{12}^3+k_{12c}\right) \left(y_{12}^3+k_{34c}\right) + O\left(\frac1\e\right),
\end{eqn}
where $\Psi_3^{(s)}$ is given by the following Witten diagram:
\begin{eqn*}
 \Psi_3^{(s)} = \scalebox{0.75}{\begin{tikzpicture}[baseline]
\draw (0, 0) circle (2.25);
\draw(-2, 1) -- (0, 0);
\draw(-2, -1) -- (0, 0);
\draw(2.25,0) -- (0, 0);
\node at (-2.25, -1.25) {$\vec k_1$};
\node at (-2.25, 1.25) {$\vec k_2$};
\node at (2.6, 0) {$\vec k_{12}$};
\end{tikzpicture}}
\end{eqn*}
By combining \eqref{WFmassless-ex-div1} and \eqref{WFmassless-ex-div2} according to \eqref{ininfromwafunction} and summing over channels,
we find an exact agreement with the computations via the Schwinger-Keldysh formalism and the dressing rules described in section \ref{sec:masslessexchSK}. See the Mathematica notebook in \cite{github} for details.

\section{Conclusions}\label{sec:conclusions}
The main purpose of this paper was to formulate a set of dressing rules which, when applied to flat space Feynman diagrams, calculate cosmological correlators on de Sitter space. We derived these rules from the shadow formalism and compared them to the Schwinger-Keldysh formalism as well as the definition of in-in correlators obtained from squaring the wavefunction for many explicit examples. In essence, these rules allow us to express cosmological correlators in terms of flat space Feynman integrals with certain theory-dependent auxiliary propagators attached to each vertex. After integrating over the energies flowing through the auxiliary propagators, the transcendentality of the in-in correlator can become greater than that of the flat space Feynman integral. 

Remarkably, in all the examples where we have been able to carry out the integrals, we find that the transcendentality of the in-in correlator is given by that of the flat space Feynman diagrams plus the number of non-trivial auxiliary propagators \footnote{Here we are using a more refined notion of transcendentality which refers to the functional form of the correlator rather than overall factors of $\pi$.}. By non-trivial, we mean an auxiliary propagator that involves an $s$-integral that gives rise to a logarithm of kinematic variables (for more details see Table \ref{tab:AuxPropSummary}). In order to demonstrate the advantage of the dressing rules over more conventional approaches, we obtained a closed expression for the tree-level five-point correlator in conformally coupled $\phi^3$ theory, which to our knowledge was not previously computed. In contrast to the five-point wavefunction coefficient which contains trilogarithms \cite{Hillman:2019wgh}, we find that the in-in correlator contains at most dilogarithms and is therefore a substantially simpler function. Moreover, this property can be immediately inferred from the dressing rules, since it arises from attaching two non-trivial propagators to tree-level Feynman diagrams implying that it has transcendentality two. 



Similar dressing rules can also be derived for massless scalar theories. In this case, the correlators exhibit IR divergences which we regulate by dimensional regularization and a closely-related regulator that preserves the de Sitter isometries. For such theories, the auxiliary propagators encode the IR divergences arising from the de Sitter expansion and have a more intricate form because they are regulator-dependent. Nevertheless, we find that the transcendentality can once again be read off from the number of auxiliary propagators attached to the flat space Feynman diagrams and we confirm that our results are consistent with the Schwinger-Keldysh and the Wave function approach, reproducing both the correct IR divergences and finite parts.

While we have focused on conformally coupled and massless scalar theories, it is interesting to ask whether our results generalise to other masses. Consider a one-loop 4-point correlator in $\phi^4$ theory with $m^2=-n(n+3)$, where $n$ is a non-negative integer. From \eqref{scalingdim}, we see that this corresponds to $\Delta_+=3+n$. Such fields belong to the exceptional series of unitary de Sitter representations \cite{Sun:2021thf}. Using the shadow formalism reviewed in section \ref{sec:review}, the in-in correlator is given by a sum over two bubble diagrams, where the bulk-to-boundary propagators are given by the first line of \eqref{eq:EAdSProps} with $\nu=-(n+\frac{3}{2})$, and the bulk-to-bulk propagators are given by the second line with with $\nu=(n+\frac{3}{2})$ in one diagram and $\nu=-(n+\frac{3}{2})$ in the other. Using the identity $-i \pi J_{\nu}(z)=i^{\nu} K_{\nu}(i z)-(-i)^{\nu} K_{\nu}(-i z) $, we find that the integrand of each bubble contains a sum over four terms coming from the product of the two bulk-to-bulk propagators, but after adding them together a non-trivial cancellation takes place and we obtain an integrand with only two terms:
\begin{eqn}
\mathcal{A}_4=&\int_0^{\infty} d^3 \ell \int_{-\infty}^{\infty} dp_1 dp_2 \int_0^{\infty} \frac{dz_1dz_2}{z_1^4z_2^4} \prod_{i=1}^2\left(z_1^{3/2}k_i^{n+\frac{3}{2}}K_{n+\frac{3}{2}}(k_i z_1)\right)  \frac{1}{p_1^2+\vec{\ell}^2} \frac{1}{p_2^2+(\vec{\ell}+\vec{y}_{12})^2} \\
& \times \left( \mathcal{I}_{n+\frac{3}{2}}+\mathcal{I}_{-n-\frac{3}{2}}\right) \prod_{i=3}^4\left(z_2^{3/2}k_i^{n+\frac{3}{2}}K_{n+\frac{3}{2}}(k_i z_2)\right),
\end{eqn}
where
\begin{eqn}
&\mathcal{I}_{n+\frac{3}{2}}+\mathcal{I}_{-n-\frac{3}{2}}\\
=&-(z_1z_2)^3 p_1 p_2\big(K_{n+\frac{3}{2}}(i p_1,z_1) K_{n+\frac{3}{2}}(i p_2,z_1) K_{n+\frac{3}{2}}(i p_1,z_2) K_{n+\frac{3}{2}}(i p_2,z_2)\\
&+ K_{n+\frac{3}{2}}(i p_1,z_1) K_{n+\frac{3}{2}}(i p_2,z_1) K_{n+\frac{3}{2}}(-i p_1,z_2) K_{n+\frac{3}{2}}(-i p_2,z_2) \big).
\end{eqn}
This simplification generalises the one first found for conformally coupled scalars in \cite{Chowdhury:2023arc} and strongly suggests the mathematical simplicity of conformally coupled $\phi^4$ correlators extends to an infinite set of masses. For more generic masses, the integrands of the bulk-to-bulk propagators are not in general even functions of $p$ and there are additional interaction vertices arising from the effective action in \eqref{eq:EAdS_action_gen_pot}, so it is not immediately obvious whether similar simplifications occur. We leave a systematic analysis of this question for future work.

 
There are a number of other future directions:
\begin{itemize}
\item By giving a simple recipe to express cosmological correlators in terms of standard flat space Feynman integrals, the dressing rules in principle allow one to import tools from the study of flat space amplitudes, such as unitarity methods. An immediate task would be to generalise the optical theorem recently derived for wavefunction coefficients in \cite{Goodhew:2020hob} to in-in correlators. Indeed, from the dressing rules one can see that in-in correlators will inherit all the usual discontinuities of flat space scattering amplitudes plus additional discontinuities arising from auxiliary propagators. Cuts of IR finite in-in correlators were recently studied in \cite{Donath:2024utn,Stefanyszyn:2024msm,Ema:2024hkj}. It would therefore be interesting to see how these results arise from dressing rules and generalize them to IR divergent cases. Moreover, it would be intersting to see how the approach developed in this paper relates to recent methods based on differential equations \cite{Arkani-Hamed:2023kig,Caloro:2023cep,He:2024olr, Hang:2024xas, Baumann:2024mvm,Grimm:2025zhv}. 

\item Another important question is how to carry our renormalization in our framework. Since UV divergences probe short distances they are essentially the same as those of flat space, so there is a natural proposal to renormalize them: simply apply the dressing rules to the flat space counterterms needed to cancel the UV divergences of flat space Feynman diagrams. IR divergences are more subtle to deal with, however. Various approaches have been proposed to treat them including holographic renormalization \cite{Bzowski:2023nef, Bhowmick:2024kld}, the stochastic formalism \cite{Starobinsky:1986fx,Gorbenko:2019rza,Anninos:2024fty}, and other approaches \cite{Beneke:2012kn,Polyakov:2012uc,Benincasa:2024rfw, Kamenshchik:2024ybm}. 
We leave a systematic understanding of how to treat IR divergences using the dressing rules for future work.

\item Another direction would be to extend the dressing rules to spinning correlators \cite{Costa:2011mg, Raju:2012zr, Raju:2012zs, Ghosh:2014kba, Sleight:2017fpc, Albayrak:2019asr, Chowdhury:2024snc}. If it is possible to recast gluon and graviton in-in correlators in terms of flat space Feynman diagrams as we have done for scalar correlators, the double copy relating gluon to graviton amplitudes in flat space \cite{Bern:2019prr} would be directly inherited by in-in correlators via the dressing rules. Note that there has been a great deal of recent progress on the double copy for wavefunction coefficients in de Sitter space \cite{Farrow:2018yni,Armstrong:2020woi,Albayrak:2020fyp,Jain:2021qcl,Diwakar:2021juk,Cheung:2022pdk,Lee:2022fgr,Herderschee:2022ntr,Armstrong:2023phb,Mei:2023jkb,Lipstein:2023pih,CarrilloGonzalez:2024sto} so it would be interesting to develop a similar understanding for in-in correlators.

\item Finally, it would be interesting to generalize this story to more generic spacetimes in order to clarify the physical significance of the dressing rules. 
In general, one could imagine that flat space observables can be lifted to observables in more generic spacetimes by adding auxiliary propagators that encode the broken symmetries. A concrete target would be to look at inflationary spacetimes where a subset of the de Sitter isometries is broken, in particular boosts \cite{Cheung:2007st,Green:2020ebl,Pajer:2020wxk}. It would be interesting to see if correlators in such backgrounds can be obtained by dressing flat space diagrams with spinning auxiliary propagators. 
\end{itemize}
We hope to report further on these directions in the future.

\subsection*{Acknowledgements}
We thank Pierre Vanhove for collaboration in the early stages of this work. Also we have greatly benefited from discussion with Samuel Abreu, Martin Beneke, Paolo Benincasa, Einan Gardi, Sadra Jazayeri, Debanjan Karan, Babli Khatun, Nilay Kundu, Shota Komatsu, Alok Laddha, Paul McFadden, Radu Moga, Yuyu Mo, Enrico Pajer, Radu Roiban, Kostas Skenderis, and Lorenzo Tancredi. A.L. is supported by an STFC Consolidated Grant ST/T000708/1. C.C. is supported by the STFC consolidated grant
(ST/X000583/1) ``New Frontiers in Particle Physics, Cosmology and Gravity''. J. Marshall is supported by a Durham Doctoral Teaching Fellowship. J. Mei is supported by the European Union (ERC, UNIVERSE PLUS, 101118787), views and opinions expressed are however those of the authors only and do not necessarily reflect those of the European Union or the European Research Council Executive Agency. Neither the European Union nor the granting authority can be held responsible for them. I.S. is supported by the Excellence Cluster Origins of the DFG under Germany’s Excellence Strategy EXC-2094 390783311 and has benefited from hospitality at the Munich Institute for Astro-, Particle and BioPhysics (MIAPbP), also funded by the DFG under Germany´s Excellence Strategy – EXC-2094 – 390783311. CC would also like to thank the participants of the POSG workshop held in ICTS, Bengaluru for useful discussions.

\appendix 

\section{Derivation of propagators}\label{app:Green}
In this appendix we review the derivation of the propagators in AdS and the form of the representation used in the paper. Our discussion follows \cite{Albayrak:2020isk} while several papers have discussed a similar representation \cite{Raju:2011mp, Meltzer:2020qbr}.

The Lagrangian for a free minimally coupled scalar field in Euclidean AdS$_4$ of mass $m$ is given as
\begin{eqn}
L =  \frac{1}{2} g^{\a\b} \p_{\a} \phi \p_{\b} \phi - \frac{1}{2} m^2 \phi^2 ~,
\end{eqn}
where the metric is that of Poincare AdS. Due to the translational invariance along the boundary direction, the field can be decomposed in terms of the Fourier modes
\begin{eqn}
\phi(z, \vec x) = \int \frac{d^3 k}{(2\pi)^3}  e^{i \vec k \cdot \vec x} \phi(z, \vec k)~.
\end{eqn}
 The equation of motion for the field $\phi$ is 
\begin{eqn}
z^{4} \p_z \big( z^{-2} \p_z \phi \big)   - (z^2 k^2 + m^2)  \phi = 0~.
\end{eqn}
This differential equation is solved by any linear combination of Bessel functions, for example, 
\begin{eqn}
\phi(z, \vec k) = c_1 z^{3/2} J_{\nu}(ik  z) + c_2 z^{3/2} Y_{\nu}(ik z)~,
\end{eqn}
where $\nu = \frac{\sqrt{9 + 4 m^2}}{2} \equiv \Delta - \frac32 $ and $c_1, c_2$ are constants of integration. These constants can be fixed by demanding the solution is regular as $z \to \infty$ which leads to $c_2 = i c_1$. This linear combination then gives the well-known solution in terms of the modified Bessel function of the second kind,
\begin{eqn}\label{eq:bulk-bndy1}
\phi(z, \vec k) = z^{3/2} K_{\nu}(k z)~,
\end{eqn}
where we have set the normalization constant to $1$. This is recognized as the bulk-boundary propagator. 

We normalize the bulk-bulk propagator in Fourier space $G(z, z'; \vec k)$ such that it satisfies the following differential equation 
\begin{eqn}\label{eq:Green}
\Big\{ z^{4} \p_z \big( z^{-2} \p_z  \big)   - (z^2 k^2 + m^2) \Big\} G_\nu(z, z'; \vec k) = \sin\left(\pi \nu  \right) z^4 \delta(z- z')~.
\end{eqn}
This unconventional normalization is chosen for this to work for fields with ghost-like kinetic terms (for example, the shadow field discussed in section \ref{sec:Dressing}). This second-order differential equation can be solved using the traditional methods by analyzing the solutions for $z > z'$ and $z < z'$ and then demanding that the solution is continuous and differentiable at $z = z'$. The boundary condition at $z = 0$ is chosen such that the Green function falls off as $z^\Delta$ as $z \to 0$. This uniquely fixes the form of the solution $\forall\ \nu \in \mathbb R$ to be
\begin{eqn}\label{eq:Theta-Theta}
G_\nu(z, z'; \vec k) = -\sin\left(\pi \nu  \right)(z z')^{3/2} \Bigg\{ \Theta(z - z') K_\nu(k z) I_\nu(k z') + \Theta(z' - z) I_\nu(k z) K_\nu(k z')  \Bigg\}~.
\end{eqn}
where $I_\nu$ is the modified Bessel function of the first kind. This equation is valid for all values of $\nu$. Note that for a field with scaling dimension $\Delta$ its shadow has the scaling dimension $d - \Delta$ and the formula above is applicable for both\footnote{For the range of scaling dimensions considered in this paper, the shadow fields satisfy Neumann boundary conditions.}.  This form of the propagator is often useful for deriving several recursion relations (see for example \cite{Arkani-Hamed:2017fdk, Chowdhury:2024snc}). By using the bulk-bulk and bulk-boundary propagators we evaluate the correlation functions via the extrapolate dictionary \cite{Harlow:2011ke}.
 
\subsection{Spectral representation} \label{app:specrep}
The alternate form for the propagator \ref{eq:Theta-Theta}, used in section \ref{sec:Dressing} of the paper, can be obtained by using the closure relation for the Bessel function which is given as,
\begin{eqn}\label{eq:closure}
\intsinf x J_{\a}(u x) J_\a(v x) = \frac{1}{u} \delta(u - v)~,
\end{eqn}
 where $\a \geq -\frac12$ and $u, v \in \mathbb R$ \cite{arfken}. 
 Using this relation it is easy to verify that the Green function in equation \eqref{eq:Green} admits the following integral representation
\begin{eqn}\label{eq:Gspectral}
G_\nu(z, z'; \vec k) = -\sin\left(\pi \nu  \right)(z z')^{3/2}  \intsinf dp \, \frac{pJ_{\nu}(p z) J_{\nu}(p z')}{p^2 + k^2}~,
\end{eqn}
which is sometimes known as the spectral representation of the Green function. Since the closure relation is valid for $\a \geq -\frac12$, this means that the form of the propagator \eqref{eq:Gspectral} is also valid for $\nu \geq - \frac12$. This range includes the conformally coupled scalar and the massless scalars (the two main fields discussed in the paper). However, the shadow field corresponding to the massless scalar has $\nu = - \frac32$ which falls outside the range of validity in equation \eqref{eq:closure}. This can be seen by explicitly expanding the Bessel functions $J_{-\frac32}(p z) J_{-\frac32}(p z')$ , which gives
\begin{eqn}
\int_0^\infty dp J_{-\frac32}(p z) J_{-\frac32}(p z')&=\frac{2}{\pi} \intsinf \frac{dp}{p^2} \Big( \cos(p z) + p z \sin(p z) \Big) \Big( \cos(p z') + p z' \sin(p z') \Big) 
\end{eqn}
and this has an extra pole at $p =0$ which naively makes the integral on the contour above divergent. However this divergence can be avoided by choosing a different contour. For example, by considering the principle value of the integral, we obtain a finite result that agrees with the form of the Green function given in \eqref{eq:Theta-Theta}. Hence the formula in \eqref{eq:Gspectral} can be extended to all real $\nu$ by choosing the following contour,
\begin{eqn}\label{eq:PVform}
G_\nu(z, z'; \vec k) = -\sin\left(\pi \nu  \right)(z z')^{3/2} \ \mbox{p.v.}\intsinf dp p \frac{J_{\nu}(p z) J_{\nu}(p z')}{p^2 + k^2}~,
\end{eqn}
where p.v. denotes the principle value of the integral which is defined as, 
\begin{eqn}\label{eq:principlevalue}
\mbox{p.v.} \intsinf dx f(x) = \lim_{\delta \to 0} \int_{\delta}^{\infty} dx  f(x) ~.
\end{eqn}
By modifying the contour this way, the integral can be performed for $\nu = - \frac32$ using the method of residues where the p.v. prescription ensures that the pole at $p = 0$ is avoided.  In practice, while computing the $p$ integral over rational functions this contour is equivalent to computing residues by avoiding the pole at $p =0$. This recovers the form of the propagator given in \eqref{eq:Theta-Theta}.

It is sometimes convenient to use an alternate form which avoids this notation by rewriting \eqref{eq:PVform} as,
\begin{eqn}
G_\nu(z, z';\vec k) = -\sin\left(\pi \nu  \right) (z z')^{3/2}  \intsinf \frac{dp}{p} \Big\{1 - \lim_{p \to 0} \Big\} \frac{p^2}{p^2 + k^2} J_{\nu}(p z) J_{\nu}(p z')~,
\end{eqn}
which is valid for $\nu \geq -3/2$ since for any $\nu > - \frac32$ the second term in the curly bracket vanishes and for $\nu = - 3/2$ it recovers the propagator given in \eqref{eq:Theta-Theta}.

\subsection{Analytic regularization \label{app:analyticReg}}
Correlators evaluated using the propagators in the previous section generally break conformal invariance.  This was noticed in \cite{Chowdhury:2023arc} while evaluating correlators at loop level for theories without any IR divergence (eg: $\phi^4$ theory with $\phi$ being a conformally coupled scalar). However for theories involving massless scalars, one encounters IR divergences even at tree-level \cite{Bzowski:2022rlz, Bzowski:2023jwt} and by regulating the divergence using schemes such as dimensional regularization (dimreg) or hard-cutoff regularization breaks conformal invariance. This means that the correlators no longer satisfy the conformal ward identities due to the presence of an extra scale introduced via the regularization procedure. In hard cutoff scheme the scale is explicitly introduced via the cutoff and in dimensional regularization this is introduced via the change of the coupling constant (see equation \eqref{eq:corr333} for an example at 3-points). Thus to preserve conformal invariance, it is necessary to use a regularization scheme that avoids this issue and one such proposal for correlators in momentum space was given in \cite{Chowdhury:2023arc} and is known as analytic regularization. 

For bulk-bulk propagators in AdS, this involves deforming the propagator given in \eqref{eq:PVform}  to 
\begin{eqn}\label{eq:anareg1}
G_{\nu}^{reg}(z, z'; \vec k) = \sin\left(\pi \nu  \right) (z z')^{3/2 - \e} \ \mbox{p.v.} \intsinf dp p \frac{J_{\nu}(p z) J_{\nu}(p z')}{(p^2 + k^2)^{1 + \e}}~,
\end{eqn}
where $\e$ is the regulator. This representation of the bulk-bulk propagator was used to perform integrals for the conformally coupled scalar in \cite{Chowdhury:2023arc} which required the presence of the regulator in the denominator above for regulating the loop integrals. However for diagrams with IR divergence, the factor of $\e$ present in the powers of $z, z'$ are also important for preserving conformal invariance. This propagator can be viewed as arising from an effective Lagrangian of the form 
\begin{eqn}
L_{eff} \sim\frac12 \int \frac{dz d^3x}{z^{-2\e}} \phi\Big( \frac{1}{z} (\p_z^2 + \vec\p^2)^{1 + \e} \frac{1}{z} \Big) \phi~,
\end{eqn}
and we refer the reader to section 5 of \cite{Chowdhury:2023arc} for more details. The bulk-boundary propagator can be obtained by taking the $z'\to 0$ limit of \eqref{eq:anareg1}, which results in 
\begin{eqn}
\mathcal B_\nu(z, \vec k)= z^{3/2} \frac{2^{ -\nu -\e } k^{\nu -\e} \sin\left(\pi \nu  \right)}{\Gamma(1 + \e)\Gamma(1 + \nu)}  K_{\nu - \e }(k z)~,
\end{eqn}
where we use $\mathcal B_\nu(z, \vec k)$ to denote the bulk-boundary propagator. 

A similar modification can be performed for the Schwinger-Keldysh propagators given in section \ref{sec:SK} for them to preserve conformal invariance.
\begin{eqn}
\Lambda^{(\Delta) reg}_{TT} &=  (\eta_1 \eta_2)^{3/2 - \e} \ \mbox{p.v.} \intinf dp p \frac{H_\beta^{(2)} (p \eta) H_\beta^{(1)} (p \eta') }{(p^2 - k^2 - i\e)^{1 + \e}}, \\
\Lambda^{(\Delta) reg}_{AA} &=  (\eta_1 \eta_2)^{3/2 - \e} \ \mbox{p.v.} \intinf dp p \frac{H_\beta^{(2)} (p \eta') H_\beta^{(1)} (p \eta) }{(p^2 - k^2 - i\e)^{1 + \e}}, \\
\Lambda^{(\Delta) reg}_{TA} &=  (\eta_1 \eta_2)^{3/2 - \e} \int_{\mathcal C_1}  dp p \frac{H_\beta^{(2)} (p \eta) H_\beta^{(1)} (p \eta) }{(p^2 - k^2)^{1 + \e}}, \\
\Lambda^{(\Delta) reg}_{AT} &=  (\eta_1 \eta_2)^{3/2 - \e} \int_{\mathcal C_2}  dp p \frac{H_\beta^{(2)} (p \eta) H_\beta^{(1)} (p \eta) }{(p^2 - k^2 )^{1 + \e}}~,
\end{eqn}
where p.v. denotes the principle value prescription as given in equation \eqref{eq:principlevalue} and the contours $\mathcal C_1$ $\mathcal C_2$ in the last two terms are given below 
\begin{eqn}
\begin{tikzpicture}
\draw (-4, 0) -- (4, 0);
\draw (0, -2.5) -- (0, 2.5);
\node at (1.5, 0.5) {$\times$};
\node at (-1.5, 0.5) {$\times$};

\draw[fermion, red] (1.5, 0.5) circle (0.35);
\draw[fermion, blue] (-1.5, 0.5) circle (0.35);

\node at (1.5, -0.25) {$k$};
\node at (-1.5, -0.25) {$-k$};

\node at (2, 1) {$\mathcal C_1$};
\node at (-1, 1) {$\mathcal C_2$};

\node at (3, 2) {$\angle p$};

\end{tikzpicture}
\end{eqn}

\subsection{dS-invariant IR regularisation}

\subsubsection*{Conformally coupled fields}
The propagators described in the previous section can be implemented in the dressing rule. For the dashed propagators, there exists a very simple modification for conformally coupled fields. For example, for the $\phi^4$ theory we have \cite{Chowdhury:2023arc}, 
\begin{eqn}\label{analyticprop1}
G_{aux}^{\mbox{cc } \phi^4}(p; k)
&= \frac{\Gamma(1+\e)}{2}  \left((k+i p)^{-\e -1} + (k-i p)^{-\e -1}\right) \\
&= \intsinf ds s^{-\e - 1} \intsinf dz \cos(p z) e^{-(k + s)z} =  \intsinf ds s^{-\e - 1} \frac{(k + s)}{p^2 + (k + s)^2}~.
\end{eqn}
In the $\e \to 0$ limit this reduces to the form of the propagator given in section \ref{sec:Dressing}
\begin{eqn}
\lim_{\e \to 0} G_{aux}^{\mbox{cc } \phi^4}(p; k) =  \frac{1}{2}  \left((k+i p)^{\e -1} + (k-i p)^{\e -1}\right) = \frac{k}{k^2 + p^2}  + O(\e)~.
\end{eqn}
This shows how the $s-$integral representation is useful for obtaining a form of the propagator consistent with the conformal ward identities. As mentioned in section \ref{sec:Dressing}, for cc $\phi^4$ theory there is no other propagator. For the other theories considered in the paper we provide the $s$-integral representation which allows one to obtain expressions consistent with the ward identity.  

\subsubsection*{Massless fields}

When the fields are massless, IR divergences can also emerge from the loop momentum integration. We saw in section \ref{ssec:massless2pt} that these can be dealt with by a cutoff, but it is also possible to use the analytic regulator. This is implemented by shifting each flat-space propagator 
\begin{eqn}
    \frac{1}{p^2+\vec{y}^2}\to \frac{1}{(p^2+\vec{y}^2)^{1+\e}}.
\end{eqn}
For the example of the two-point tadpole in $\phi^4$, \eqref{eq:2ptMasslessTadpole} becomes 

\begin{eqn}
    \sim \frac{k^3}{\e^2}\,,
\end{eqn}
as one might have expected. The loop IR divergence has simply contributed another power of $\frac{1}{\e}$, and the result is manifestly de Sitter invariant.

\section{General derivation of dressing rules \label{app:MasslessAuxProp}}

In this appendix we give a general derivation of the dressing rules for the various theories considered in this paper. We will mainly the phrase the discussion in terms of massless theories, for which the dressing rules are the most nontrivial, but along the way will also explain how to adapt the derivation to conformally coupled theories.

\subsection{Vanishing propagators \label{appsec:vanishingProps}}

We would like to phrase the discussion in terms of a generic polynomial interaction vertex. To achieve a description of a general vertex, let's suppose that we are dealing entirely with $\phi_-$ fields at the vertex. It is not immediately obvious we should do this, and furthermore it seems antithetical to our stated goal of finding a general form for the auxiliary propagator. We shall argue in what follows that there are only the parity of the vertex determines the type of auxiliary propagator which will emerge; in other words, flipping two of the legs from $\phi_-$ to $\phi_+$ at most gives an overall sign. 


Up to overall factors which we strip off, the massless bulk-boundary propagator with external energy $k$ is (see  \eqref{eq:bulk-bndy1}) 
\begin{eqn}
    \mathcal{B}_{-3/2}(z,\vec{k}) \sim (1+k z)e^{-k z}.
\end{eqn}
Having stripped off the $k^{2\Delta_--d}=k^{-3}$, which can be restored after computing any diagram. We note that this structure goes to unity when $k\to0$. Therefore, we might contemplate attaching as many of these propagators as we wish to a vertex since we can always remove them at any step by taking this limit. A product of $N_b$ such propagators can be written
\begin{eqn}
    \prod_{i=1}^{N_b}(1+k_i z)e^{-k_i z}= \sum_{i=1}^{N_b} b_i z^i e^{-b_1 z}
    \label{eq:BbSum}
\end{eqn}

In order to be self-contained, let us repeat the definition of $b_i$, and also $a_j$ which will appear shortly.
\be 
a_j= \sum_{l_1<\ldots <l_j} p_{l_1}\ldots p_{l_j},\,\,\,\,
b_j= \sum_{l_1<\ldots <l_j} k_{l_1}\ldots k_{l_j}\,,
\label{eq:abParametersDef}
\ee 
Note that $b_1$ is what we would call $k_{ext}$ in this language. We also define
\be 
a_M = \prod_{j=1}^M p_j
\ee 
for convenience. $M$ is the number of internal propagators at the vertex, and therefore the largest number $j=M$ for which $a_j$ is non-zero.

Similarly, consider the bulk-bulk propagator (see \eqref{eq:Gspectral}),

\begin{eqn}
    G_{-3/2}(z_1, z_2,\vec{y}) \sim \int dp \frac{1}{(p^2+\vec{y}^2)}\frac{\cos(p z_1)+p z_1 \sin(p z_1)}{p} \frac{\cos(p z_2)+p z_2 \sin(p z_2)}{p}.
    \label{eq:besselJm32}
\end{eqn}

The flat space structure, $\frac{1}{p^2+\vec{y}^2}$, will be captured by the fact we aim to dress a flat-space diagram and therefore, because we aim to obtain an auxiliary propagator by performing the $z$ integration, can be stripped off. We aim to look at a single vertex (we will discuss in the following section how properties of the rest of the diagram matter in allowing certain manipulations), so let us take only the part with $z_1$ dependence. We would like to, in analogy with the bulk-boundary case, be able to remove this propagator by setting the energy to zero. Clearly this cannot work due to the $p$ in the numerator, but we can collect such factors from all bulk-bulk propagators at the vertex into an overall $1/a_M$, which we insert into the definition of the auxiliary propagator. Recall also that the $p$ integration contour is defined to avoid this pole (see \eqref{eq:PVform}). 

We now have the required property that
\begin{eqn}
    \lim_{p\to 0}\cos(p z_1)+p z_1 \sin(p z_1)\to 1.
\end{eqn}
Note that it was crucial we used $\phi_-$, since for $\phi_+$ the relevant term is $\sin(p z_1) - p z_1 \cos(p z_1)$ which goes to zero in the limit. We would like to obtain an equivalent of \eqref{eq:BbSum} for bulk-bulk contributions, but we must first take a detour to consider whether this approach even makes sense.

\subsection{Vertex factors \label{appsec:VertexFactors}}

Recall from the shadow action, \eqref{eq:EAdS_action_gen_pot}, that the shadow formalism potential is 
\begin{eqn}
    V_s(\phi_+, \phi_-) = 2\Re\bigg(e^{i\frac{d-1}{2}}V\big(e^{-i \frac{\pi}{2}\Delta_+} \phi_+ + e^{-i\frac{\pi}{2}\Delta_-}\phi_-\big)\bigg)
    \label{eq:SPotential}
\end{eqn}
where $V$ is the dS potential. For the sake of brevity we will set $\epsilon=0$ to make this point. It should be clear that pieces ending up with a factor of $i$ will be suppressed vertices and those without will be unsuppressed, due to taking the real part. With $d=3$, $\Delta_+=3$, $\Delta_-=0$, and $V(\phi)\sim\phi^v$ the term with $l$ $\phi^+$ fields in \eqref{eq:SPotential} becomes 
\begin{eqn}
    &\sim e^{-i \frac{\pi}{2}\Delta_+ l}e^{-i \frac{\pi}{2}\Delta_-(v-l)}\phi_+^l \phi_-^{v-l} \\
    &= (i)^l \phi_+^l \phi_-^{v-l}.
\end{eqn}

Clearly changing an even number of fields at the vertex from $\phi^+$ to $\phi^-$ or vice-versa results in a minus sign, whereas exchanging an odd number flips the vertex between suppressed and unsuppressed. We shall see in what follows that this sign change is a feature rather than a bug, and is necessary to bestow the dressed diagrams with energy conservation. This implies\footnote{The sceptical reader is encouraged to repeat the above analysis using $\Delta_+=3+\e$, $\Delta_-=\e$, and $d=3+2\e$. They ought to find that the full vertex factor, after taking the real part, can be written $2\bigg(\cos\big(\frac{\pi}{2}l\big)\cos\big(\frac{\pi}{2}(v-2)\e\big)-\sin\big(\frac{\pi}{2}l\big)\sin\big(\frac{\pi}{2}(v-2)\e\big)\bigg)$ from which it is clear that $l\to l+1$ changes the type of vertex and $l\to l+2$ flips the sign but keeps the type the same. Furthermore the claim made in footnote \ref{fn:dimShift}, that a shift $d\to 3+\frac{2}{v-2}\epsilon$ is needed for $\phi^v$, is clear. If one considers diagrams with multiple types of vertex, a consistent shift should be chosen. } that to derive the two types of auxiliary propagator, dotted and dashed, one can consider the behaviour at a vertex with zero $\phi^+$ legs, and a vertex with one $\phi^+$ leg. These possibilities correspond to suppressed and unsuppressed vertices in the shadow language. All other vertices enter only via adding loops, as we now explain.

\subsection{From shadow diagrams to dressed diagrams \label{appsec:AddingLoops}}

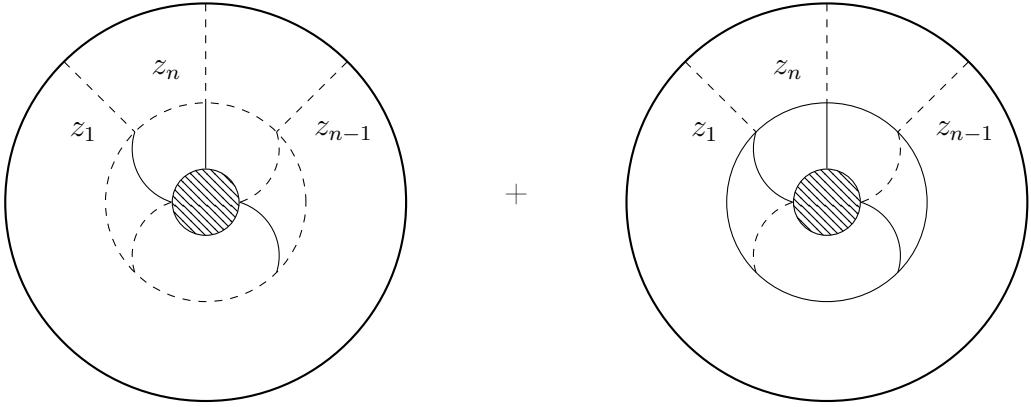
\begin{figure}
    \centering
    \begin{tikzpicture}[baseline=(b.base)]
        \begin{feynman}
            \vertex (a);
            \vertex [right= 75pt of a, blob, minimum size=25pt] (b) {};
            \vertex [right=75pt of b](c);
            \vertex [above = 75pt of b](zno);
            \vertex [above = 37.5pt of b](zni);
            \vertex [above right = 53.03pt and 53.03pt of b](zn1o);
            \vertex [above right = 26.52pt and 26.52pt of b](zn1i);
            \vertex [above left = 53.03pt and 53.03pt of b](z1o);
            \vertex [above left = 26.52pt and 26.52pt of b](z1i);
            \vertex [left = 10pt of z1i](z1l){\(z_1\)};
            \vertex [above left = 5pt and 5pt of zni](znl){\(z_n\)};
            \vertex [right = 10pt of zn1i](zn1l){\(z_{n-1}\)};
            \vertex [below left = 26.52pt and 26.52pt of b](ze1);
            \vertex [below right = 26.52pt and 26.52pt of b](ze2);

            \diagram* {
                (z1o) --[scalar] (z1i),
                (zno) --[scalar] (zni),
                (zn1o)  --[scalar] (zn1i),
                (z1i) --[quarter right] (b),
                (zni) -- (b),
                (zn1i) -- [scalar,quarter left](b),
                (ze1) -- [scalar,quarter left](b),
                (ze2) -- [quarter right](b),
            };
            \draw[thick] (b) circle [radius=75pt];
            \draw[dashed] (b) circle [radius=37.5pt];
            
        \end{feynman}
    \end{tikzpicture}
    \hspace{1cm}
    +
    \hspace{1cm}
    \begin{tikzpicture}[baseline=(b.base)]
        \begin{feynman}
            \vertex (a);
            \vertex [right= 75pt of a, blob, minimum size=25pt] (b) {};
            \vertex [right=75pt of b](c);
            \vertex [above = 75pt of b](zno);
            \vertex [above = 37.5pt of b](zni);
            \vertex [above right = 53.03pt and 53.03pt of b](zn1o);
            \vertex [above right = 26.52pt and 26.52pt of b](zn1i);
            \vertex [above left = 53.03pt and 53.03pt of b](z1o);
            \vertex [above left = 26.52pt and 26.52pt of b](z1i);
            \vertex [left = 10pt of z1i](z1l){\(z_1\)};
            \vertex [above left = 5pt and 5pt of zni](znl){\(z_n\)};
            \vertex [right = 10pt of zn1i](zn1l){\(z_{n-1}\)};
            \vertex [below left = 26.52pt and 26.52pt of b](ze1);
            \vertex [below right = 26.52pt and 26.52pt of b](ze2);

            \diagram* {
                (z1o) --[scalar] (z1i),
                (zno) --[scalar] (zni),
                (zn1o)  --[scalar] (zn1i),
                (z1i) --[quarter right] (b),
                (zni) -- (b),
                (zn1i) -- [scalar,quarter left](b),
                (ze1) -- [scalar,quarter left](b),
                (ze2) -- [quarter right](b),
            };
            \draw[thick] (b) circle [radius=75pt];
            \draw (b) circle [radius=37.5pt];
            
        \end{feynman}
    \end{tikzpicture}
    \caption{A depiction of two diagrams which can be added together to successfully dress every vertex in the loop. For simplicity we show the all-dashed and all-solid loops being added, but propagators in the loop can be any combination of dashed and solid. 
    Once this pairwise addition is shown to work, the entire diagram can be dressed by summing over all possible pairs. For $L$ loops there are $2^L$ diagrams which add to a single dressing. }
    \label{fig:GeneralDerivationLoop}
\end{figure}

We aim to show that the properties of these other vertices, and the diagram topology, can be reduced to a simple rule which can be applied in the valuation of the $z$ integral at any vertex. Let us illustrate the point by considering the conformally coupled case. A similar proof also goes through for the massless case as discussed in  section \ref{appsec:MasslessAddition}.
The conformally coupled bulk-bulk propagator goes as 
\begin{eqn}
    G_{-1/2}(z_1, z_2,\vec{y}) \sim \int dp \frac{1}{(p^2+\vec{y}^2)}(z_1 z_2) \cos(p z_1) \cos(p z_2)
    \label{eq:besselJm12}
\end{eqn}
for $\phi^-$, and for $\phi^+$ one has sines rather than cosines. Now consider a loop with $n$ vertices as shown on the left of Fig.\ \ref{fig:GeneralDerivationLoop}. We assume only that none of the vertices in the loop are connected by a single propagator other than those which comprise the loop\footnote{This assumption can be relaxed by making use of the result which follows. See section \ref{appsec:iceCream} for an explicit worked example.}, and make no assumptions about the valence of the vertices. We can consider both $\phi^+$ and $\phi^-$ simultaneously by using 
\begin{eqn}
    \cos(p z-\frac{\pi}{2} t) = \begin{cases}
        \cos(p z) & t=0 \\
        \sin(pz) & t=1.
    \end{cases}
    \label{eq:generalCCTrig}
\end{eqn}

For an $n-$vertex loop, the product of propagators has the trigonometric structure

\begin{eqn}
        \intinf \prod_{j=1}^n dp_j  \cos(p_j z_{j-1} -\frac{\pi}{2}t_j) \cos(p_j z_{j}-\frac{\pi}{2}t_j).
        \label{eq:CCBBIntegrand}
\end{eqn}
Now let us consider the $z_j$\textsuperscript{th} vertex. Using a standard trigonometric addition identity, we may write
\begin{eqn}
    &\intinf dp_j \cos(p_j z_{j} -\frac{\pi}{2}t_j) \cos(p_{j+1} z_{j}-\frac{\pi}{2}t_j) \\
    &= \intinf dp_j\frac{1}{2} \bigg(\cos\big((p_j+p_{j+1})z_j-\frac{\pi}{2}(t_j + t_{j+1})\big)+\cos\big((p_j-p_{j+1})z_j-\frac{\pi}{2}(t_j-t_{j+1})\big)\bigg)\\
    &=\intinf  dp_j\cos\big((p_j+p_{j+1})z_{j}-\frac{\pi}{2}(t_j+t_{j+1})\big)
\label{eq:generalCCTrigReduction}
\end{eqn}
where the equality holds under the integral and using the symmetry properties of $\cos(p_{j+1} z_{j+1}-\frac \pi 2 t_{j+1})$. We can repeat this trick for any other bulk-bulk propagators at the vertex (i.e. ones which do not compose part of the loop) because of the assumption that they do not connect directly to any other vertex in the loop. We obtain under the integral 
\begin{eqn}
      \cos(p_j z_{j-1} -\frac{\pi}{2}t_j) \cos(p_j z_{j}-\frac{\pi}{2}t_j)\cdots\to \cos\big(p^{j}_{\rm tot} z_{j}-\frac{\pi}{2}t^{j}_{\rm tot}\big)
\end{eqn}
where $\cdots$ refers to other trigonometric factors at the vertex (i.e. from propagators which are not part of the loop), $p^{j}_{\rm tot}$ is the sum of all internal energies at the vertex, and $t^{j}_{\rm tot}$ is the associated sum of $t$ variables\footnote{Ultimately $t_{\rm tot}$ just describes whether the the vertex is suppressed (if it is even, for conformal coupling) or unsuppressed (if it is odd).} The ellipses indicate any other bulk-bulk propagators at the $z_j$ vertex. 

We can repeat this process for \textit{almost} every propagator in the loop. But when it comes to the vertex $z_n$ we hit a problem; the $\cos(p_1 z_1 - \frac\pi 2t_1)$ term is no longer in the product because we have already applied our trick to reduce it. The result is 
\begin{eqn}
(\text{\ref{eq:CCBBIntegrand}})=\intinf \bigg(\prod_{j=1}^{n-1} dp_j \cos\big(p^{j}_{\rm tot} z_{j}-\frac{\pi}{2}t^{j}_{\rm tot}\big)\bigg)\,dp_n\,\cos(p_n z_{n} -\frac{\pi}{2}t_n) \cos(p_{1} z_{n}-\frac{\pi}{2}t_{1})
    \label{eq:CCBBReduced}
\end{eqn}
We now add it to the diagram with all propagators in the loop swapped $\phi^+ \leftrightarrow \phi^-$, implemented by $t_j\to 1-t_j$, as shown on the right of Fig.\ \ref{fig:GeneralDerivationLoop}. This means 
\begin{eqn}
    t^j_{\rm tot}=t_j+t_{j+1}+s \to 2 - t_j - t_{j+1}+s= 2(1-t_j-t_{j+1})+t^j_{\rm tot}
\end{eqn}
where $s$ is the sum of $t$ variables at the vertex which do not belong to propagators in the loop. It turns out that each term under the product in \eqref{eq:CCBBReduced} becomes the following

\begin{eqn}
    \cos\big(\theta +\pi(1-t_j-t_{j+1})\big)=(-1)^{1+t_j+t_{j+1}}\cos(\theta).
\end{eqn}
Recall that one of the two shadow fields is always ghostlike so we pick up a minus sign for each propagator flipped. 
Additionally there is a minus sign associated with every vertex, following the discussion in section \ref{appsec:VertexFactors}.  
Overall this second diagram can be written
\begin{eqn}
   &(-1)^n (-1)^{n-1} (-1)^n (-1)^{t_n+t_1} \intinf \bigg(\prod_{j=1}^{n-1} dp_j \cos\big(p^{j}_{\rm tot} z_{j}-\frac{\pi}{2}t^{j}_{\rm tot}\big)\bigg)\\
   &\times dp_n\,\underbrace{\cos(p_n z_{n} -\frac{\pi}{2}(1-t_n))}_{(-1)^{t_n}\sin(p_n z_{n}-\frac{\pi}{2}t_n)} \underbrace{\cos(p_{1} z_{n}-\frac{\pi}{2}(1-t_{1}))}_{(-1)^{t_1}\sin(p_n z_{n}-\frac{\pi}{2}t_1)}.
\end{eqn}
Finally, we find that the sum of the two diagrams\footnote{Recall that everything else in the diagram is kept exactly the same, only propagators in the loop change so any other factors are common to both terms} is 
\begin{eqn}\label{appeq:confdress1}
    &\intinf \bigg(\prod_{j=1}^{n-1} dp_j \cos\big(p^{j}_{\rm tot} z_{j}-\frac{\pi}{2}t^{j}_{\rm tot}\big)\bigg)\\
    &\times \bigg(\intinf dp_n \, \cos(p_n z_n -\frac{\pi}{2}t_n)\cos(p_1 z_1- \frac{\pi}{2}t_1)-(-1)^n\sin(p_n z_n -\frac{\pi}{2}t_n)\sin(p_1 z_1- \frac{\pi}{2}t_1)\bigg)\\
    &=\intinf \prod_{j=1}^{n} dp_j \cos\big(p^{j}_{\rm tot} z_{j}-\frac{\pi}{2}t^{j}_{\rm tot}\big)
\end{eqn}
where on the last line we have also applied the trick to any other bulk-bulk propagators at the $z_n$ vertex\footnote{Note that if $n$ is odd then $p_{\rm tot}^n\sim p_1 - p_n+\sum (\rm other)$ and likewise for $t_{tot}^n$. This is not cause for concern, and the energy $p_{\rm tot}$ which will come to be associated with each auxiliary propagator can be determined exactly by imposing energy conservation on a dressed diagram. To see the energy-conserving structure, it is useful to shift the variables: $p_n \to \omega$, $p_i \to -p_{i-1}+ E_i $.  }. 

In the above derivation, it was assumed that none of the vertices in the loop were directly connected to each other by a single propagator which was not a part of the loop. Cases in which this is not true work out in the same way by applying this result recursively. 

\subsubsection{Conformally coupled dressing rules}

Following the result in \eqref{appeq:confdress1}, let us complete the derivation of the dressing rules for conformally coupled fields and verify that it matches those found by examples in section \ref{sec:Dressing}. 

Consider a vertex with $N_b$ bulk-bulk and $N_B$ bulk-boundary propagators attached. We take all legs to be $\phi_-$, apart from one which is controlled by the parameter $t$. We wish to perform the $z$ integral. By the above analysis, we can always write the product of trigonometric terms at the vertex as

\begin{eqn}
    z^{N_B} \cos(p_{\text{tot}} z - \frac{\pi}{2} t)
\end{eqn}
where we have restored a factor of $z$ as appeared in \eqref{eq:besselJm12}. Conformally coupled bulk-boundary propagators go as 
\begin{eqn}
    \mathcal{B}_{-1/2}(z,\vec{k}) \sim z e^{-k z}.
\end{eqn}
so a product of $N_b$ of them is simply $z^{N_b} e^{-k_{\text{ext}} z}$. 
Therefore, including the measure and accounting for the dimensional regularisation parameter we get the integral 
\begin{eqn}
    -2\cos\left(\frac{\pi}{2}\e - \frac{\pi}{2}(t+v)\right)\int dz \, z^{v + \e - 4} e^{-k_{\text{ext}} z}\cos(p_{\text{tot}} z - \frac{\pi}{2} t)
    \label{eq:CCAuxPropIntegral}
\end{eqn}
where $v=N_b+N_B$ is the valency of the interaction and therefore the total number of propagators at the vertex and we have inserted the vertex factor for conformally coupled theories\footnote{Whether the all-$\phi^-$ vertex is suppressed or not depends on the valency. For massless theories it is the all-$\phi^+$ vertex with this property so the valency will not appear in the vertex factor for that case. }. Note that if $v\geq 4$ then the power of $z$ is not negative and therefore there is no IR divergence (even at intermediate steps). In such cases we may take $\e \to 0$ and it is evident that one of the auxiliary propagators vanishes. This is precisely what we observed for conformally coupled $\phi^4$, where there was a single auxiliary propagator. Furthermore, for the vertex factor to be non-zero, $t+v$ must be even. This means that for $\phi^5$, it is the $t=1$ case which is non-zero, whereas for $\phi^6$ it is $t=0$. The effect on the $z$ integral of increasing the valency by one is that the power of $z$ increases by one and the cosine gets phase shifted. Both of these effects are encapsulated by taking a derivative with respect to $p_{\text{tot}}$.

For $\phi^4$ theory we obtain the auxiliary propagator for $t=0$ by 
\begin{eqn}
    -2\int_0^\infty dz \, e^{-k_{\rm ext}}\cos(p_{\rm tot} z) = -\frac{2k_{\rm ext}}{k_{\rm ext}^2+p_{\rm tot}^2}.
\end{eqn}
Therefore, from \eqref{eq:CCAuxPropIntegral}  the auxiliary propagator for $\phi^v$ with $v \geq 4$ can be written via absorbing the powers of $z$ through derivatives in $p_{\rm tot}$ and is given by,
\begin{eqn}
    \frac{d^{v-4}}{d p_{\rm tot}^{v-4}}\left( -\frac{2k_{\rm ext}}{k_{\rm ext}^2+p_{\rm tot}^2}\right).
\end{eqn}
For odd $v$, these can only be attached in an even number (since we have all-$\phi^-$ correlators and if $v$ is odd then the all-$\phi^-$ vertex is zero) which immediately implies that the contact diagram for odd-power interactions is zero. Note that this is consistent with the expectation via computing it from the wave function. This is because, for such theories, the contact diagram is IR finite and therefore leads to an imaginary wave function coefficient via the analytical continuation discussed in section \ref{sec:WF}. Hence, the corresponding in-in correlator is zero, as it only receives contributions from the real part.

The exception to this is $\phi^3$ which must be considered separately\footnote{From the wave function perspective, this is related to the fact that the on-shell action in this theory is IR divergent. See section \ref{phi34pt} for a discussion on obtaining this result by analytically continuing the wave function. }. In this special case, there are two auxiliary propagators. One can verify these by evaluating \eqref{eq:CCAuxPropIntegral} for $v=3$ and both $t=0,1$ in the $\e \to 0$ limit to obtain \eqref{eq:cc3SuppAuxProp} and \eqref{eq:CC3UnsuppAuxProp} up to an overall phase. For the $s-$integral form shown in \eqref{eq:CC3UnsuppAuxProp}, we set $\e=0$ then use 

\begin{eqn}
    z^{-1}e^{-k_{\rm ext}z}= \int_0^\infty ds \, e^{-(s+k_{\rm ext})z}.
\end{eqn}

\subsubsection{Other diagrams\label{appsec:iceCream}}
Here we briefly consider a case in which the assumption of section \ref{appsec:AddingLoops} is violated and show how the result holds. 
For any generic loop diagram, let us label the loop we start with as ``primary'' and the loop induced by this additional propagator ``secondary''. We then sum over diagrams in which the secondary loop is flipped to propagators of the other type. The full symmetry of that loop has been dealt with and the assumptions of the previous section are valid for the primary loop. Any additional loops could be dealt with analogously. 
Note that we are effectively using the fact that any propagator directly connecting vertices in the primary loop forms a two-vertex loop, for which the above result is perfectly valid. 

We consider an example at two loop to provide clarity, namely the ice cream diagram which was discussed in \cite{Chowdhury:2023arc}. Figure \ref{fig:iceCreamSum} shows the four diagrams which must be added together to obtain a full dressing. Fig.\ \ref{fig:iceCream_Loops} shows a convention for labelling the three possible loops in this two-loop diagram. Considering each of these as the ``primary'' loop leads to a different pair of diagrams to be summed.

\begin{figure}[H]
    \centering
    \begin{subfigure}{0.45\textwidth}
    \centering
    \begin{tikzpicture}[baseline=(b.base)]
        \begin{feynman}
            \vertex (a);
            \vertex [right = 50pt of a](b);
            \vertex [right = 50pt of b](c);
            \vertex [right=20pt of a](l);
            \vertex [ left=20pt of c](r);
            \vertex [ below = 30pt of b](d);
            \vertex [below right = 49.85pt and 3.92pt of b](dr);
            \vertex [below left = 49.85pt and 3.92pt of b](dl);

            \diagram* {
                (a) --[scalar](l),
                (c) --[scalar](r),
                (dr)--[scalar](d),
                (dl)--[scalar](d),
                (l) --[scalar](r),
                (l) --[scalar,half left](r),
                (l) --[scalar, quarter right](d),
                (d) --[scalar, quarter right](r),
                
            };
            \draw[thick] ($(a)!0.5!(c)$) circle [ 
            radius=50pt
            ];
            
        \end{feynman}
    \end{tikzpicture}
    \caption{\label{sfig:iceCreama}}
    \end{subfigure}
    \begin{subfigure}{0.45\textwidth}
    \centering
    \begin{tikzpicture}[baseline=(b.base)]
        \begin{feynman}
            \vertex (a);
            \vertex [right = 50pt of a](b);
            \vertex [right = 50pt of b](c);
            \vertex [right=20pt of a](l);
            \vertex [ left=20pt of c](r);
            \vertex [ below = 30pt of b](d);
            \vertex [below right = 49.85pt and 3.92pt of b](dr);
            \vertex [below left = 49.85pt and 3.92pt of b](dl);

            \diagram* {
                (a) --[scalar](l),
                (c) --[scalar](r),
                (dr)--[scalar](d),
                (dl)--[scalar](d),
                (l) --(r),
                (l) --[half left](r),
                (l) --[scalar, quarter right](d),
                (d) --[scalar, quarter right](r),
                
            };
            \draw[thick] ($(a)!0.5!(c)$) circle [ 
            radius=50pt
            ];
            
        \end{feynman}
    \end{tikzpicture}
    \caption{\label{sfig:iceCreamb}}
    \end{subfigure}
    \par\bigskip
    \begin{subfigure}{0.45\textwidth}
    \centering
    \begin{tikzpicture}[baseline=(b.base)]
        \begin{feynman}
            \vertex (a);
            \vertex [right = 50pt of a](b);
            \vertex [right = 50pt of b](c);
            \vertex [right=20pt of a](l);
            \vertex [ left=20pt of c](r);
            \vertex [ below = 30pt of b](d);
            \vertex [below right = 49.85pt and 3.92pt of b](dr);
            \vertex [below left = 49.85pt and 3.92pt of b](dl);

            \diagram* {
                (a) --[scalar](l),
                (c) --[scalar](r),
                (dr)--[scalar](d),
                (dl)--[scalar](d),
                (l) --(r),
                (l) --[scalar,half left](r),
                (l) --[ quarter right](d),
                (d) --[ quarter right](r),
                
            };
            \draw[thick] ($(a)!0.5!(c)$) circle [ 
            radius=50pt
            ];
            
        \end{feynman}
    \end{tikzpicture}
    \caption{\label{sfig:iceCreamc}}
    \end{subfigure}
    \begin{subfigure}{0.45\textwidth}
    \centering
    \begin{tikzpicture}[baseline=(b.base)]
        \begin{feynman}
            \vertex (a);
            \vertex [right = 50pt of a](b);
            \vertex [right = 50pt of b](c);
            \vertex [right=20pt of a](l);
            \vertex [ left=20pt of c](r);
            \vertex [ below = 30pt of b](d);
            \vertex [below right = 49.85pt and 3.92pt of b](dr);
            \vertex [below left = 49.85pt and 3.92pt of b](dl);

            \diagram* {
                (a) --[scalar](l),
                (c) --[scalar](r),
                (dr)--[scalar](d),
                (dl)--[scalar](d),
                (l) --[scalar](r),
                (l) --[half left](r),
                (l) --[ quarter right](d),
                (d) --[ quarter right](r),
                
            };
            \draw[thick] ($(a)!0.5!(c)$) circle [ 
            radius=50pt
            ];
            
        \end{feynman}
    \end{tikzpicture}
    \caption{\label{sfig:iceCreamd}}
    \end{subfigure}
    \caption{The four diagrams which must be summed to fully dress the ``ice-cream'' diagram. This diagram violates the assumptions made in the above sections, because in any loop one can isolate, the two vertices are connected by another propagator which is not part of that loop. Nonetheless these four diagrams can be found by choosing one loop, flipping all the propagators, and adding that diagram.}
    \label{fig:iceCreamSum}
\end{figure}
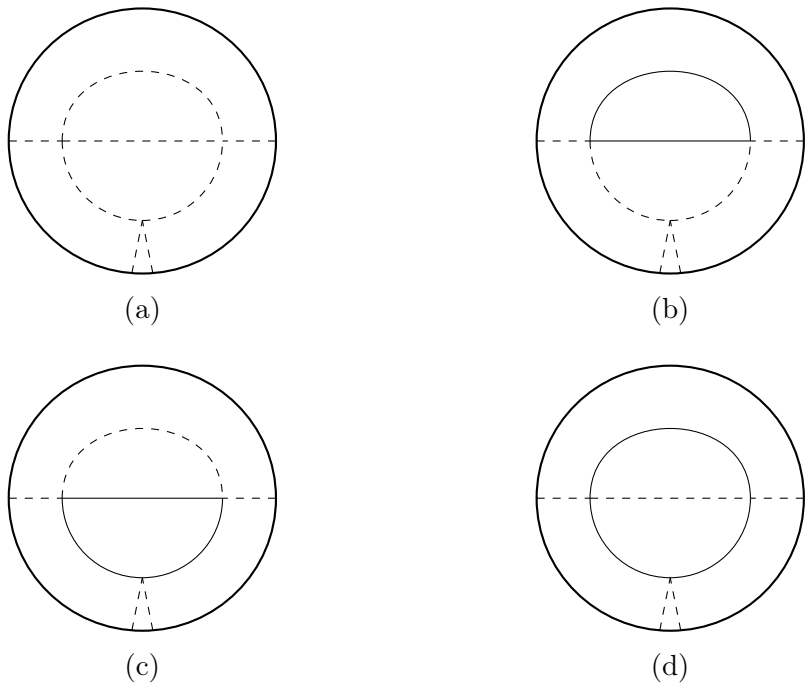

\begin{figure}[H]
    \centering
    \scalebox{1.5}{\begin{tikzpicture}[baseline=(b.base)]
        \begin{feynman}
            \vertex (a);
            \vertex [right = 50pt of a](b);
            \vertex [right = 50pt of b](c);
            \vertex [right=20pt of a](l);
            \vertex [ left=20pt of c,](r);
            \vertex [ below = 30pt of b](d);
            \vertex [below right = 49.85pt and 3.92pt of b](dr);
            \vertex [below left = 49.85pt and 3.92pt of b](dl);
            \vertex [above=30pt of b](t);

            \diagram* {
                (l) --(r),
                (l) --[half left](r),
                (l) --[ quarter right](d),
                (d) --[ quarter right](r),
                (l) --[half left, momentum'=$2$,CBOrange,draw=none](r),
                (r) --[half left, momentum'=$1$,CBBlue,draw=none](l),
                (d) --[half left, momentum=$3$,CBPurple,draw=none](t),
                
            };
            
        \end{feynman}
    \end{tikzpicture}}
    \caption{Our labelling convention for the different loops one could consider in the ice cream diagram. The three loops each correspond to effectively ignoring one of the propagators in the diagram.  }
    \label{fig:iceCream_Loops}
\end{figure}
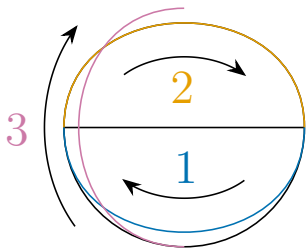


Let us take loop 1 to be the primary and loop 2 the secondary. Then, starting from Fig.\ \ref{sfig:iceCreama}, dealing with the secondary loop corresponds to taking the sum of diagrams in the top row. After doing so, we can must then add the diagrams where the primary loop is flipped. This generates the two remaining diagrams, and indeed the sum of four produces a single dressing. In general, for a diagram with $L$ loops, $2^L$ shadow diagrams will add to a single dressed diagram of the form given in \eqref{appeq:confdress1}!

\subsection{Massless dressing rules \label{appsec:MasslessAddition}}

Let us consider how the derivation above changes for massless fields. Recall from \eqref{eq:besselJm32} that the massless $\phi^-$ bulk-bulk propagator trigonometric structure is 
\begin{eqn*}
    \cos(p z) + p z \sin(p z)
\end{eqn*}
so when there are two at a vertex we get
\begin{eqn}
    &\int dp_2\bigg(\cos(p_1 z) + p_1 z \sin(p_1 z)\bigg)\bigg(\cos(p_2 z) + p_2 z \sin(p_2 z)\bigg) \\
    &= \int dp_2\cos(p_1 z) \cos(p_2 z) + p_1 z \sin(p_1 z) \cos(p_2 z) + p_2 z \cos(p_1 z) \sin(p_2 z)\\
    &\qquad+  p_1 p_2 z^2 \sin(p_1 z) \sin(p_2 z)\\
    &= \int dp_2\big(1+p_1 p_2 z^2\big)\cos\big((p_1 +p_2)z\big)+(p_1+p_2)z\sin\big((p_1+p_2)z\big).
\end{eqn}
Again the second equality holds under the $p_2$ integral. Following the procedure explained for the conformally coupled case, one can show that indeed adding together shadow loops is always sufficient to do this form of reduction at every vertex in the loop. Furthermore, one can show that for a vertex with $N_B$ bulk-bulk propagators attached each controlled by a parameter $t$\footnote{In analogy with the conformally coupled case \eqref{eq:CCAuxPropIntegral}, the trigonometric structure of the massless propagator can be written \begin{eqn}
    \cos\left(p z - \frac{\pi}{2} t\right)+p z \sin\left(p z- \frac{\pi}{2} t\right)
\end{eqn}
} the result is 

\begin{eqn}
    \sum_{j=0}^{N_B} z^j a_j \cos\big(p_{\rm tot}-\frac{\pi}{2}(t_{\rm tot}+j)\big).
    \label{eq:MasslessReducedFactor}
\end{eqn}

We now have all the tools required to derive the general auxiliary propagators for a massless theory.

\subsection{Massless Auxiliary propagators \label{appsec:GAP}}
In this section we perform the $z$-integrals in the structure above and obtain the final form of the auxiliary propagators for a massless $\phi^v$ theory. 

Consider a vertex with $N_b$ bulk-bulk propagators and $N_B$ bulk-boundary propagators attached. We multiply \eqref{eq:BbSum} and \eqref{eq:MasslessReducedFactor} to obtain the final $z$ integral at this vertex up to an overall sign
\begin{eqn}
&\cos\big(\frac{\pi}{2}(\e -t)\big)\int_0^\infty dz \, \frac{z^{-4+\epsilon}}{a_M}\sum_{i=0}^{N_b} \sum_{j=0}^{N_B}a_j b_i z^{i+j}  e^{-k_{\rm ext} z} \cos\big(p_{\rm tot}z - \frac{\pi}{2}(t+j)\big) \\
&=\cos\big(\frac{\pi}{2}(\e -t)\big)\int_0^\infty dz \, \frac{z^{-4+\epsilon}}{a_M}\sum_{i=0,j=0}^{i+j=v} a_j b_i z^{i+j}  e^{-k_{\rm ext} z} \cos\big(p_{\rm tot}z - \frac{\pi}{2}(t+j)\big)
\label{eq:AuxPropDerivationStart}
\end{eqn}
where the equality follows by noting that $N_b + N_B = v$ and terms vanish when $i+ j > v$, because the sum of the number of internal and external legs cannot exceed the total number of legs.

Next we remove negative powers of $z$ by introducing the $s-$integrals via,
\be 
z^{-x} e^{-k z} = \int_0^\infty ds \, \frac{s^{x-1} e^{- (s+k) z}}{\Gamma(x)}
\ee 
to obtain 
\begin{eqn}
\cos\big(\frac{\pi}{2}(\e -t)\big) \intsinf  \frac{ds \,s^{3-i-j-\e}}{a_M\Gamma(4-i-j-\e)}  \sum_{i=0,j=0}^{i+j=v} a_j b_i \int_0^\infty dz e^{-(s+k_{\rm ext}) z} \cos\big(p_{\rm tot} z - \frac{\pi}{2}(t+j)\big).
\label{eq:GeneralSInt}
\end{eqn}
Performing the $z$ integration,
\begin{eqn}
    \int_0^\infty dz \cos\big(p z- \frac{\pi}{2}(t+j)\big)e^{-(s+k_{\rm ext}) z} = \frac{(s+k_{\rm ext}) \cos\big(\frac{\pi}{2}(t+j)\big)+p_{\rm tot}\sin\big(\frac{\pi}{2}(t+j)\big)}{p^2+(s+k)^2},
\end{eqn}
gives the final form of the propagator,
\begin{eqn}
&\int_{0}^\infty ds \frac{\cos\big(\frac{\pi}{2}(\e -t)\big)}{(s+k_{\rm ext})^2+p_{\rm tot}^2}\frac{\sum_{i,j=0}^{i+j=v} s^{3-i-j+\epsilon}a_j b_i \cos\big(\frac{\pi}{2}(t+j)\big)+p_{\rm tot}\sin\big(\frac{\pi}{2}(t+j)\big)} {a_M \Gamma(4-i-j-\epsilon)} \\
&= \begin{cases}
   \cos\big(\frac{\pi}{2} \e\big)\int_{0}^\infty ds\frac{\mathcal{N}_v}{(s+k_{\rm ext})^2+p_{\rm tot}^2} 
   & t=0, \quad \rm{Dashed \; Propagator} \\ \\
    \sin\big(\frac{\pi}{2} \e\big)\int_{0}^\infty ds \frac{\widetilde{\mathcal{N}}_v}{(s+k_{\rm ext})^2+p_{\rm tot}^2} 
    & t=1, \quad \rm{Dotted \; Propagator}
\end{cases}
\end{eqn}
where the numerators are precisely those defined in \eqref{eq:t0Integrand} and \eqref{eq:t1Integrand}.

\section{Finite massless correlators}\label{app:Integrals}
In this Appendix we demonstrate the use of the $s$-integral representation in order to compute correlators in massless theories. While the $s$-integral representation was introduced from the perspective of the dressing rule in section \ref{sec:Dressing}, it is also useful from the perspective of the wavefunction and the Schwinger-Keldysh approach \cite{Arkani-Hamed:2018kmz, Benincasa:2019vqr}  
since the $s$-integral representation trivializes $z$ integrals 
and allows one to express them in terms of correlators in flat space. Thus, the 4-point correlator of massless fields in $\phi^3$ theory (see section \ref{sec:masslessexchSK}) can be expressed as a combination of the following integrals\footnote{For further discussion of such integrals, including GKZ systems and positivity properties, see refs.\ \cite{Grimm:2024tbg,henn2024positivity}.},
\begin{eqn}
\sum_{\a, \b} c_{\a \b}\intsinf ds_1 ds_2 \frac{s_1^\a s_2^\b}{(E_L + s_1)(E_R + s_2)(E + s_1 + s_2)} 
\end{eqn}
where $c_{\a,\b}$ are coefficients which are functions of the external momenta, $\a, \b$ are real numbers and $E_{L/R}, E$ are partial and total energies respectively, and in this context are positive real numbers. This integral can be viewed as a  generalization of equation (3.1) of \cite{Arkani-Hamed:2023kig} to a 2-parameter deformation of the standard cosmological integral and can be expressed in terms of the Gauss Hypergeometric function,
\begin{eqn}
&\intsinf ds_1 ds_2 \frac{s_1^\a s_2^\b}{(E_L + s_1)(E_R + s_2)(E + s_1 + s_2)} \\
&= -\frac{\pi  }{\sin(\pi \beta) \Gamma(-\beta ) E_L \left(E_L+E_R-E\right)}\left(\frac{1}{E-E_R}\right) \\
&\times\Bigg[E^{1 + \a + \b} \Gamma(\alpha +1) \Gamma (-\alpha -\beta -1) \\
&\hspace{-1cm}\quad\times\Big\{\left(E-E_R\right) \, _2F_1\left(1,\alpha +1;\alpha +\beta +2;\frac{E}{E_L}\right)-E_L \, _2F_1\left(1,\alpha +1;\alpha +\beta +2;\frac{E}{E-E_R}\right)\Big\}\\
& +\pi  \Gamma(-\beta ) E_L \left(E-E_R\right) \\
&\hspace{-1cm}\quad \times \Big\{\frac{E_R^\b}{\sin\pi \a} \left( E_L^\a - (E - E_R)^\b\right) - \frac{1}{\sin(\pi(\a + \b))} \left(E_L^{\alpha } \left(E_L-E\right){}^{\beta }  -(-1)^\b E_R ^{\beta} (E - E_R)^\a \right)\Big\}\Bigg].
\end{eqn}
Typically the parameters $\a, \b$ are integers and also contain the regulator $\e$. This allows us to perform a series expansion of such functions near $\e = 0$ and are conveniently implemented in Mathematica using the package \reference{HypExp} \cite{Huber:2005yg}. We provide the explicit forms of the expressions obtained using SK/dressing and the wave function below. 


\subsection{$\phi^2 \chi$ exchange}

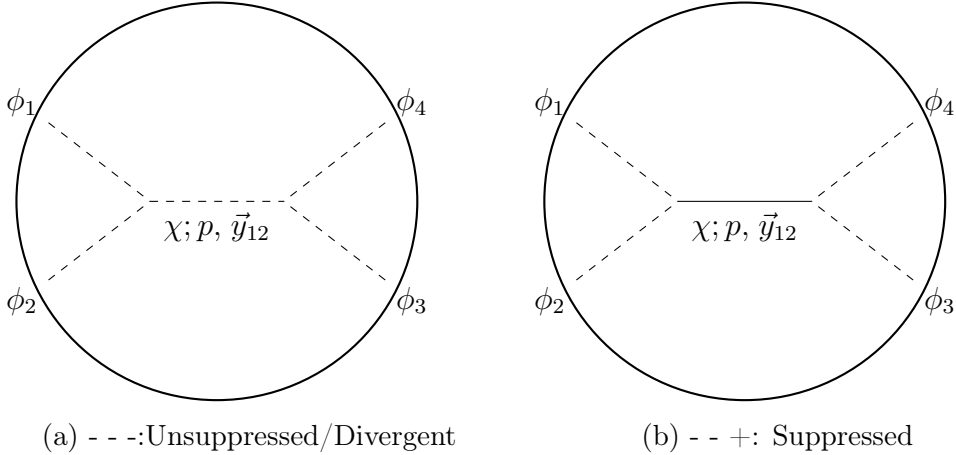
\begin{figure}[H]
    \centering
    \begin{subfigure}{0.45\textwidth}
    \begin{tikzpicture}[baseline=(b.base)]
        \begin{feynman}
            \vertex (a) ;
            \vertex [ right = 50pt of a       ] (i1) ;
            \vertex [ right = 50pt of i1      ] (i2);
            \vertex [right = 75pt of a      ] (b);
            \vertex [right = 75pt of b      ] (c);
            \vertex [above left = 28pt and 63pt of b](l1) {\(\phi_1 \)};
            \vertex [below left = 28pt and 63pt of b](l2) {\( \phi_2\)};
            \vertex [below right = 28pt and 63pt of b](r1) {\(\phi_3\)};
            \vertex [above right = 28pt and 63pt of b](r2) {\(\phi_4\)};
            \vertex [above = 50pt of b](e);
    
            \diagram* {
                (l1) -- [scalar] (i1),
                (l2) --[scalar] (i1),
                (r1)  --[scalar](i2),
                (r2)  --[scalar](i2),
                (i1) --[scalar,edge label'=$\chi;p {,}\;\vec{y}_{12}$] (i2),
            };
        \end{feynman}

        \draw[thick] ($(a)!0.5!(c)$) circle [
        radius=75pt
        ];
    \end{tikzpicture}
    \caption{- - -:Unsuppressed/Divergent \label{sfig:phi2chiUnsupp}}
    \end{subfigure}
    \begin{subfigure}{0.45\textwidth}
    \begin{tikzpicture}[baseline=(b.base)]
        \begin{feynman}
            \vertex (a) ;            
            \vertex [ right = 50pt of a       ] (i1) ;
            \vertex [ right = 50pt of i1      ] (i2);
            \vertex [right = 75pt of a      ] (b);
            \vertex [right = 75pt of b      ] (c);
            \vertex [above left = 28pt and 63pt of b](l1) {\(\phi_1 \)};
            \vertex [below left = 28pt and 63pt of b](l2) {\( \phi_2\)};
            \vertex [below right = 28pt and 63pt of b](r1) {\(\phi_3\)};
            \vertex [above right = 28pt and 63pt of b](r2) {\(\phi_4\)};
            \vertex [above = 50pt of b](e);
    
            \diagram* {
                (l1) -- [scalar] (i1),
                (l2) --[scalar] (i1),
                (r1)  --[scalar](i2),
                (r2)  --[scalar](i2),
                (i1) --[edge label'=$\chi;p {,}\;\vec{y}_{12}$] (i2),
            };

            \draw[thick] ($(a)!0.5!(c)$) circle [
            radius=75pt
            ];
        \end{feynman}
    \end{tikzpicture}
    \caption{- - +: Suppressed \label{sfig:phi2chiSupp}}
    \end{subfigure}
    \caption{The shadow diagrams for four-point $\phi^2 \chi$ exchange. The right diagram does not contribute as it is suppressed and not divergent. The external legs are conformally coupled and the exchanged leg is massless.\label{fig:phi2chi}}
\end{figure}

To demonstrate clearly how one uses the dressing rules in practise for the massless exchange, we consider a simplified model. The so-called $\phi^2 \chi$ vertex has two conformally coupled legs ($\phi$) and one massless leg ($\chi$). The two potential diagrams for the exchange diagram are shown in Fig.\ \ref{fig:phi2chi}.

The diagram in Fig.\ \ref{sfig:phi2chiSupp} turns out not to contribute as we shall now argue and briefly explain how we derive dressing rules for this interaction based on the massless ones. 

Plugging the $\phi^2 \chi$ interaction into the shadow action and noting that we only care about $(\phi^-)^2$ terms, we find the vertex factors are
\be 
\begin{split}
    &\phi^- \phi^- \chi^-: \qquad 2 \lambda \cos(\frac{\pi}{2}\epsilon) \\
    &\phi^- \phi^- \chi^+: \qquad 2 \lambda \sin(\frac{\pi}{2}\epsilon)
\end{split}
\ee
which indeed match the massless case in terms of $\pm$ field types so we can start from the existing auxiliary propagators. Next, we note that 
\begin{eqn}
    &z^{\frac{3}{2}}K_\frac{1}{2} \propto z e^{-k z}\\
    &z^{\frac{3}{2}}K_{\frac{3}{2}} \propto (1+ k z)e^{-k z}
\end{eqn}
and, since we ignore overall factor of $k$ anyway, if we discard the term without a power of $z$ in the massless bulk-boundary propagator we get the conformally coupled one. In terms of the massless auxiliary propagators, \eqref{eq:t0Integrand} and \eqref{eq:t1Integrand}, this precisely corresponds to fixing the index $i=2$ in the sum. Hence, the dressing rules for this theory can be obtained as a special case of the massless $\phi^3$ dressing rules by picking out a particular term in the sum. If one does this to \eqref{eq:M3UnsuppAuxProp} and performs the $s$-integral, it turns out the result is IR finite and, due to the $\sin$ prefactor, vanishes in the $\epsilon \to 0$ limit. Therefore the only contribution comes from the dressed diagram with two dashed auxiliary propagators, which derives from the diagram in Fig.\ \ref{sfig:phi2chiUnsupp}.

In the notebook {\reference{phi2chi\_Exchange.nb}}, found at \cite{github}, we perform this integral which follows precisely the same steps as the fully massless calculation without obscuring the procedure with additional complexity.
The result is 
\begin{eqn}
    \mathcal{A}_{\phi^2 \chi}^{reg}&=-\frac{k_{12}k_{34}}{k^3}\frac{1}{\epsilon^2}+ \frac{k_{12}k_{34}\big(2(\gamma_E-1)+\log E_L +\log E_R \big)-k k_{1234}}{k^3}\frac{1}{\epsilon}\\
    &+ \frac{1}{2k^3}\Bigg(k_{12} k_{34} \Bigg[ \text{Li}_2\left(\frac{E - E_L}{E}\right)+\text{Li}_2\left(\frac{E - E_R}{E}\right) -\log^2E_L -\log^2E_R- \log E_L \log E_R \\
    &+\log^2k_{1234}+\log(E_L E_R)\left(4(1-\gamma_E)-\log k_{1234} \right)-6+4 \gamma_E\left(2-\gamma_E\right) \Bigg] \\
    &+2 k k_{1234} \left(-2 + 2 \gamma_E   +  \log(k_{1234})\right) +2k \left( k_{34}\log E_L+k_{12}\log E_R \right)\Bigg)  +\mathcal{O}(\epsilon)
\end{eqn}
which can be verified by repeating the calculation in the Schwinger-Keldysh formalism.

\subsection{Massless $\phi^3$ 4-point correlator }\label{app:shadowmassless}
Here we present the finite part of the massless exchange diagram in $\phi^3$ theory obtained using the dressing rules and matched with the Schwinger-Keldysh formalism. 
This is also given in the Mathematica notebook \reference{wavefunctionshadow.nb} found in \cite{github}. Since the equation is very long we break it into a few different terms, 
\begin{eqn}
\mathcal A_4  = \mathcal A_4^{(0)}+ \mathcal A_4^{(1)}  + \mathcal A_4^{(2)}  + \mathcal A_4^{(3)}
\end{eqn}
where 
{\small \begin{eqn}
&\mathcal A_4^{(0)}\\
&= \frac{\pi^2}{864 k k_1^3 k_2^3 k_3^3 k_4^3}\Bigg[  k^6 - \left(k_1^3+ k_2^3+k_3^3+k_4^3\right) k^3 -3 \left(k_1^3+k_2^3\right) \left(k_3^3+k_4^3\right) \Bigg] \\
&+\frac{\gamma_E}{216 k k_1^3 k_2^3 k_3^3 k_4^3} \Bigg[7 k^3 \left(k_{12 c}+k_{34 c}\right)+6 k_{12} k_{34} k^2 \left(\frac{k_{12 c}}{k_{12}}+\frac{k_{34 c}}{k_{34}}\right)+\frac{6 E k k_{12 c} k_{34 c}}{k_{12} k_{34}} +16 k_{34 c} k_{12 c}   \\
&\qquad\qquad\qquad + 6 k_3 k_4 k_{34} k_{12 c} +6 k_1 k_2 k_{12} k_{34 c}\Bigg] \\
&+\frac{1}{1296 k k_1^3 k_2^3 k_3^3 k_4^3} \Bigg[18 k^5E-18 k^4 k_{12} k_{34} + k^4\big\{-23 k_{12 c}-23 k_{34 c}+6 E k_{34} k_{12}+6 k_1 k_2 k_{12}+6 k_3 k_4 k_{34} \big\} \\
& - 6k^2 \big\{ 8 k_{34} k_{12 c}+8 k_{12} k_{34 c}+3 k_1 k_2 k_{12} k_{34}+3 k_3 k_4 k_{12} k_{34} \big\} \\
&\qquad - \frac{6k E}{k_{12} k_{34}} \big\{ 3 k_3 k_4 k_{34} k_{12 c}+3 k_1 k_2 k_{12} k_{34 c} +8 k_{34 c} k_{12 c}\big\}\\
&- 6 \big\{ 8 k_3 k_4 k_{34} k_{12 c}+8 k_1 k_2 k_{12} k_{34 c}+14 k_{12 c} k_{34 c}+3 k_1 k_2 k_3 k_4 k_{12} k_{34} \big\}\Bigg] \\
\end{eqn}}
{\small\begin{eqn}
 &\mathcal A_4^{(1)} \\
 &= + \frac{1}{144 k k_1^3 k_2^3 k_3^3 k_4^3} \Bigg[ \big( k_{12 c} k_{34 c}-k^6\big) \log^2 E -  \left(k^3+k_{12c}\right) k_{34c} \log^2 E_L -  \left(k^3+k_{34c}\right) k_{12c}\log^2 E_R \\
& \qquad +\left(k^3+k_{12c}\right) \left(k^3-k_{34c}\right) \log E_L \log E +\left(k^3+k_{34c}\right) \left(k^3-k_{12c}\right) \log E_R \log E \\
&- \left(k_{12 c}+k^3\right) \left(k_{34 c}+k^3\right) \log E_L \log E_R \Bigg]\\
&-\frac{1}{216 k_1^3 k_2^3 k_3^3 k_4^3} \Bigg[ -3 E\frac{ k_{12 c} k_{34 c}}{k_{12} k_{34}}+3 Ek^4+\left(k_{12c}+k_{34c}\right) k^2 \Bigg] \log E  \\
\end{eqn}}
{\begin{eqn}
&\small\mathcal A_4^{(2)} \\
&= + \frac{1}{216 k k_1^3 k_2^3 k_3^3 k_4^3} \Bigg[ -6\left(k_{12 c}+k^3\right) k_{34 c} \gamma_E+\left(3 k_3 k_4 k_{34} k_{12 c}+8 k_{34 c} k_{12 c}+3 k_1 k_2 k_{12} k_{34 c}\right)  \\
&\qquad + k^3 \left(8 k_{34 c}+3 k_3 k_4 k_{34}\right)+3 k^2 \left(k_{34} k_{12 c}+k_{12} k_{34 c}\right)+\frac{3 k k_{12 c} k_{34 c}}{k_{12}}+3 k_{34} k^5\Bigg] \log E_L\\
&+ \frac{1}{216 k k_1^3 k_2^3 k_3^3 k_4^3} \Bigg[ -6\left(k_{34 c}+k^3\right) k_{12 c} \gamma_E+\left(3 k_1 k_2 k_{12} k_{34 c}+8 k_{12 c} k_{34 c}+3 k_3 k_4 k_{34} k_{12 c}\right)  \\
&\qquad + k^3 \left(8 k_{12 c}+3 k_1 k_2 k_{12}\right)+3 k^2 \left(k_{12} k_{34 c}+k_{34} k_{12 c}\right)+\frac{3 k k_{12 c} k_{34 c}}{k_{34}}+3 k_{12} k^5\Bigg] \log E_R\\
\end{eqn}}
{\small\begin{eqn}
&\mathcal A_4^{(3)} \\
&= -\frac{1}{144 k k_1^3 k_2^3 k_3^3 k_4^3} \Bigg[ \left(k_{12 c}+k^3\right) \left(k^3-k_{34 c}\right) \text{Li}_2\left(\frac{E-E_L}{E}\right) +  \left(k_{34 c}+k^3\right) \left(k^3-k_{12 c}\right)\text{Li}_2\left(\frac{E - E_R}{E}\right) \Bigg]
\end{eqn}}
where $k_{12c} = k_{1}^3 + k_2^3$ and $k_{34c} = k_{3}^3 + k_4^3$ and $E_L = k_{12} + k$, $E_R = k_{34} +k$~.

\subsection{Massless $\phi^3$ 4-point wavefunction}\label{app:WFmassless}
Here we present the wavefunction coefficient for the massless exchange diagram in $\phi^3$ theory whose leading divergence was presented in \eqref{WFmassless-ex-div1} and \eqref{WFmassless-ex-div2}. The full result including the derivation can be found in
\cite{github}., 
\begin{eqn}
&\Psi_4^{(s)} \\
&= \frac{1}{\e^2} \frac{2 k^3+k_{1234c}}{72 } \\
&+ \frac{1}{\e} \frac{1}{216 } \Big[ -12 \gamma_E  k^3+16 k^3-6 \bigl(k^3+k_{12c}^3\bigr) \log(E_L E_R )\\
&+6 (k_1 k_2^2+ k_3 k_4^2+ k_1^2 k_2+ k_3^2 k_4 ) +6 k^2 k_{1234} +6 k (k_1^2 + k_2^2 + k_3^2 + k_4^2)\\
&-6 ( k_1 k_2 + k_3 k_4)k + (-6 \gamma_E+9) k_{1234c}
\Big]\\
&  -\frac{1}{1296 k^3 }\Bigg[ -3 \bigl(8 \gamma_E  (-8+3 \gamma_E )+3 \pi ^2+56\bigr) k^6+12 (-11+6 \gamma_E ) k_{1234} k^5\\
&+12 \bigl((-8+6 \gamma_E ) k_1^2+(5-6 \gamma_E ) k_2 k_1+2 (-4+3 \gamma_E ) k_4^2+2 (-4+3 \gamma_E ) \bigl(k_2^2+k_3^2\bigr)+(5-6 \gamma_E ) k_3 k_4\bigr) k^4\\
&-\bigl(\bigl(36 \gamma_E  (-3+\gamma_E )+3 \pi ^2+122\bigr) k_1^3+12 \bigl((9-6 \gamma_E ) k_2+4 \bigl(k_3+k_4\bigr)\bigr) k_1^2\\
&-12 \bigl((-9+6 \gamma_E ) k_2^2+\bigl(k_3+k_4\bigr) k_2-4 k_3^2-4 k_4^2+k_3 k_4\bigr) k_1+2 \bigl((18 \gamma_E  (-3+\gamma_E )+61) k_2^3\\
&+24 \bigl(k_3+k_4\bigr) k_2^2+6 \bigl(4 k_3^2-k_4 k_3+4 k_4^2\bigr) k_2+(18 \gamma_E  (-3+\gamma_E )+61) k_3^3+(18 \gamma_E  (-3+\gamma_E )+61) k_4^3\\
&+18 (3-2 \gamma_E ) k_3 k_4^2+18 (3-2 \gamma_E ) k_3^2 k_4\bigr)+3 \pi ^2 \bigl(k_2^3+k_{34 c}\bigr)\bigr) k^3\\
&-36 \bigl(k_1^2-k_2 k_1+k_2^2\bigr) \bigl(k_3^2-k_4 k_3+k_4^2\bigr) k^2+18 \text{Li}_2\bigl(1-\frac{E_L}{E}\bigr) \bigl(k^3+k_{12 c}\bigr) \bigl(k^3-k_{34 c}\bigr)\\
&+3 \pi ^2 k_{12 c} k_{34 c}+18 \text{Li}_2\bigl(1-\frac{E_R}{E}\bigr) \bigl(k^3-k_{12 c}\bigr) \bigl(k^3+k_{34 c}\bigr)\\
&+6 \Bigl(-3 \log ^2\bigl(E_L\bigr) \bigl(k^3+k_{12 c}\bigr) k^3+\log \bigl(E_R\bigr) \bigl(2 (8-6 \gamma_E ) k^5+6 \bigl(k_3+k_4\bigr) k^4\\
&+6 \bigl(k_1^2-k_2 k_1+k_2^2+k_3^2+k_4^2-k_3 k_4\bigr) k^3+2 \bigl((8-6 \gamma_E ) k_3^3+3 k_4 k_3^2+3 k_4^2 k_3+2 (4-3 \gamma_E ) k_4^3\bigr) k^2\\
&-3 \log \bigl(E_R\bigr) \bigl(k^3+k_{34 c}\bigr) k^2+6 \bigl(k_1^2-k_2 k_1+k_2^2\bigr) k_{34 c}\bigr) k+3 \log ^2(E) \bigl(k^6-k_{12 c} k_{34 c}\bigr)\\
&+\log (E) 
\bigl(6 k_{1234}k^5+2 \bigl(\bigl(k_1+k_2\bigr)^3+\bigl(k_3+k_4\bigr){}^3\bigr) k^3\\
&-6 \bigl(k_1^2-k_2 k_1+k_2^2\bigr) \bigl(k_1+k_2+k_3+k_4\bigr) \bigl(k_3^2-k_4 k_3+k_4^2\bigr) k-3 \log \bigl(E_L\bigr) \bigl(k^3+k_{12 c}\bigr) \bigl(k^3-k_{34 c}\bigr)\\
&-3 \log \bigl(E_R\bigr) \bigl(k^3-k_{12 c}\bigr) \bigl(k^3+k_{34 c}\bigr)\bigr)+\log \bigl(E_L\bigr) \bigl(2 (8-6 \gamma_E ) k^6+6 \bigl(k_1+k_2\bigr) k^5\\
&+6 \bigl(k_1^2-k_2 k_1+k_2^2+k_3^2+k_4^2-k_3 k_4\bigr) k^4+2 \bigl((8-6 \gamma_E ) k_1^3+3 k_2 k_1^2+3 k_2^2 k_1+2 (4-3 \gamma_E ) k_2^3\bigr) k^3\\
&+6 \bigl(k_3^2-k_4 k_3+k_4^2\bigr) k_{12 c} k-3 \log \bigl(E_R\bigr) \bigl(k^3+k_{12 c}\bigr) \bigl(k^3+k_{34 c}\bigr)\bigr)\Bigr)-9 \pi^2\big(k^3+k_{12c}\big)\big(k^3+k_{34c} \big)  \Bigg]
\end{eqn}
\begin{eqn}
&\re \Psi_3 \re \Psi_3 = \frac{1}{\e^2} \frac{\bigl(k^3+k_1^3+k_2^3\bigr) \bigl(k^3+k_3^3+k_4^3\bigr)}{72  }\\
&\quad+ \frac{1}{\e} \frac{1}{8 } \Bigg[ \frac{1}{9} (2-2 \gamma_E ) \bigl(k^3+k_1^3+k_2^3\bigr) \bigl(k^3+k_3^3+k_4^3\bigr)\\
&+\frac{1}{9} \bigl(k^3+k_1^3+k_2^3\bigr) \bigl(\bigl(k_3+k_4\bigr) k^2+\bigl(k_3^2-k_4 k_3+k_4^2\bigr) k+k_3 k_4 \bigl(k_3+k_4\bigr)\bigr)\\
&+\bigl(k^3+k_3^3+k_4^3\bigr) \bigl(\bigl(k^3+k_1^3+k_2^3\bigr) \bigl(\frac{2}{27}-\frac{1}{9} \log \bigl(\bigl(k+k_1+k_2\bigr) \bigl(k+k_3+k_4\bigr)\bigr)\bigr)\\
&+\frac{1}{9} \bigl(\bigl(k_1+k_2\bigr) k^2+\bigl(k_1^2-k_2 k_1+k_2^2\bigr) k+k_1 k_2 \bigl(k_1+k_2\bigr)\bigr)\bigr) \Bigg] \\
&\quad + \frac{1}{8  } \Bigg[ \frac{1}{9} k \bigl(k^3+k_1^3+k_2^3\bigr) k_3 k_4+\frac{1}{54} \bigl(18-24 \gamma_E +12 \gamma_E ^2+\pi ^2\bigr) \bigl(k^3+k_1^3+k_2^3\bigr) \bigl(k^3+k_3^3+k_4^3\bigr)\\
&+\bigl(k^3+k_3^3+k_4^3\bigr) \bigl(\bigl(k^3+k_1^3+k_2^3\bigr) \bigl(\frac{1}{18} \log ^2\bigl(\bigl(k+k_1+k_2\bigr) \bigl(k+k_3+k_4\bigr)\bigr)\\
&-\frac{2}{27} \log \bigl(\bigl(k+k_1+k_2\bigr) \bigl(k+k_3+k_4\bigr)\bigr)+\frac{1}{27}\bigr)\\
&+\bigl(\bigl(k_1+k_2\bigr) k^2+\bigl(k_1^2-k_2 k_1+k_2^2\bigr) k+k_1 k_2 \bigl(k_1+k_2\bigr)\bigr) \bigl(\frac{2}{27}-\frac{1}{9} \log \bigl(\bigl(k+k_1+k_2\bigr) \bigl(k+k_3+k_4\bigr)\bigr)\bigr)\\
&+\frac{1}{9} k k_1 k_2\bigr)+\bigl(\bigl(k_3+k_4\bigr) k^2+\bigl(k_3^2-k_4 k_3+k_4^2\bigr) k+k_3 k_4 \bigl(k_3+k_4\bigr)\bigr) \bigl(\bigl(k^3+k_1^3+k_2^3\bigr) \bigl(\frac{2}{27}\\
&-\frac{1}{9} \log \bigl(\bigl(k+k_1+k_2\bigr) \bigl(k+k_3+k_4\bigr)\bigr)\bigr)+\frac{1}{9} \bigl(\bigl(k_1+k_2\bigr) k^2+\bigl(k_1^2-k_2 k_1+k_2^2\bigr) k+k_1 k_2 \bigl(k_1+k_2\bigr)\bigr)\bigr)\\
&+(2-2 \gamma_E ) \bigl(\frac{1}{9} \bigl(k^3+k_1^3+k_2^3\bigr) \bigl(\bigl(k_3+k_4\bigr) k^2+\bigl(k_3^2-k_4 k_3+k_4^2\bigr) k+k_3 k_4 \bigl(k_3+k_4\bigr)\bigr)\\
&+\bigl(k^3+k_3^3+k_4^3\bigr) \bigl(\bigl(k^3+k_1^3+k_2^3\bigr) \bigl(\frac{2}{27}-\frac{1}{9} \log \bigl(\bigl(k+k_1+k_2\bigr) \bigl(k+k_3+k_4\bigr)\bigr)\bigr)\\
&+\frac{1}{9} \bigl(\bigl(k_1+k_2\bigr) k^2+\bigl(k_1^2-k_2 k_1+k_2^2\bigr) k+k_1 k_2 \bigl(k_1+k_2\bigr)\bigr)\bigr)\bigr) \Bigg]
\end{eqn}
The s-channel contribution to the 4-point in-in correlator is then given by
\begin{eqn}
\mathcal A_4^{(s)} = \re\Psi_4^{(s)} - \frac{2}{k^3} \re\Psi_3 \re\Psi_3.
\end{eqn}

\subsection{Conformally coupled 5-point}
We provide some details on how 
to compute the 5-point wavefunction coefficient in conformally coupled $\phi^3$ theory 
given in section \ref{5ptcc}. 
Using the propagators given in section \ref{sec:WF}, we have the following expression for a particular channel of the wavefunction coefficient $\Psi_5$:
\begin{eqn}
\Psi_5 &= \intsinf \frac{dz_1 dz_2 dz_3}{z_1 z_2 z_3} e^{- k_{12} z_1} e^{- k_{5} z_2} e^{- k_{34} z_3} \\
&\quad \times\intinf \frac{dp_1}{p_1^2 + y_{12}^2} \sin(p_1 z_2) \sin(p_1 z_2) \intinf \frac{dp_2}{p_2^2 + y_{34}^2} \sin(p_2 z_2) \sin(p_2 z_3) .
\end{eqn}
Using the $s$-integral notation we can absorb the $\frac{1}{z}$ factors into the exponentials and obtain 
\begin{eqn}
\Psi_5 &= \intsinf ds_1 ds_2 ds_3 \intsinf dz_1 dz_2 dz_3 e^{- (k_{12} + s_1) z_1} e^{- (k_{5} + s_2) z_2} e^{- (k_{34} + s_3) z_3} \\
&\quad \times\intinf \frac{dp_1}{p_1^2 + y_{12}^2} \sin(p_1 z_2) \sin(p_1 z_2) \intinf \frac{dp_2}{p_2^2 + y_{34}^2} \sin(p_2 z_2) \sin(p_2 z_3) .
\end{eqn}
The $z$ integrals can then be performed easily and give rational functions, 
\begin{eqn}
&\Psi_5 = \intsinf ds_1 ds_2 ds_3 \frac{1}{k_{12345} + s_1 + s_2 + s_3} \frac{1}{y_{12} + y_{34} + k_5 + s_2} \\
&\times \Bigg[ \frac{1}{y_{12} + k_{345} + s_2 + s_3} \frac{1}{k_{12} + y_{12} + s_1}\frac{1}{y_{34} + k_{34} + s_3}  + (k_{12} \leftrightarrow k_{34}, s_1 \leftrightarrow s_3, y_{12} \leftrightarrow y_{34})
\Bigg].
\end{eqn}
The $s$-integrals are now expressed in a form where they can be compared with well-known recursion relations for Polylogs and each $s$-integral increases the transcendentality by 1. Hence the final answer for this computation is a Polylog of weight 3 as also found in \cite{Hillman:2019wgh} using symbols. In equation \eqref{psi5Li3} we have only quoted the terms with Li$_3$ and the rest of the terms can be found in equation 5.8 of \cite{Hillman:2019wgh}.  

\section{Some loop-level correlators}\label{app:explicit}
In this appendix we demonstrate how the dressing rules allow one to explicitly evaluate some loop level integrals and also shed light on the simpler analytic structure of the in-in correlator. 
\subsection{Bubble}
We evaluate the four-point function $\braket{\phi(\vec k_1) \cdots \phi(\vec k_4)}$ in the $z^2 \phi^2 \chi^2$ theory where $\phi$ is conformally coupled and $\chi$ is massless. The four point function of $\phi$ in this theory has massless fields flowing in the loop and hence captures the non-trivial nature of the of the fully massless $\phi^4$ theory as well. This relation is made explicit by using the ``Weight shifting operators developed in \cite{Chowdhury:2024snc}.
We therefore evaluate the four point function $\braket{\phi(\vec k_1) \cdots \phi(\vec k_4)}$ in the $z^2 \phi^2 \chi^2$ theory using a dressing representation similar to the one in section \ref{sec:Dressing}. This leads to the following integral, 
\begin{eqn}
&\braket{\phi(\vec k_1) \cdots \phi(\vec k_4)}_{z^2 \phi^2 \chi^2} \\
&=  \frac{4k_{12} k_{34}}{k_1k_2k_3k_4} \int d^3 l \intinf \frac{dp_1 dp_2}{(p_1^2 + l^2) \big( p_2^2 + (\vec l + \vec y_{12})^2 \big)} \\
&\times \Bigg[ \frac{1}{ \big( k_{12}^2 + (p_1 + p_2)^2 \big)^2} \left( \frac{3(p_1 + p_2)^2 - k_{12}^2}{(p_1 + p_2)^2 + k_{12}^2} + \frac{3(p_1 + p_2)^2 + k_{12}^2}{2 p_1 p_2 } \right) \times (k_{12} \leftrightarrow k_{34}) \Bigg]
\end{eqn}
where we have explicitly replaced $L^0$ with $p_1$ in order to perform the integrals.

The final result after performing the $p_1, p_2$ and loop integrals (in dimensional regularization) can be expressed in the following form,
\begin{eqn}
\braket{\phi(\vec k_1) \cdots \phi(\vec k_4)}_{z^2 \phi^2 \chi^2}  = \p_{k_{12}}^2 \p_{k_{34}}^2 \braket{\phi(\vec k_1) \cdots \phi(\vec k_4)}_{\phi^4}  + \mbox{rational terms}
\end{eqn}
where the correlator  $\braket{\phi(\vec k_1) \cdots \phi(\vec k_4)}_{\phi^4}$ denotes the 4-point bubble diagram in $\phi^4$ theory and is evaluated in many places, for example see equation (5.48) of  \cite{Chowdhury:2023arc}. The logarithmic dependence on the momenta and the UV divergence comes from the this term alone and the rational terms are enumerated below,
\begin{eqn}
 \mbox{rational terms}&= \frac{\pi ^2}{4 k^2 k_{12} k_{34}\left(y_{12}+k_{12}\right)^2  \left(y_{12}+k_{34}\right)^2 k_{1234}^3}\\ 
&\times \Big\{k_{12} k_{34} k_{1234}^3  -2 y_{12}^5-4 k_{1234} y_{12}^4-2 k_{12} k_{34} y_{12}^3+4 k_{1234}^3 y_{12}^2+2 k_{1234}^4 y_{12}\Big\}~.
\end{eqn}
This hints at yet another simplicity for the in-in correlator as it shows that the non-rational terms are completely determined from the conformally coupled piece. This is in contrast to the wavefunction coefficient, where one obtains non-rational terms for the massless bubble from pieces not present in the conformally coupled bubble diagram (see appendix C of \cite{Chowdhury:2024snc} for an explicit example). 

\subsubsection{Restoring Conformal Invariance}\label{confinvreg}
In the example above we have evaluated the integral using a regularization scheme that breaks conformal invariance. We provide a simple modification of the integral below that allows one to restore scale invariance for $\braket{\phi(\vec k_1) \cdots \phi(\vec k_4)}_{\phi^4}$. Recall that from the dressing rule in sectin \ref{sec:Dressing} we obtain the following integral representation 
\begin{eqn}
  \braket{\phi(\vec k_1) \cdots \phi(\vec k_4)}_{\phi^4} = k_{12} k_{34} \intinf \frac{dp}{(p^2 + k_{12}^2)(p^2 + k_{34}^2)} \int \frac{d^4 L}{L^2(L + K)^2}
\end{eqn}
Since the loop integral is UV divergent it has to be regulated. For this example we use the propagators given in \eqref{analyticprop1} in order to obtain an integral that preserves conformal invariance. Under this regularization scheme we obtain 
\begin{eqn}\label{bubbleanareg}
  &\braket{\phi(\vec k_1) \cdots \phi(\vec k_4)}^{\rm an. reg}_{\phi^4} =\frac{\Gamma(1-\e)^2}{4} \\
  &\times\intinf dp \big( (k_{12} + i p)^{\e - 1} + (k_{12} - i p)^{\e - 1} \big)\big( (k_{34} + i p)^{\e - 1} + (k_{34} - i p)^{\e - 1} \big)\int \frac{d^4 L}{\big(L^2(L + K)^2\big)^{1 + \e}}
\end{eqn}
where we have also modified the powers in the denominators of the propagators in the loop integral according to standard convention of analytic regularization. Performing this integral results in equation 5.48 of ref.\ \cite{Chowdhury:2023arc}.

\subsection{Triangle}\label{sec:tria}
The integral representation for the in-in correlator of the triangle diagram with conformally coupled scalars has been obtained in section 4.3 of \cite{Chowdhury:2023arc}. For simplicity we work with the six-point function in $\phi^4$ theory. Using the dressing rules given in the previous section we obtain,
 \begin{eqn}
\braket{\phi(\vec k_1) \cdots \phi(\vec k_6)} = \intinf \frac{dp dp'}{(p^2 + k_{1}^2) (p'^2 + k_{2}^2) \big( (p + p')^2 + k_3^2 \big)}  \int \frac{d^4 L}{L^2 (L + P_1)^2 (L + P_2)^2}
 \end{eqn}
 where 
 \begin{eqn}
 L^\mu &= (p_3, \vec l), \qquad
 P_1^\mu = (p_1 - p_3, \vec k_{12}), \qquad
  P_2^\mu = (p_2 - p_3, - \vec k_{56}). 
 \end{eqn}
 and 
 \begin{eqn}
 p &= p_1 - p_3, \qquad
 p' = p_2 - p_1~.
 \end{eqn}
 While we do not obtain an analytic expression for the final integral it is possible to comment on the transcendentality of the integral by analyzing the form of the loop integrand. We note that by integrating out $p, p'$ and $L^0$ we obtain the 3-dimensional loop integral which has the following poles\footnote{The expression is evaluated explicitly in equation B.9 of \cite{Chowdhury:2023arc}.}
 \begin{eqn}\label{in-in-Tria1}
\frac{1}{k_{123}} \int d^3 l \frac{1}{y_{12} (k_{12} + y_{12} + y_{23})(k_{34} + y_{23} + y_{31}) (k_{56} + y_{31} + y_{12}) (k_{1234} + y_{12} + y_{23})}
 \end{eqn}
where $y_{12} = |\vec l|, \ y_{23} = |\vec l + \vec k_{12}|, \ y_{31} = |\vec l - \vec k_{56}|$~. Such integrals have been analyzed in \cite{Benincasa:2024ptf} by studying the differential equations these integrals satisfy. Through this analysis, the class of integrals in \eqref{in-in-Tria1} is expected to be a Polylogarithmic function of external momenta. This is in contrast to the wave function coefficient which is argued to be an Elliptic Polylog \cite{Benincasa:2024ptf}\footnote{While the final integral has not been evaluated analytically, it can be evaluated order by order in the squeezed limit and can be written as an infinite sum of Polylogs (see \cite{Chowdhury:2023khl} for a discussion of this).}. The difference can be traced back to the singularities of the loop integrand. The wave function coefficient and the in-in correlators contain an overlapping set of poles with one main difference. Instead of the pole $\frac{1}{y_{12}}$  that appears in the in-in correlator \eqref{in-in-Tria1}, the wave function coefficient contains $\frac{1}{k_{1\cdots 6} + 2 y_1}$. Due to integration measure $d^3 l$, the pole $\frac{1}{y_{12}}$ disappears and therefore results in a simpler structure for the integral. A similar difference also occurs for the case of the bubble diagram\footnote{The integrands for the in-in correlator and the wave function coefficient for the bubble are explicitly given in equation (4.21) of \cite{Chowdhury:2023arc} and equation (3.23) of \cite{Chowdhury:2023khl} respectively. } and is responsible for the wave function depending on the momenta as Li$_2$ in comparison to the $\log$ dependence of the in-in correlator.

\section{Dressing rules for the wavefunction}\label{app:WFdress}
The dressing rules for in-in correlators are presented in section \ref{sec:Dressing} and it is shown how they can be obtained by starting with Feynman diagram for a scattering amplitude in flat space and dressing it with appropriate factors. It is a natural question to ask if there exists any such rule for wavefunction coefficients (or by an analytical continuation, AdS correlators). In this Appendix we provide similar rules for wavefunction coefficients in dS and also point out an important difference between this structure and that of in-in correlators. 

These rules make use of the spectral representation of the propagator given in appendix \ref{app:Green}. We first illustrate the derivation with an example of the bubble diagram for conformally coupled scalars and then provide a summary of the rules and provide further examples.

\subsection{Derivation}
Consider a bubble diagram for $\phi^4$ theory where $\phi$ is a conformally coupled scalar. This is explicitly given as 
\begin{eqn}
\begin{tikzpicture}[baseline]
\draw[very thick] (0, 0) circle (2);
\draw ({2*cos(150)}, {2*sin(150)}) -- (-1, 0);
\draw ({2*cos(-150)}, {2*sin(-150)}) -- (-1, 0);
\draw ({2*cos(30)}, {2*sin(30)}) -- (1, 0);
\draw ({2*cos(-30)}, {2*sin(-30)}) -- (1, 0);
\draw (0, 0) circle (1);

\node at ({2.25*cos(-150)}, {2.25*sin(-150)}) {$1$};
\node at ({2.25*cos(150)}, {2.25*sin(150)}) {$2$};
\node at ({2.25*cos(30)}, {2.25*sin(30)}) {$3$};
\node at ({2.25*cos(-30)}, {2.25*sin(-30)}) {$4$};
\end{tikzpicture}
= \int d^3 l \intsinf \frac{dz_1}{z_1^2} \frac{dz_2}{z_2^2} e^{- k_{12} z_1} e^{- k_{34} z_2} G_{1/2}(z_1, z_2, \vec l) G_{1/2}(z_1, z_2, \vec l + \vec k)~.
\end{eqn}
By explicitly substituting the propagators in the spectral representation (see appendix \ref{app:specrep}) we obtain
\begin{eqn}\label{eq:WFbubble1}
\scalebox{0.4}{\begin{tikzpicture}[baseline]
\draw[very thick] (0, 0) circle (2);
\draw ({2*cos(150)}, {2*sin(150)}) -- (-1, 0);
\draw ({2*cos(-150)}, {2*sin(-150)}) -- (-1, 0);
\draw ({2*cos(30)}, {2*sin(30)}) -- (1, 0);
\draw ({2*cos(-30)}, {2*sin(-30)}) -- (1, 0);
\draw (0, 0) circle (1);

\node at ({2.25*cos(-150)}, {2.25*sin(-150)}) {$1$};
\node at ({2.25*cos(150)}, {2.25*sin(150)}) {$2$};
\node at ({2.25*cos(30)}, {2.25*sin(30)}) {$3$};
\node at ({2.25*cos(-30)}, {2.25*sin(-30)}) {$4$};
\end{tikzpicture}}
= \int d^3 l \intsinf dz_1 dz_2 e^{- k_{12} z_1} e^{- k_{34} z_2}\intinf \frac{dp_1 \sin(p_1 z_1) \sin(p_1 z_2) }{p_1^2 + l^2} \intinf \frac{dp_2 \sin(p_2 z_1) \sin(p_2 z_2)}{p_2^2 + |\vec l + \vec k|^2} 
\end{eqn}
The dressing rule for the wavefunction is obtained by first performing the $z$ integral. At each vertex we have an integral of the form
\begin{eqn}\label{eq:WFdress0}
\intsinf dz e^{- k z} \sin(p_1 z) \sin(p_2 z) = \frac{k}{2} \left\{\frac{1}{k^2 + (p_1 - p_2)^2} - \frac{1}{k^2 + (p_1 + p_2)^2} \right\} \equiv \mathcal P_{1/2}(k; p_1, p_2)~.
\end{eqn}
Hence by interchanging the order of integrals and performing the $z$ integrals first we obtain the following form for \eqref{eq:WFbubble1}:
\begin{eqn}
\scalebox{0.4}{\begin{tikzpicture}[baseline]
\draw[very thick] (0, 0) circle (2);
\draw ({2*cos(150)}, {2*sin(150)}) -- (-1, 0);
\draw ({2*cos(-150)}, {2*sin(-150)}) -- (-1, 0);
\draw ({2*cos(30)}, {2*sin(30)}) -- (1, 0);
\draw ({2*cos(-30)}, {2*sin(-30)}) -- (1, 0);
\draw (0, 0) circle (1);

\node at ({2.25*cos(-150)}, {2.25*sin(-150)}) {$1$};
\node at ({2.25*cos(150)}, {2.25*sin(150)}) {$2$};
\node at ({2.25*cos(30)}, {2.25*sin(30)}) {$3$};
\node at ({2.25*cos(-30)}, {2.25*sin(-30)}) {$4$};
\end{tikzpicture}}
= \int d^3 l \intinf dp_1 dp_2 \frac{1}{(p_1^2 + l^2)(p_2^2 + (\vec l + \vec k)^2)}  \mathcal P_{1/2}(k_{12}; p_1, p_2)  \mathcal P_{1/2}(k_{34}; p_2, p_1)~.
\end{eqn}
We now define the 4-vectors $L^\mu = (p_1, \vec l)$ and $K^\mu = (p_2 - p_1, \vec k) $ such that $d^4 L = dp_1 d^3 l$ which allows to write the above equation as 
\begin{eqn}\label{eq:WFdress1}
\scalebox{0.4}{\begin{tikzpicture}[baseline]
\draw[very thick] (0, 0) circle (2);
\draw ({2*cos(150)}, {2*sin(150)}) -- (-1, 0);
\draw ({2*cos(-150)}, {2*sin(-150)}) -- (-1, 0);
\draw ({2*cos(30)}, {2*sin(30)}) -- (1, 0);
\draw ({2*cos(-30)}, {2*sin(-30)}) -- (1, 0);
\draw (0, 0) circle (1);

\node at ({2.25*cos(-150)}, {2.25*sin(-150)}) {$1$};
\node at ({2.25*cos(150)}, {2.25*sin(150)}) {$2$};
\node at ({2.25*cos(30)}, {2.25*sin(30)}) {$3$};
\node at ({2.25*cos(-30)}, {2.25*sin(-30)}) {$4$};
\end{tikzpicture}}
= \intinf dp_2 \int d^4 L \frac{1}{L^2 (L + K)^2} \mathcal P_{1/2}(k_{12}; L^0, p_2 - L^0)  \mathcal P_{1/2}(k_{34}; p_2 - L^0, L^0)~.
\end{eqn}
This clearly resembles a dressed-up Feynman propagator in flat space, i.e., 
\begin{eqn}\label{eq:WFdress2}
\begin{tikzpicture}[baseline]
\draw[very thick] (0, 0) circle (2);
\draw ({2*cos(150)}, {2*sin(150)}) -- (-1, 0);
\draw ({2*cos(-150)}, {2*sin(-150)}) -- (-1, 0);
\draw ({2*cos(30)}, {2*sin(30)}) -- (1, 0);
\draw ({2*cos(-30)}, {2*sin(-30)}) -- (1, 0);
\draw (0, 0) circle (1);

\node at ({2.25*cos(-150)}, {2.25*sin(-150)}) {$1$};
\node at ({2.25*cos(150)}, {2.25*sin(150)}) {$2$};
\node at ({2.25*cos(30)}, {2.25*sin(30)}) {$3$};
\node at ({2.25*cos(-30)}, {2.25*sin(-30)}) {$4$};
\end{tikzpicture} 
= 
\begin{tikzpicture}[baseline]
\draw ({2*cos(150)}, {2*sin(150)}) -- (-1, 0);
\draw ({2*cos(-150)}, {2*sin(-150)}) -- (-1, 0);
\draw ({2*cos(30)}, {2*sin(30)}) -- (1, 0);
\draw ({2*cos(-30)}, {2*sin(-30)}) -- (1, 0);
\draw (0, 0) circle (1);

\node at ({2.25*cos(-150)}, {2.25*sin(-150)}) {$1$};
\node at ({2.25*cos(150)}, {2.25*sin(150)}) {$2$};
\node at ({2.25*cos(30)}, {2.25*sin(30)}) {$3$};
\node at ({2.25*cos(-30)}, {2.25*sin(-30)}) {$4$};

\draw[thick, purple, dotted] (-1, 0) -- (0, -2);
\draw[thick, purple, dotted] (1, 0) -- (0, -2);

\node at (-0.75, -1.25) {$\frac12$}; 
\node at (0.75, -1.25) {$\frac12$}; 
\end{tikzpicture}
\end{eqn}
where the purple lines with the label $\frac{1}{2}$ denote the factors of $\mathcal P_{1/2}$ appearing in equation \eqref{eq:WFdress1}. 

While these dressing rules are simple to state, there are important differences with the dressing rules obtained for in-in correlators in the section \ref{sec:Dressing}. This example already shows that the dressing rule for a conformally coupled field introduces non-Lorentz invariant objects in the loop integrand. This is in contrast to the dressing rules for the in-in correlators of the conformally coupled scalars (or for the examples studied in \cite{Chowdhury:2023arc}) for which the loop integrals are always over Lorentz-invariant objects. Since the dressing factors for the in-in correlators of massless scalars contain non-Lorentz invariant numerators (see section \ref{sec:dressmassless}), it might be suggestive that the dressing rules of the wavefunction are closer to those. However, the dressing factors of the wave function also contain non-Lorentz invariant denominators as shown in \eqref{eq:WFdress0}. This is a general feature for the wave function and also holds for massless scalars as will be shown in the next section. Due to the presence of non-Lorentz invariant poles, wavefunction coefficients generally contain more singularities than in-in correlators.

\subsection{Summary}
We provide a summary of the dressing rules for conformally coupled fields with $\phi^4$ interaction and massless fields with $\phi^4$ interaction. The derivation for the latter follows the conformally coupled case where one integrates over a product of $J_{3/2}$ instead of $J_{1/2}$ as done in \eqref{eq:WFdress0}. In each case below, the dressed propagators are attached to each vertex

\subsubsection*{conformally coupled $\phi^4$}
For vertices with 2 external legs and 2 internal legs, we have the following dressing factor\footnote{The case of a 3 external legs is obtained from the formula \eqref{eq:WFconf22} by setting $p_2 = p_1 = \frac{p}2$. }
\begin{eqn}\label{eq:WFconf22}
\mathcal P_{1/2}(k; p_1, p_2) =  \frac{k}{2} \left\{\frac{1}{k^2 + (p_1 - p_2)^2} - \frac{1}{k^2 + (p_1 + p_2)^2} \right\}
\end{eqn}

The dressing factor for graphs with 3 external legs and one internal leg (eg: a sunset graph) is given as, 
\begin{eqn}
&\mathcal P_{1/2}(k; p_1, p_2, p_3) \\
&= \frac{1}{4} \Bigg\{ \frac{p_1+p_2-p_3}{k^2+\left(p_1+p_2-p_3\right){}^2}+\frac{p_1-p_2+p_3}{k^2+\left(p_1-p_2+p_3\right){}^2}+\frac{-p_1+p_2+p_3}{k^2+\left(-p_1+p_2+p_3\right){}^2}\\
&\qquad\qquad -\frac{p_1+p_2+p_3}{k^2+\left(p_1+p_2+p_3\right){}^2} \Bigg\}
\end{eqn}

\subsubsection*{massless $\phi^4$}
For the case of 3 external legs attached to one internal leg we have
\begin{eqn}
&\mathcal P_{3/2}(k; p) = \frac{2p^3}{(p^2 + k^2)^3}
\end{eqn}
The expression quickly gets more complicated but is still manageable for the case of 2 external legs attached to one internal leg,
\begin{eqn}
&\mathcal P_{3/2}(k; p_1, p_2) = 
\frac{16 k p_1^3 p_2^3 \left(5 k^4+2 k^2 \left(p_1^2+p_2^2\right)-3 \left(p_1^2-p_2^2\right){}^2\right)}{\left(k^2+\left(p_1-p_2\right){}^2\right){}^3 \left(k^2+\left(p_1+p_2\right){}^2\right){}^3}
\end{eqn}
For this case, we obtain cubic terms in the denominator. This shows how the dressing rules for the wavefunction get increasingly complicated as we consider fields with increasing conformal dimension. Hence these rules demonstrate the simplicity of the dressing rules for the in-in correlator.

\subsection{Examples}
We provide some simple examples below demonstrating the use of the dressing rules given above. The bubble diagram containing conformally coupled scalars was already studied previously. This can be easily generalized to the triangle diagram as shown below, 
\begin{eqn}
\begin{tikzpicture}[baseline]
\draw[very thick] (0, 0) circle (2);
\draw ({2*cos(200)}, {2*sin(200)}) -- ({1*cos(210)}, {1*sin(210)});
\draw ({2*cos(220)}, {2*sin(220)}) -- ({1*cos(210)}, {1*sin(210)});

\draw ({2*cos(-20)}, {2*sin(-20)}) -- ({1*cos(-30)}, {1*sin(-30)});
\draw ({2*cos(-40)}, {2*sin(-40)}) -- ({1*cos(-30)}, {1*sin(-30)});

\draw ({2*cos(80)}, {2*sin(80)}) -- ({1*cos(90)}, {1*sin(90)});
\draw ({2*cos(100)}, {2*sin(100)}) -- ({1*cos(90)}, {1*sin(90)});

\draw (0, 0) circle (1);

\node at ({2.25*cos(220)}, {2.25*sin(220)}) {$1$};
\node at ({2.25*cos(200)}, {2.25*sin(200)}) {$2$};

\node at ({2.25*cos(100)}, {2.25*sin(100)}) {$3$};
\node at ({2.25*cos(80)}, {2.25*sin(80)}) {$4$};

\node at ({2.25*cos(-20)}, {2.25*sin(-20)}) {$5$};
\node at ({2.25*cos(-40)}, {2.25*sin(-40)}) {$6$};

\end{tikzpicture}
= 
\begin{tikzpicture}[baseline]
\draw ({2*cos(200)}, {2*sin(200)}) -- ({1*cos(210)}, {1*sin(210)});
\draw ({2*cos(220)}, {2*sin(220)}) -- ({1*cos(210)}, {1*sin(210)});

\draw ({2*cos(-20)}, {2*sin(-20)}) -- ({1*cos(-30)}, {1*sin(-30)});
\draw ({2*cos(-40)}, {2*sin(-40)}) -- ({1*cos(-30)}, {1*sin(-30)});

\draw ({2*cos(80)}, {2*sin(80)}) -- ({1*cos(90)}, {1*sin(90)});
\draw ({2*cos(100)}, {2*sin(100)}) -- ({1*cos(90)}, {1*sin(90)});

\draw (0, 0) circle (1);

\node at ({2.25*cos(220)}, {2.25*sin(220)}) {$1$};
\node at ({2.25*cos(200)}, {2.25*sin(200)}) {$2$};

\node at ({2.25*cos(100)}, {2.25*sin(100)}) {$3$};
\node at ({2.25*cos(80)}, {2.25*sin(80)}) {$4$};

\node at ({2.25*cos(-20)}, {2.25*sin(-20)}) {$5$};
\node at ({2.25*cos(-40)}, {2.25*sin(-40)}) {$6$};

\draw[thick, dotted, purple] ({1*cos(90)}, {1*sin(90)}) -- (0, -2);
\draw[thick, dotted, purple] ({1*cos(-20)}, {1*sin(-20)}) -- (0, -2);
\draw[thick, dotted, purple] ({1*cos(220)}, {1*sin(220)}) -- (0, -2);
\end{tikzpicture}
\end{eqn}
By using the propagator \eqref{eq:WFconf22}, the explicit expression for this is given as 
\begin{eqn}
\scalebox{0.5}{\begin{tikzpicture}[baseline]
\draw[very thick] (0, 0) circle (2);
\draw ({2*cos(200)}, {2*sin(200)}) -- ({1*cos(210)}, {1*sin(210)});
\draw ({2*cos(220)}, {2*sin(220)}) -- ({1*cos(210)}, {1*sin(210)});

\draw ({2*cos(-20)}, {2*sin(-20)}) -- ({1*cos(-30)}, {1*sin(-30)});
\draw ({2*cos(-40)}, {2*sin(-40)}) -- ({1*cos(-30)}, {1*sin(-30)});

\draw ({2*cos(80)}, {2*sin(80)}) -- ({1*cos(90)}, {1*sin(90)});
\draw ({2*cos(100)}, {2*sin(100)}) -- ({1*cos(90)}, {1*sin(90)});

\draw (0, 0) circle (1);

\node at ({2.25*cos(220)}, {2.25*sin(220)}) {$1$};
\node at ({2.25*cos(200)}, {2.25*sin(200)}) {$2$};

\node at ({2.25*cos(100)}, {2.25*sin(100)}) {$3$};
\node at ({2.25*cos(80)}, {2.25*sin(80)}) {$4$};

\node at ({2.25*cos(-20)}, {2.25*sin(-20)}) {$5$};
\node at ({2.25*cos(-40)}, {2.25*sin(-40)}) {$6$};

\end{tikzpicture}}
&= \intinf dp_2 dp_3 \int \frac{d^4 L}{L^2 (L + K_1)^2 (L + K_2)^2} \\
&\hspace{0.4cm}\times \mathcal P_{1/2}(k_{12}; L^0, p_2 - L^0) \mathcal P_{1/2}(k_{34}; p_2 - L^0, p_3 - L^0) \mathcal P_{1/2}(k_{56}; p_3 - L^0, L^0)~.
\end{eqn}
By integrating out $L^0, p_2, p_3$ one obtains the $d^3l$-integrand for the triangle diagram given in \cite{Chowdhury:2023khl, Benincasa:2024ptf}. The implication of the pole structure of this integrand is discussed in \cite{Benincasa:2024ptf} and briefly reviewed in section \ref{sec:tria}.

\bibliography{references}
\bibliographystyle{JHEP}
\end{document}